# Towards the convergent therapeutic potential of GPCRs in autism spectrum disorders

Anil ANNAMNEEDI[1,2#], Caroline GORA[1], Ana DUDAS[1], Xavier LERAY[1], Véronique BOZON[1], Pascale CREPIEUX[1], Lucie P. PELLISSIER[1#]

[1] Team biology of GPCR Signaling systems (BIOS), CNRS, IFCE, INRAE, Université de Tours, PRC, F-37380, Nouzilly, France.
[2] LE STUDIUM Loire Valley Institute for Advanced Studies, 45000, Orléans, France.

# Corresponding authors:
Anil ANNAMNEEDI, PhD. Email: anil.annamneedi@inrae.fr; orcid.org/0000-0002-6743-8627
Lucie P. PELLISSIER, PhD. Email: lucie.pellissier@inrae.fr; orcid.org/0000-0001-7085-3242

**ABSTRACT**

Changes in genetic and/or environmental factors to developing neural circuits and subsequent synaptic functions are known to be a causative underlying the varied socio-emotional behavioural patterns associated with autism spectrum disorders (ASD). Seven transmembrane G protein-coupled receptors (GPCRs) comprising the largest family of cell-surface receptors, mediate the transfer of extracellular signals to downstream cellular responses. Disruption of GPCR and their signalling have been implicated as a convergent pathologic mechanism of ASD. Here, we aim to review the literature about the 23 GPCRs that are genetically associated to ASD pathology according to Simons Foundation Autism Research Initiative (SFARI) database such as oxytocin (OXTR) and vasopressin ($V_{1A}$, $V_{1B}$) receptors, metabotropic glutamate ($mGlu_5$, $mGlu_7$) and gamma-aminobutyric acid ($GABA_B$) receptors, dopamine ($D_1$, $D_2$), serotoninergic ($5\text{-}HT_{1B}$ and additionally included the $5\text{-}HT_{2A}$, $5\text{-}HT_7$ receptors for their strong relevance to ASD), adrenergic ($\beta_2$) and cholinergic ($M_3$) receptors, adenosine ($A_{2A}$, $A_3$) receptors, angiotensin ($AT_2$) receptors, cannabinoid ($CB_1$) receptors, chemokine ($CX_3CR1$) receptors, orphan (GPR37, GPR85) and olfactory (OR1C1, OR2M4, OR2T10, OR52M1) receptors. We discussed the genetic variants, relation to core ASD behavioural deficits and update on pharmacological compounds targeting these 23 GPCRs. Of these OTR, $V_{1A}$, $mGlu_5$, $D_2$, $5\text{-}HT_{2A}$, $CB_1$, and GPR37 serve as the best therapeutic targets and have potential towards core domains of ASD pathology. With a functional crosstalk between different GPCRs and converging pharmacological responses, there is an urge to develop novel therapeutic strategies based on multiple GPCRs to reduce the socio-economic burden associated with ASD and we strongly emphasize the need to prioritize the increased clinical trials targeting the multiple GPCRs.

## INTRODUCTION

Autism spectrum disorder (ASD) is a neurodevelopmental disorder diagnosed on core clinical symptoms defined by the Diagnostic and Statistical Manual of Mental Disorders, Fifth Edition (DSM-V), social interaction and communication deficits and stereotyped, restrained or compulsive behaviours. ASD often associates with comorbid symptoms, such as anxiety, epilepsy, sleep disturbances, motor coordination impairment, gastro-intestinal disorders, intellectual disability. Whereas its worldwide prevalence is around 1/100 child birth, its aetiology remains not fully understood as 70% of cases remains sporadic, highlighting the polygenic complexity of the disease. Currently, there is no pharmacological drug treatment for the core symptoms of autism. Clinical trials did not lead to successful outcome due to important placebo effect, lack of efficacy and potentially the large aetiological diversity of patients. Therefore, two main gaps remain in autism research: the identification of convergent and robust therapeutical targets and the development of potent drugs to succeed in clinical trials probably only in sub-class of patients, determined based on the complex multigenic profile, to achieve successful clinical trials. New advances need to overcome these challenges.
Important recent sequencing studies in large ASD cohorts robustly identified hundreds of *de novo* candidate genes that converge in two main categories 'gene expression regulation' or 'neuronal communication, signalling or plasticity' [1-3]. These genes are mostly expressed in developing excitatory and inhibitory cortical neurons of the human cortex [1]. Rather than a single mutation, accumulation of risk alleles and increase or decrease copy number variants may underly the pathological process among affected individuals [4]. More than a thousand of candidate genes are listed in the Simons Foundation Autism Research Initiative (SFARI) database (https://gene.sfari.org). Currently, one of the main research areas in ASD is to transform the large number of identified candidate and risk alleles in convergent neurobiological mechanism [4]. Genomic and transcriptomic data from patients converge to identify common intracellular signalling pathways, with enriched Wnt/β-catenin or Notch signalling in early embryonic development cluster and ERK, AKT, mTOR or Wnt/β-catenin signalling in postnatal development cluster [2, 3, 5].
Interestingly, G protein-coupled receptors (GPCRs) are key master regulators of these intracellular pathways, neuronal communication and plasticity and gene expression. In fact, numerous genes identified in ASD meta-analysis modules are either their neurotransmitter/ligands (Wnt), GPCRs themselves (5-$HT_{2A}$), their direct effectors (Protein Kinase Cβ), scaffolding partners (SHANK), downstream signalling pathways (ERK), or transcription factors (the cAMP response element binding protein CREB) [3]. In this review, we focus on the convergent potential of GPCRs in ASD and why this receptor family fulfil all criteria of promising therapeutical targets for ASD. Based their fine tune pharmacology and their diversity, GPCRs represent the greatest therapeutical options for ASD to lead successful clinical trials. Here, we will describe all the GPCR pathogenic (copy number, loss of function, missense, frameshift, stop gained) variants associated to ASD so far, but also emphasize that they are the most dysregulated genes in ASD, they are fine tune targets with innovative pharmaceutical agents and they control multiple signalling and biological processes that are convergent in this neurodevelopmental disorder.

### *1) GPCRs receptor family are therapeutical targets for ASD*

Canonical GPCRs display seven transmembrane (TM) helices and an extracellular domain composed of the N-terminus and three extracellular loops 1-3 (EL1-3) involved in ligand binding, conformational changes and activation transmission to the intracellular signalling domain (**Figure 1**). The intracellular domain, composed of the three intracellular loops 1-3 (IL1-3) and the C-terminus with an $8^{th}$ helix parallel to the plasma membrane, induces intracellular signalling molecular recruitment and activation (*e.g.,* heterotrimeric G protein and β-arrestin). Whereas their versatile nature was a major issue for a long time, many GPCR structures in inactive, intermediate and active conformations are now available with the development of cryo-microscopy. GPCRs are translated inside the membrane of the endoplasmic reticulum (ER), thus actively exported to the plasma membrane. Due to their molecular complexity, they are often prone to misfolding or lack of ER export, which may result in cell toxicity [6]. Most of GPCRs, splicing isoforms and their signalling molecules are expressed in the Central Nervous System (CNS) [7, 8], located at the plasma membrane of soma, dendrites and/or axons (in pre-, post- and/or peri-synaptic compartments) of neurons, but also in astrocytes, oligodendrocytes and microglia. Some GPCRs display widespread expression profiles throughout the brain (*e.g.,* glutamate or GABA metabotropic receptors or GPR85) and others have a discrete and brain structure or cell-type specific pattern (e.g., GPR37, GPR88, vasopressin $V1_B$ receptor) [9, 10]. Transcriptomic data from prefrontal cortex tissue showed that GPCRs are more dysregulated compared to other family genes in ASD and in other neurodevelopmental and psychiatric disorders [11]. At global level, around 200 GPCR genes have been shown to be potentially linked with autism and among them, 25% were shown to be dysregulated. Among them, 74% belong to the class A with adenosine *ADORA1* and adrenergic *ADRA1D* being the most down-regulated and 24% are orphan receptors. Among the downstream signalling partners, the most affected G protein are $G\alpha_i$ (also found in the SFARI gene list) and $G\alpha_{12/13}$ proteins. In 2018, Babu and colleagues' asses the natural genetic landscape of GPCRs targeted by a drug in nearly 70 000 individuals from the 1000 genomes project [12] and available in the GPCRdb database [13]. They identified missense or copy number variants that may influence ligand binding or signalling molecules recruitment. It included GPCR candidate genes in ASD or their other family members: vasopressin $V_{1B}$, dopamine $D_1$, $D_2$, $D_3$ and $D_5$, serotonin 5-$HT_{2A}$, 5-$HT_{2B}$, 5-$HT_{2C}$, 5-$HT_4$ and $\beta_1$- and $\beta_2$-adrenergic and $GABA_B$. This new area of research based on receptor bias is known as pharmacogenomics. Considering the major impact of GPCR signalling, a slight modification in a GPCR in combination with another ASD gene variants or GPCRs would lead to a neuronal pathogenic process. Application of pharmacogenomics remains an outstanding hypothesis to decipher the processes underlying this neurodevelopmental disorder. Despite these evidences, further studies are needed to fully understand the convergent role of GPCRs and which GPCR or GPCR-dependent signalling molecules are involved in ASD aetiology.

- **GPCRs display the most diversity of pharmacological agents**

GPCRs respond to various ligands ranging from photons, amino acids, peptides up to large glycosylated proteins (**Figure 1A**). With around 800 GPCRs, they are the main receptor family in the CNS and are subdivided into five classes [7]: the largest rhodopsin-like class A, the secretin/class B, the glutamate/class C, the Frizzled class, the adhesion class. Sensory GPCRs (e.g., olfactory, vision, taste and pheromone receptors), which accounts for most GPCR genes, are comprised mostly in class A and few in class C. Although they act as detector of scents in the olfactory epithelium, the function and expression of olfactory GPCRs in the brain are not well documented. Till date, hundreds of GPCRs (including olfactory receptors) remains without any identified ligand and are called orphan receptors.

In addition to their diverse natural ligands, drugs can modulate GPCR activity, with diverse pharmacological profiles (**Figure 1B**). Those drugs can either be chemical compounds, peptides, large autoantibodies and more recently, a new class has appeared, the biologicals (*e.g.,* antibody fragments) [14], which can be easily encoded and vectorized. Orthosteric agonists, inverse agonists and antagonists bind in the natural ligand binding pocket and respectively activate, inactivate the receptor and/or prevent the binding of the endogenous ligand. In contrast, through binding in allosteric sites, positive or negative allosteric modulators enhance or decrease the GPCR activity only in presence of its natural ligand. Finally, GPCRs form cell-specific homo- or hetero-oligomers depending of the GPCR composition and subcellular localisation in a particular cell type. Each GPCR in the oligomer or expressed in the same cell may influence the pharmacology of the other GPCRs, opening a new area of the GPCR pharmacology. Despite their name and classification in subgroups according to their preferential G protein coupling (*e.g.,* $G\alpha_{s/olf}$, $G\alpha_{i/o}$, $G\alpha_{q/11}$ or $G\alpha_{12/13}$), they couple to several G proteins. In addition, they recruit $\beta$-arrestin, leading to specific intracellular signalling pathways or kinetics, such as ERK kinases and endocytosis. Therefore, different ligands may favour one coupling over another, which is called signalling bias (**Figure 1**). This pharmacological property of GPCR is of great interest for therapeutical applications, as one biased drug can control a wanted signalling cascade, preventing unwanted effects. Considering oligomerisation status, drug pharmacological profile, receptor or ligand signalling bias, GPCRs are the most specific and fine-tuned targets.

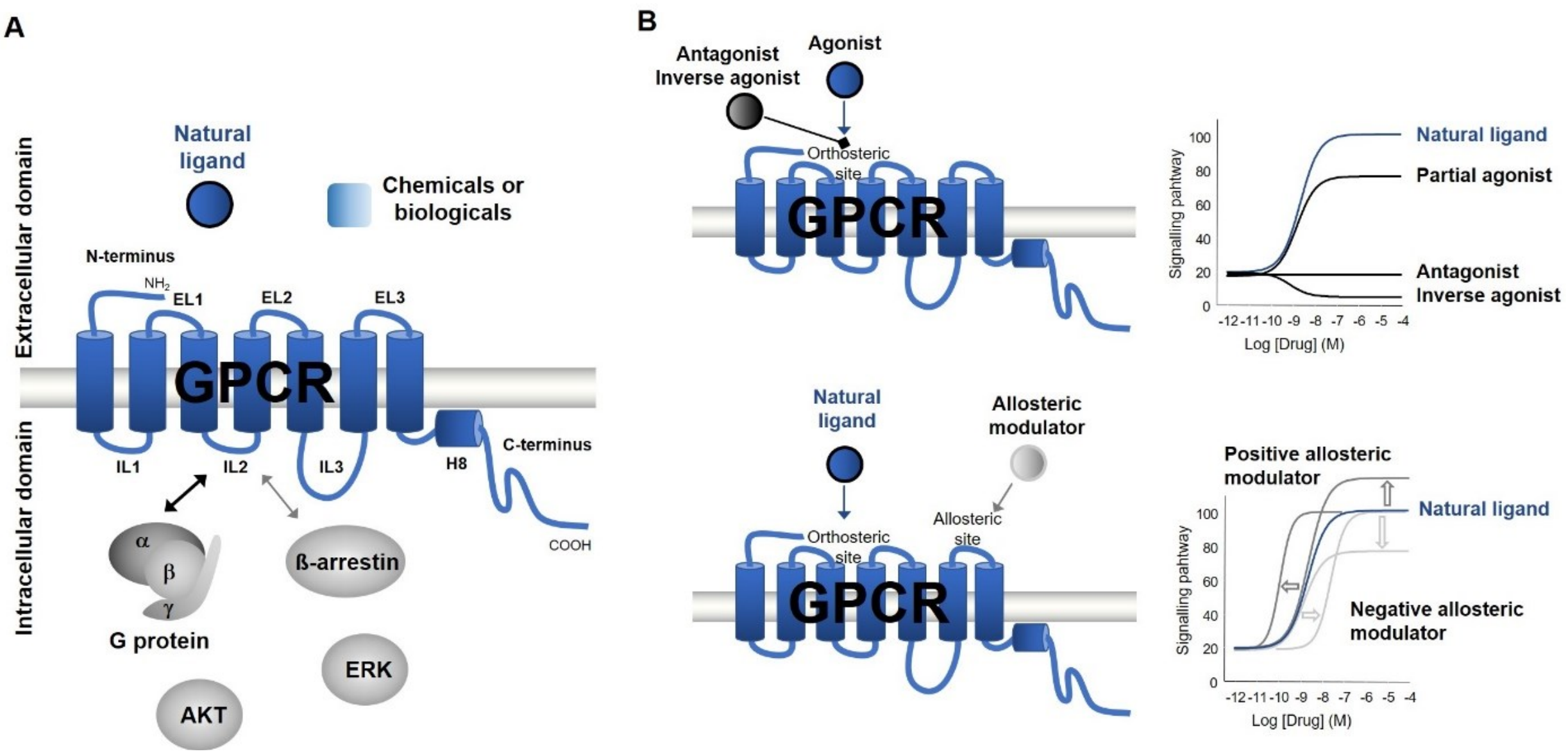

**Figure 1 GPCR signalling and pharmacology**

**A**) GPCR are composed of seven transmembrane domains connecting the extracellular domain (N-terminus, Extracellular loops (EL) 1-3) where various ligands bind the receptor (*e.g.,* natural ligands and chemicals/biologicals) to the intracellular domain (Intracellular loops (IL) 1-3, helix H8, C-terminus) that couple and recruit the direct signalling effectors, which activates the downstream signalling pathways (Akt, ERK) and cellular processes. A ligand named biased ligand has the particular pharmacological profile to activate one over the several signalling pathways for a given receptor. Here, we represented a G protein biased ligand that favour G protein coupling (black two headed arrow) over β-arrestin recruitment (grey two headed arrow). **B**) GPCRs display a rich pharmacopeia of ligands, with agonists, antagonists and inverse agonists that bind to the orthosteric binding site (e.g., the binding site of the natural ligand) to respectively activate, prevent the agonist binding and inactivate the receptor. In addition, another class of ligand bind allosteric sites that increase or decrease the efficacy or efficiency of the natural ligand or agonists, respectively called positive or negative allosteric modulators.

- **GPCRs are the most druggable targets for ASD**

With 800 genes in the human genome, GPCRs are the most abundant receptor family and more than half of the GPCRs are expressed in the brain [15, 16]. More than 30% of the drugs approved on the global market, or in clinical trials, targets a GPCR in various disorders [7, 17, 18]. Nevertheless, 20% of them are used for neurological conditions, including ASD. This small proportion will increase as hundreds of GPCRs in the CNS remains orphan. Although their function remains to be fully elucidated, they represent advantageous target in neurological disorders [10] and especially ASD. So far, no pharmaceutical agent reaches the market to improve primary symptoms of ASD. The therapeutical potential of GPCRs in ASD has started with class C $mGlu_5$ and $GABA_B$ GPCRs up to phase 2 clinical trials for monogenic Fragile X syndrome using a negative allosteric modulator and a selective agonist, respectively [19], in order to restore the glutamate/GABA unbalance. The lack of efficacy of these treatments might result from the wide expression of these receptors throughout the brain. Since then, various compounds targeting more discrete GPCRs has been tested. Oxytocin peptide and vasopressin $V_{1A}$ chemical antagonist administration failed to improve social abilities over placebo in phase 3 clinical trials whereas vasopressin peptide remains to be tested in larger cohort of ASD patients [20-22]. Regarding their prosocial effects, the oxytocin/vasopressin family remains interesting targets for the autism community, but further developments are needed to overpass placebo effects. Cannabidiol, cannabidivarin or endocannabinoid mix targeting CB1 receptors passed safety phase 1 clinical trials [23] (https://clinicaltrials.gov/). Drugs targeting serotonin 5-$HT_{1A}$ and 5-$HT_{2A}$ and dopamine $D_2$ and $D_3$ receptors are approved in the market for schizophrenia [17]. Considering the ASD-schizophrenia continuum, some of these drugs could be tested for potential improvement in ASD. Interestingly, the few GPCRs that have been tested so far are all candidate genes that are listed in the SFARI (**Tables 1-2**). Overall, the therapeutical potential of GPCRs has only began to be explored.

- **GPCRs are master regulators of intracellular signalling pathway**

Upon activation, GPCRs couple to heterotrimeric G protein ($\alpha$, $\beta$ and $\gamma$ subunits) and recruit $\beta$-arrestin. $G\alpha_{s/olf}$ protein activates adenylyl cyclases that hydrolyse ATP in cytoplasmic cAMP production, which is regulated by phosphodiesterases. cAMP activates exchange protein activated by cAMP (EPAC), calcium ion channels opening (CNGC) and the protein kinase A (PKA), which in turn phosphorylates all their downstream effectors (*e.g.,* ERK, PI3K/AKT and CREB, DARP32, ionotropic glutamatergic receptors). Conversely, $G\alpha_{i/o}$ inhibits adenylyl cyclases and cAMP production. *GNAI1* gene coding $G\alpha_{i1}$ is a high confidence gene (SFARI gene list), whereas *GNAS*, *GNB2*, *ADCY3* and *ADCY5* encoding $G\alpha_s$, $G\beta_2$ and adenylyl cyclase 3 and 5 are strong candidate genes. $G\alpha_{q/11}$ proteins activate phospholipase $C\beta$, which hydrolyses phosphatidylinositol-4,5-bisphosphate into diacylglycerol (DAG) and inositol-1,4,5-triphosphate ($IP_3$). Released $IP_3$ binds to their ryanodine receptors on the endoplasmic reticulum, which leads to calcium release from this cellular stock. Both calcium and DAG activates Protein Kinase C (PKC) and its effectors, such as Akt (also known as Protein kinase B, PKB), or Extracellular signal-regulated kinases (ERK). ERK1/2 also known as mitogen actived protein kinases (MAPK) and Akt are two convergent kinases reported in ASD [3, 5, 24]. $G\alpha_{12/13}$ activates Rho Guanine exchange Factor (GEF) and RhoA, which acts on the cytoskeleton promoting neurites formation. $G\beta\gamma$ proteins also participate in downstream signalling through their activation of G protein-coupled inwardly-rectifying potassium channels (GIRK) or other channels. Downstream intracellular signalling cascades lead to downstream cellular process, translation of specific mRNAs [25-28] and gene transcription. Little is known about their actual effect on translation or which specific transcription factor or genes are under the control of GPCRs, with the exception of the cAMP-responsive CREB, which allow expression for example of *BDNF*, *Fos* or *Jun* genes. However, signalling cascades, and scaffolding protein partners (*e.g.,* Shank1-3), ion channels and interaction and transactivation of tyrosine kinase receptors are more documented for some GPCRs. In order to estimate the full therapeutic potential of GPCRs as master regulator of downstream signalling targets, we used the last SFARI list of 1045 candidate genes (release on June, 6th 2022). Combining Kyoto Encyclopedia of Genes and Genomes (KEGG) gene ontology (biological process 'GPCR Signalling Pathway'), literature and our knowledge, we identified 23 GPCRs and 129 genes linked to GPCRs (**Table 1**, **Figure 2** and **Data and Software Availability** section).

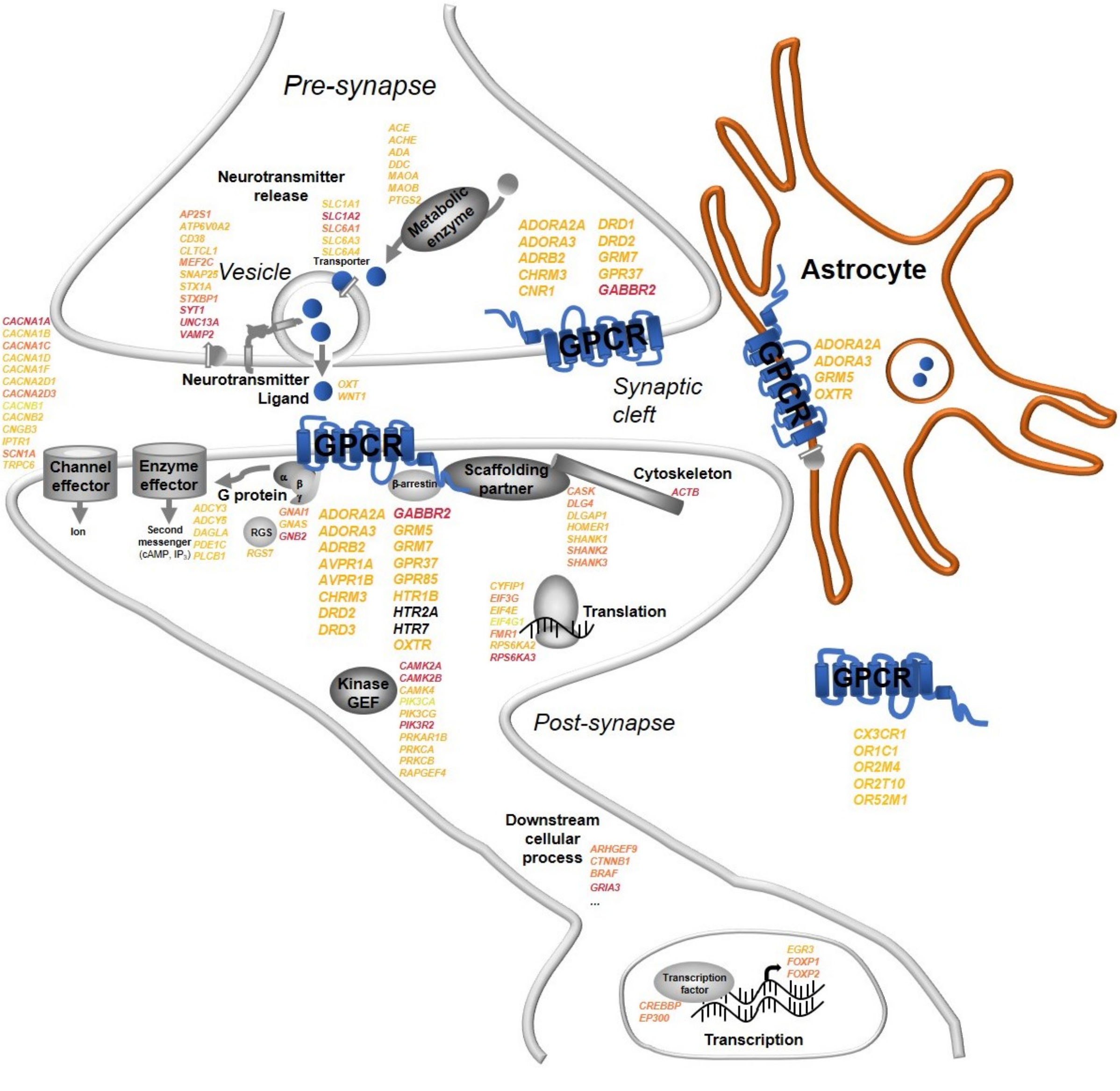

**Figure 2 Nearly 15% of SFARI genes participate in GPCR activity and signalling processes**
GPCR ligand are synthesised by metabolic enzymes, loaded by their transporters into synaptic vesicle that will fuse with the presynaptic membrane upon increase of intracellular calcium, leads to neurotransmitter release in the synaptic cleft. There, these neurotransmitters/ligands will either be recaptured by membrane transporters, be degraded, and/or bind and activate their cognate GPCR. Even in absence of ligand, GPCRs are present in preformed higher complexes with scaffolding partners and signalling partners. Upon activation, signalling partners activate enzyme and channel effectors to produce second messengers. These second messengers activate major kinases that will tune up or down downstream cellular processes, including signalling, translation and transcription. All the syndromic, category 1 (high confidence), 2 (strong candidate) and 3 (suggestive evidences) genes extracted from the gene list (**Table 1**) are in red, dark orange, light orange and green colours according to SFARI colour code. Further, the list of GPCRs not included in SFARI are colour coded in black. The GPCRs are mentioned according to their localization at pre, post-synaptic sides or at astrocyte.

Therefore, nearly 15% of the SFARI candidate genes highlight convergent GPCR involvement for ASD. We listed and analysed the functional effect of the ASD-associated variants in this 23 GPCR genes in addition to the oxytocin peptide ligand (**Table 2**). Interestingly, we extracted 2 orphan GPR37 and GPR85 and 4 olfactory OR1C1, OR2M4, OR2T10, OR52M1 receptors with relative unknown functions in the brain. All receptors are strong candidates (gene score of 2) in the SFARI database. Therefore, in this review, we explore the therapeutical target values of these 23 GPCRs in addition to interesting relative family member for ASD. First, we review their general particularities, their pharmacological landscape (available drugs on the market or in clinical trials), their downstream signalling, their localisation profile (e.g., wide vs specific expression) and their global functions. Then, we evaluate if dedicated animal models recapitulate validity criteria for ASD [29], e.g., face and predictive validity (induction of the core and/or comorbid symptoms), construct validity (biochemistry/cell biology, anatomy/connectivity and/or genetic pathological mechanisms), and predictive remission (improvement of phenotypes using ligands targeting these GPCRs or their available structures for potential novel *in silico* drug design or specificity) (**Table 3**). Finally, we review the different variants associated to ASD, their potential involvement in the aetiology of ASD (level of expression) and the results of clinical trials. Based on these evidences, we conclude on the therapeutical potential of these GPCRs.

### 2) Oxytocin and vasopressin receptors

First described in the 1960s for its effects on reproduction and maternal behaviours [30], the oxytocin (OT) modulates all social behaviours [31-33]. OT and its paralog arginine vasopressin (AVP) prepropeptides are processed and maturated in cyclic neuropeptides. OT and AVP neurons are located in the paraventricular and supraoptic nuclei of the hypothalamus, and OT neurons also in the accessory nuclei, tuberal nucleus and in bed nucleus of the stria terminalis [34-36]. Both peptides are released into the brain and systemic circulation via the posterior pituitary to act as hormones [37]. Extremely conserved in mammals with only 2 of their 9 amino acids, OT and AVP are closely related in both structure, receptors and signalling, making them very complex to study individually. In the CNS, both oxytocin and vasopressin, rarely studied together, are involved in cognition and in many social processes such as social recognition and memory [38], defensive behaviours [39], trust [40], empathy [41] and maternal attachment [42, 43] in mammals, including humans [44]. These two neuropeptides induce similar and opposite effects depending on the brain structures and species [45].
Mice lacking OT peptide (*Oxt* KO mice) may display impairment in social memory and aggressive and anxious-like behaviours (**Table 3**), but these phenotypes were inconsistent across different labs. Social memory deficits were observed [46, 47] whereas other labs found normal social approach or social novelty preference behaviours [48]. Studies found both reduction and increase of aggression in *Oxt* KO mice [49-52] which could be explained by parental genotype [53]. Oxytocin administration in the lateral ventricles or in the medial amygdala restored anxious-like behaviour [54] and social recognition [47]. However, dams have problems with milk ejection [55], which makes this mouse line difficult to breed and study. Differences in the CNS phenotype might also be explained by the compensatory effect of AVP [35] and their receptors. Brattleboro rats, which carry a homozygous frameshift deletion in the *Avp* gene, leading to the lack of vasopressin synthesis display social preference deficit [56-58]. Genetic deletion of the *Avp* gene constitutively in mice leads to lethality, in the absence of peripheral AVP administration [59].
*OXT* and *AVP* genes are located around 10kb apart on human chromosome 20 and encode 125 and 164 amino acid prepropeptide comprising also neurophysin I and II respectively, which act as carrier protein for the two peptides. Several studies have associated oxytocin-vasopressin systems and autism spectrum disorders mainly because, in one hand, they regulate social behaviour [60-66], and, on the other hand, plasma levels of oxytocin positively correlates with the severity of autism disorders [67-69]. Furthermore, several variants in the *OXT* gene have been associated with ASD (**Table2**). Logically, many studies have investigated the therapeutic potential of oxytocin or vasopressin in improving ASD symptoms. Indeed, a study has shown that an intranasal administration of oxytocin at low dose improves emotion recognition during the Reading the Mind in the Eyes Task in young men with autistic condition [70]. During a Social ball-tossing game, adults with Asperger syndrome showed an increase of trust and interaction with the good player after oxytocin inhalation [71]. According to the yale-brown obsessive-compulsive scale, oxytocin also reduced the severity of repetitive behaviours in adults with autism spectrum disorders [72]. Conversely, oxytocin peptide administration failed to improve social abilities over placebo in phase 2 clinical trials [20]. Finally, administration of oxytocin shows no improvement and often side effects in several studies in humans [73-76]. Administration of vasopressin in humans improves the encoding of happy and angry faces compared with placebo (based on The Profile of Mood States scale) [77]. A phase 2 clinical trial showed that four weeks intranasal administration of vasopressin in 30 ASD children improve their social skills (on the Social Responsiveness Scale) and reduced anxiety and repetitive behaviour, with minimal side effects [22]. Whereas vasopressin remains to be tested in larger cohort of patients, the first results indicate that vasopressin might be more efficient than oxytocin to provide pro-social effects. oxytocin and vasopressin-induced receptor activation and signalling pathways might explain this difference. Further studies are needed to understand the common and specific effect of these neuropeptides. Despite, the mitigated results of clinical trials targeting this family, the autism scientific community is still convinced about the potential value of targeting the oxytocin/vasopressin system to ameliorate autism-like symptoms, regarding their major prosocial effects. Furthermore, oxytocin administration may not be effective in all patients, but rather only in subfamily of patients with oxytocin-related aetiology. Finally, oxytocin and vasopressin ligand might not be idealistic treatment as both peptides bind and activate with nanomolar affinity class A GPCRs, oxytocin receptor (OTR), vasopressin $V_{1A}$ and $V_{1B}$ respectively in the CNS (and $V_2$ in the periphery) [78]. In the following section, we review why OTR, $V_{1A}$ and $V_{1B}$ receptors, all three been associated to ASD remain of interest to define new treatment targeting this family.

- **OT receptor**

Oxytocin receptor (OTR), vasopressin $V_{1A}$, $V_{1B}$ and $V_2$ are highly conserved through evolution, sharing 70% similarity between them and more particularly 70-80% homology in their transmembrane domain), and over 95% homology between rodents and humans. This clearly impacted the discovery of specific ligand discovery. Decades of research lead to the discovery of several agonists for OTR [79-82], in addition to oxytocin and vasopressin, peptide compounds such as $Thr^4Gly^7$-OT (TGOT) [83] or carbetocin, with a potential biased agonism towards $G\alpha_q$ although controversial [84, 85] and the first chemical agonist LIT001 [86]. In contrast, several antagonists have been developed, such as the synthetic compound L-372,662 [87] or atosiban, an approved drug for prelabour term. However, except L-372,662, all these ligands also bind to vasopressin receptors. As OTR structure was recently resolved in both antagonist-bound inactive and oxytocin-bound active state, comparing OTR and $V_2$ structures might boost the discovery of selective ligands [88-90]. OTR, along with $V_{1A}$ and $V_{1B}$ preferentially activate $G\alpha_{q/11}$ pathways, but also potentially recruit other G proteins and β-arrestin [78, 81, 84, 91]. OTR is able to induce NF-κB p65, ERK1/2, and p38 kinase activation [92] and hippocampal synaptic plasticity through local translation of the protein kinase Mζ [93, 94]. OTR also act through the transactivation of tyrosine kinase receptors and epidermal growth factor receptor (EGFR) [93, 95]. OTR is expressed in critical CNS regions for the regulation of social behaviour and emotion such as the cortex, hippocampus, hypothalamus, amygdala and olfactory bulb [34, 96, 97]. OTR are located at synapses, axons and in astrocytes [36, 96]. In neurons, OTR is preferentially in GABAergic neurons (somatostatin interneurons and $D_1$ medium spiny neurons in the nucleus accumbens and Cyp26b1 neurons in the bed nucleus of the stria terminalis) [36]. OTR expression is sexually dimorphic depending on the brain region and species. In rats, higher oxytocin receptor binding density is found in males compared to females in the posterior bed nucleus stria terminalis, nucleus accumbens, dorsal caudate putamen, hippocampal CA1 region and medial amygdala, whereas females have higher levels in medial caudate putamen [98]. Sexual dimorphism for oxytocin receptor distribution and binding was also observed for two mouse backgrounds, but not for C57Bl/6J mice [99] and humans [100]. OTR receptor activation is involved in complex social behaviours, like maternal care [101, 102], social recognition [47, 103], aggression [104, 105], mating [106] but also in pair bonding [32, 107, 108], empathy [41, 109] and could induce anxiolytic effects [110, 111].

Deletion of the oxytocin receptor in rodents (*Oxtr* KO mice) revealed an autism-like behavioural phenotype (**Table 3**; [51, 112, 113], including both social deficits and stereotyped behaviours. *Oxtr* KO mice display severe social impairments, such as reduced frontal approaches, following or huddling behaviours, nose contacts with an unfamiliar mouse in reciprocal social interaction or 3-chamber tests [113, 114]. It also affects social memory, increases aggression, and reduces pup vocalizations following maternal separation [51, 112]. Whereas increased self-grooming, anxious-like behaviours and cognitive inflexibility has been observed compared to WT mice, it is not consistent through labs and/or *Oxtr* mouse lines [51, 112, 114, 115]. *Oxtr* heterozygous mice express social deficits, but neither aggressive behaviour and cognitive inflexibility [115]. Interestingly, intraventricular administration of oxytocin or vasopressin restores the social deficits in *Oxtr* KO mice via $V_{1A}$ receptors[51, 112, 115]. This finding highlights the compensatory mechanism inside this GPCR family. Interestingly, genetic ablation of OTR receptor using CRISPR/Cas9 in monogamous prairie voles lead to deficit in social novelty and an increase in repetitive behaviour, but no deficit in the reciprocal social interaction, vocalization or maternal behaviour [116].

Oxytocin receptor gene (*OXTR*) spans over 4 exons and encode for 5 splicing transcript variants that differ in their 5' untranslated region and only one receptor OTR. More than twenty variants in the *OXTR* gene have been associated with ASD (**Table 2**), mostly located outside the receptor coding region, leading to potential receptor expression dysregulation. Regarding the effect of oxytocin deletion and this finding, severe variants in the coding region could impair milk ejection and pup survival in mammals. Liu et al. identified 2 SNPs of the *OXTR* (rs2268491, rs2254298) in association with ASD in Japanese population [117]. Another study made on a panel of Finnish and English families whose parents have a neurotypical child and a child with ASD showed a significant correlation between the rs237887 polymorphism located on intron 3 of the oxytocin receptor gene and performance on the Face Recognition Memory test [118]. Taken together, these data support the idea that the oxytocin receptor may be involved in the aetiology and as risk factor of autism in different ethnic groups.

- **Vasopressin $V_{1A}$ and $V_{1B}$ receptors**

Vasopressin, well known as the antidiuretic hormone via the activation of $V_2$ receptors, and oxytocin also bind to $V_{1A}$ and $V_{1B}$ receptors. Only $V_2$ receptor structure has been resolved so far [90]. Only $V_{1A}$ and $V_{1B}$ express in the brain cells and couple mostly to $G\alpha_{q/11}$ proteins [119]. **$V_{1A}$ receptor** is activated by a selective agonist F-180 with high affinity [120]. Two antagonists, balovaptan [82, 121, 122] and RG7713 [123] have been tested in clinical trials for ASD. Moreover, administration of the $V_{1A}$ antagonist [d(CH2)5Tyr(Me)AVP] into the medial amygdala of rats, but not the ventricular system, affect maternal memory [124]. $V_{1A}$ receptor transcripts are expressed throughout the brain in different structures with higher density in hippocampus, dentate gyrus-granule cell layer, ventral subiculum. Relatively lower transcripts are expressed in central amygdala, hypothalamus and granule cells in cerebellum. Transcripts are further expressed in retina, and in the periphery, in liver and reproductive organs and smooth muscle cells where it induces the vasoconstrictive effects of the hormone [125]. $V_{1A}$ receptor expression is sexually dimorphic as densities are higher in males in the piriform and somatosensory cortex, the posterior medial bed nucleus stria terminalis, and the anterior ventral thalamic nucleus [45]. $V_{1A}$ is involved in maternal care [126, 127], social recognition [128, 129], affiliative behaviour and pair bonding [130, 131]. Mice deleted for the $V_{1A}$ receptor (*Avpr1a* KO mice) display reduced anxiety-like behaviours [132], social memory and interaction deficits [128, 133] but normal aggression [134] (**Table 3**). Using CRISPR/Cas9 gene editing, genetic deletion of the *Avpr1a* gene in hamster lead interestingly to an increase in aggressive behaviour and social communication via odour marking.
In contrast to *OXTR, AVPR1A* gene encodes only one transcript variant from two exons and the $V_{1A}$ receptor. Several studies pointed the importance of the repeat length of the RS1 or RS3 promoter and 5' untranslated region in the *AVPR1A* gene associated with ASD risk, as most variants are identified in this region (**Table 2**; [135-137]. Indeed, these variants whether influence human relationships and altruism [138, 139], personality in primates [140], or social behaviour in rodents [141], the shorter the RS3 (or RS1) promoter repeats are, the lower the gene expression of the receptor is, concomitant with great social deficits. Finally, the first attempt of clinical trials with a specific compound has been done. A first study administrating $V_{1A}$ antagonist RG7713 intravenously revealed improvement in an eye-tracking test in patients [123]. Subsequent phase 2 trial using oral administration of balovaptan, a specific $V_{1A}$ antagonist daily for 12 weeks showed improved socialization and communication (in Vineland-II Adaptive Behavior Scale, but not Social Responsiveness Scale) in 223 men with the condition [121, 122] but not in children [142]. Whereas these promising results, vasopressin $V_{1A}$ chemical antagonist administration for 24 weeks failed to improve social abilities over placebo (in Vineland-II 2DC score) in 64 women and 257 men in phase 3 clinical trials [21]. Indeed, it is not yet clear if $V_{1A}$ receptors should be activated or inhibited. Further investigations are still required to understand the therapeutical potential of $V_{1A.}$

**$V_{1B}$ receptor** is activated by various modifications of deamino-[Cys1] arginine vasopressin at positions 4 and 8 leading to $V_{1B}$ agonists (d[Cha4,Lys8]VP, d[Cha4,Dab8]VP, d[Leu4,Lys8]VP, d[Leu4,Dap8]VP) [143]. $V_{1B}$ antagonists has been also developed, such as SSR149415, which significantly reduced aggressive and chasing behaviours in rats [144-147]. $V_{1B}$ receptors expression in the brain is more discrete, with the highest levels in the CA2 region of the hippocampus, where they control preference and social memory, in the supraoptic nucleus of the hypothalamus, piriform and entorhinal cortex and olfactory bulb [9, 148, 149], but also in the anterior pituitary regulating the corticotropic stress axis [150]. Their expression could be dimorphic, but it was only assessed in male rats so far [9]. $V_{1B}$ receptors has been mainly studied for their modulation of aggression [144, 151-153] and stress and anxious-like responses [154-157]. Deletion of $V_{1B}$ receptors in mice (*Avpr1b* receptor KO) leads to an increase of dominance behaviours [158], a decrease in aggression [159, 160], vocalizations [161], loss of motivation, impaired social and episodic memory [153, 162] (**Table 3**).

In humans, the *AVPR1B* gene encodes one transcript spanning over two exons and the $V_{1B}$ receptor. Only two studies associated three *AVPR1B* gene variants in the coding sequence with autism [163, 164] (**Table 2**). The SNP, rs28373064 in the *AVPR1B* gene is associate with empathy and prosociality [165] and in contrast, other SNP, rs35369693 is shown to be associated with to mood disorders and increased aggression in females and children, respectively [163, 166]. In addition, Babu's lab showed that *AVPR1B* belong to the top ten GPCR genes that display the highest density and number of distinct missense variants in individuals sequenced in the 1000 genomes project [12, 13].

In conclusion, all these evidences in animal models and humans clearly establish OTR and $V_{1A}$ receptors as major therapeutical targets for autism spectrum disorders, and potentially $V_{1B}$ receptors as well. Nevertheless, so far, clinical trials failed to bypass the placebo effect observed in patients. Regarding the complexity, the conservation and the cross-talk of the oxytocin vasopressin and receptors, further investigations are needed to unravel the oxytocin-vasopressin system and develop new ligands or identify the exact targets (e.g., one receptor and one signalling pathway or one hetero-oligomer) in this system. Indeed, OTR, $V_{1A}$ and $V_{1B}$ receptors form homo- and hetero-oligomers, which could represent more specific targets [167-170].

### *3) Metabotropic glutamate and gamma-aminobutyric acid receptors*

Glutamate and gamma-aminobutyric acid (GABA) are the major neurotransmitters at the excitatory and inhibitory synapses in the mammalian CNS respectively. The metabotropic glutamate mGluRs and $GABA_B$ receptors belong to class C GPCRs and regulate neuronal excitability [171, 172]. In contrast to class A GPCRs, the endogenous ligands bind in their large extracellular N-terminal domain called Venus Fly Trap that characterize the class C GPCRs, which closes upon activation. In addition, they form constitutive (homo- or hetero-) oligomers which lead to specific rearrangement of subunits during activation. This class is composed of orphan receptors (*GPC5B*, *GPC5C*, *GPC5D*, *GPC6A*, *GP156*, *GP158*, *GP179*, *RAI3*), sweet and umami taste receptors (*TS1R1*, *TS1R2*, *TS1R3*), mGluRs (*GRM1*, *GRM2*, *GRM3*, *GRM4*, *GRM5*, *GRM6*, *GRM7*, *GRM8*), $GABA_B$ receptors (*GABBR1*, *GABBR2*), calcium-sensing receptors (*CASR*) [13, 173]. mGluRs and $GABA_B$ receptors are mainly expressed in pre- and postsynaptic compartments in the brain (except $mGlu_8$ only in the retina) as well as peripheral tissue. Only $mGlu_5$, $mGlu_7$ and $GABA_{B2}$ subunits are present in the SFARI list.

- **$mGlu_5$ receptors**

Metabotropic glutamate receptors are further sub-grouped based on their G protein coupling, sequence similarities and pharmacological properties [171]. Group I mGluRs comprising $mGlu_1$ and $mGlu_5$ are coupled to $G\alpha_{q/11}$ proteins and selectively activated by 3,5-dihydroxyphenylglycine (DHPG) and (RS)-2-Chloro-5-hydroxyphenylglycine (CHPG, specific for $mGlu_5$) [82, 174, 175]. Both the DHPG and CHPG, potentiates the depolarization of the CA1 hippocampal neurons induced by the N-methyl D-aspartate (NMDA). Selective antagonist mavoglurant (AFQ-056) and the allosteric modulator basimglurant (RG-7090, RO-4917523) targeting $mGlu_5$ receptor, initially developed for a monogenic ASD condition [19, 176-178], are still in clinical trials for other applications, dyskinesia, obsessive–compulsive disorder and depression. Further, activation of $mGlu_5$ receptors leads to induction of LTP and translation along with downstream signalling including mTOR activation [179, 180]. Alterations in $mGlu_5$ receptor signalling, including receptor expression affects the excitability ($mGlu_5$-dependent LTD), inhibitory neurotransmission modulation, or synaptic and neuronal developmental that have been linked to ASD or Schizophrenia [181, 182]. Group I metabotropic glutamate receptors are primarily localized to post-synapse and also found on astrocytes and microglia. $mGlu_5$ receptors are distributed throughout the brain regions including cerebral cortex, hippocampus, amygdala, basal ganglia, thalamus, hypothalamus, cerebellum and brainstem with higher expression being observed in septum and olfactory bulb [183-185]. Genetic ablation of $mGlu_5$ in mice (*Grm5* KO mice) leads to impairment in ASD related traits (**Table 3**): deficits in social interaction and digging and marble burying behaviour, hyperactivity and reduced anxiety [186]. Further *Grm5* KO mice display severe deficits in sensorimotor gating, irrespective of the genetic background strain [187].

*GRM5* gene encodes two splice variants of $mGlu_5$ ($mGlu_{5a}$ and $mGlu_{5b}$), both co-expressed in the brain and with $mGlu_{5b}$ receptor, *per se* expressed predominantly during the adult stage. Genetic studies have identified de novo variants in *GRM5* gene in individuals with ASD condition (**Table 2**), highlighting *GRM5* as one of the most susceptible gene in ASD [188, 189]. It was also reported that copy number loss variants (CNVs) in the eight genes coding for mGlu receptors increase the prevalence of syndromic ASD in children [190]. Further, SNPs in glutamate transporter genes such as *SLC1A1* and *SLC1A2* are associated with ASD, especially rs301430 in the *SLC1A1* gene linked to repetitive behaviour and anxiety in children with ASD [191-193]. A pharmacogenomic study showed that *GRM5* belongs to the GPCR genes that display the highest density of missense variants in individuals sequenced in the 1000 genomes project [12, 13]. Abnormal expression of $mGlu_5$ receptors and their signalling have been linked with ASD and associated comorbidities such as intellectual disability [181]. Increased levels of $mGlu_5$ receptor protein and its signalling were reported in different brain regions including cerebellar vermis region and superior frontal cortex in children with autism compared to the healthy counterparts [194, 195]. Higher $mGlu_5$ receptor protein expression was also reported in prefrontal cortex of patients with monogenic Fragile X syndrome (FXS) condition [196]. In contrary, Chana and colleagues have reported a decrease in $mGlu_5$ receptor expression at genetic and protein levels in dorsolateral prefrontal cortex (DLPFC) of ASD patients [197]. The observed increase in $mGlu_5$ signalling explain why clinical trials with the antagonist mavoglurant and the allosteric modulator basimglurant of the $mGlu_5$ receptor, were successful in correcting the broad range of phenotypes in *Fmr1* KO [178]. However, administration of these compounds failed to show similar therapeutic benefits based on the normalized scores using Wechsler Intelligence Scale (WISC-III) compared to FMRP levels in FXS patients in phase 2b/3 clinical trials [19, 176, 177].

Altogether these data argue for $mGlu_5$ receptor role in ASD pathogenesis and why it is one of the first GPCR targeted for ASD. However, targeting $mGlu_5$ even with a selective negative allosteric modulator did not pass the placebo effect in patients. This might be attributed to differences in promoter methylation, mRNA and/or the wide range of receptor expression.

- **$mGlu_7$ receptors**

The $mGlu_7$ receptors along with $mGlu_4$, $mGlu_6$ and $mGlu_8$ belongs to Group III mGluR that are predominantly localized to presynaptic active zone and regulate the neurotransmitter release [198, 199]. Comparatively, $mGlu_7$ receptors have less affinity towards glutamate within the Group III metabotropic glutamate receptor members, which makes the $mGlu_7$ an "emergency brake". The $mGlu_7$ receptors acts as auto- or hetero-receptors at active zone sites regulating the release of neurotransmitters like glutamate and GABA [171, 198, 200, 201]. Agonist 2-amino-4-phosphonobutyrate (L-AP4) activates this group III mGluRs [82], which negatively regulates glutamate or GABA release. Conversely, the selective positive allosteric modulator N,N'-dibenzhydrylethane-1,2-diamine dihydrochloride (AMN082) affects positively the extracellular glutamate levels and negatively GABA levels [202-204]. MMPIP and ADX71743 are selective negative modulators or antagonists of $mGlu_7$ receptors respectively [205]. Group III mGluRs couple mainly to $G\alpha_{i/o}$, but $mGlu_7$ might also couple to $G\alpha_{q/11}$ protein [206]. Further, activation of $mGlu_7$ can trigger $G\beta\gamma$-mediated opening of GIRK channels and $K^+$ influx, and activation of PKC via its scaffolding partner protein interacting with C-kinase type 1 (PICK1) and inhibition of N and P/Q-type of calcium channels resulting in $mGlu_7$-dependent regulation of synaptic transmission [206-208]. The $mGlu_7$ receptors are expressed during neurodevelopment in order to augment the synapse formation and stabilization during this critical period [209]. $mGlu_7$ receptors are widely expressed in the CNS, with relative higher expression in amygdala, hippocampus and hypothalamus [210].

Mice lacking the $mGlu_7$ (*Grm7* KO) display social memory deficits but intact sociability and anxiolytic behaviour compared to control mice (**Table 3**) [211, 212]. In addition, *Grm7* KO mice exhibit severe aversive learning deficits in context and conditioned cue. Furthermore, these mice also show comorbid symptoms with motor coordination impairment and seizures [211]. Similarly, mice carrying Ile154Thr *GRM7* mutation (*Grm7* mutant) also recapitulate the contextual fear memory deficits, motor disabilities and seizures [213].

In humans, *GRM7* gene encode for two major isoforms ($mGlu_{7a}$, $mGlu_{7b}$) of $mGlu_7$ receptors that differ at the C-terminus, which can lead to different protein-protein interactions and receptor coupling [8]. SNPs and CNVs in *GRM7* gene encoding for $mGlu_7$ receptors, have been linked to different neurodevelopmental disorders including ASD (**Table 2**), epilepsy, microcephaly, attention deficit hyperactive disorder-ADHD [214-220]. Further, several SNPs in *GRM7* gene have been associated with increased risk for ASD [221]. Using bioinformatic tools, other *GRM7* gene mutation Gly471Arg have been predicted to affect glutamate binding [222]. Song and colleagues studied the pathogenic effects of the Ile154Thr, Arg658Trp and Thr675Lys mutations in primary neuronal cultures. All mutants lead to $mGlu_7$ degradation through proteasomal and autophagy pathways, lack axonal growth due to altered cAMP-PKA-ERK signalling, and reduced numbers of synapses, which can be rescued by AMN082. Further, both Arg658Trp and Thr675Lys mutations results in reduced surface receptor expression [223]. This is in line with reduced expression of $mGlu_7$ receptor in post-mortem motor cortex samples from patients with Rett syndrome (RTT) and mRNA levels of $mGlu_7$ in a mouse model of RTT (i.e., *Mecp2* KO mice) [224, 225]. These findings suggest a role of $mGlu_7$ levels of expression in the pathogenesis of this neurodevelopmental disorder

In conclusion, $mGlu_7$ receptors are promising targets as they are involved in ASD aetiology, they are expressed in brain structures involved in cognitive and emotional process and selective agonists or positive allosteric modulators (e.g., AMN082) have been developed. Their administration could normalize behaviours, especially repetitive and stereotyped behaviours, in mouse models of ASD via the regulation glutamate and GABA release and, and eventually in patients.

- **$GABA_B$ receptors**

Metabotropic $GABA_B$ receptors are formed by obligatory hetero-oligomer of $GABA_{B1}$ and $GABA_{B2}$ subunits through their C-terminus coiled-coiled interaction domain. GABA binds to $GABA_{B1}$ and following activation couple to downstream $G\alpha_{i/o}$ protein signalling molecules through $GABA_{B2}$. (R) Baclofen (β-[4-chlorophenyl], GABA (STX209), a selective agonist of the $GABA_B$ receptors [226] is approved as a skeletal muscle relaxant. It has been also used as a therapeutic agent in treating neuropathological conditions driven by GABAergic dysfunctions, including ASD. The $GABA_B$ receptors are present at both pre- and post-synaptic compartments, and through $G\alpha_{i/o}$ and $G\beta\gamma$ proteins, they inhibit $Ca^{2+}$ channels at the presynaptic sites and activate GIRK potassium channels, resulting in $K^+$ influx and hyperpolarization of the neuronal membrane [227-230]. $GABA_B$ receptors normalize the levels of phosphorylation of NF-κB and ERK signalling pathways [231]. The inhibitory presynaptic $GABA_B$ receptors function as auto-receptors. Whereas, the excitatory presynaptic $GABA_B$ receptors function as heteroreceptors and supresses the neurotransmitter release. The postsynaptic $GABA_B$ receptors are responsible for induction of slow inhibitory postsynaptic currents, which shunts the excitatory currents [232], thus modulating the neuronal excitability. These receptors are also play a role in dendritic spike generation and excitatory postsynaptic potentials amplification.
Studies using $GABA_B$ receptor deletion in mice (both $GABA_{B1}$ and $GABA_{B2}$ subunits or *Gabbr1-Gabbr2* double KO) have revealed emotional behavioural disturbances, increased anxious-like behaviour and antidepressant-like behaviour [233, 234] (**Table 3**). These antidepressant properties of $GABA_B$ receptors are potentially mediated by interactions between these receptors, serotonergic system and BDNF [235, 236]. Further, $GABA_{B1a}$ and $GABA_{B1b}$ KO mice exhibit altered stress-induced social withdrawal behaviour [237]. The effect of *Gabbr2* deletion in animal models on social or stereotyped behaviour has not been reported. However, administration of arbaclofen resulted in reversal of social behavioural and other pathogenetic phenotypes observed in FXS and other ASD mouse models such as BTBR and C58 inbred strains [238-240].
Compelling evidences suggest the association between mutations in *GABBR2* (gene encoding $GABA_{B2}$ subunits) and altered $GABA_B$ receptor function and neurodevelopmental disorders including ASD (**Table 2**), epileptic encephalopathy, intellectual disability and RTT [241-247]. In RTT patients, different pathogenic mutations in *GABBR2* were observed [246]. The RTT patients included in the same study, were *MECP2* (prime responsible gene in RTT) mutation-negative (without any *MECP2* gene mutations) and authors have found *de novo* mutations in novel genes. They found two unrelated patients have the same variant in *GABBR2* gene i.e., Ala567Thr, which was also reported also in a Portuguese patient having Rett like syndrome [248]. This mutation disrupts receptor activation, when expressed in HEK cell lines. Further, the frog (*Xenopus tropicalis*) tadpole larvae carrying the Ala567Thr mutation have shown increased seizure like behaviour and altered swimming patterns which were partially rescued by agonist baclofen [246]. In the same study, authors have shown the severe effects on $GABA_B$ receptor dysfunction by expressing the mutants of *GABBR2* gene related to epileptic encephalopathy (Ser695Ile and Ile705Asn) [245] in HEK cells as well as the effect on behavioural phenotype in *X. tropicalis* tadpoles. Clinical trials at the level of phase 2 and open label trials using arbaclofen demonstrated a beneficial effect treating the patients with FXS and ASD and the treatment well tolerated [19, 249-251]. Later, these clinical trials were discontinued for the lack of efficacy [19]. Nevertheless, it is currently tested as an adjuvant therapy to risperidone for irritability (Aberrant Behavior Checklist-Community Edition scale)[252]. Furthermore, reduced expression of $GABA_B$ receptor subunits were observed in cerebellum, cortical areas related to working memory (BA9, BA40 areas) and socio-emotional behaviours (cingulate cortex) and fusiform gyrus- a brain area important for facial identification from ASD patient brains [253, 254].
In conclusion, $GABA_B$ receptors remains a promising therapeutical target for ASD. However, their efficacy might be greater in combination with the administration of other ligand, such as risperidone or targeting mGluR, for example.

### *4) Biogenic amine receptors*

Biogenic amine receptors are class A GPCR family that interact with endogenous aminergic ligands such as dopamine and serotonin. Both the dopaminergic and serotonergic (or 5-HTR) receptors play crucial role in controlling variety of behavioural, physiological functions related to central and peripheral nervous system [255]. So far, only dopamine $D_1$, $D_2$ and $D_3$ and serotonin 5-$HT_{1B}$ have been listed as strong candidate genes in the SFARI database.

- **Dopamine receptors**

All dopaminergic receptors belong to class A GPCRs with 5 receptors, $D_1$-$D_5$ subdivided in $D_1$-like and $D_2$-like class receptors based on their G protein coupling [256]. They can activate their signalling cascades through both G-protein-dependent and -independent mechanisms. Several studies reported that dopamine receptors form hetero-oligomers with each other, but also with other GPCRs [257]. In particular, the formation of hetero-oligomers between different subtypes of dopaminergic receptors was found between $D_1$ and $D_2$ receptors [257, 258] or $D_3$ receptors [259, 260] and $D_2$ and $D_4$ receptors [261, 262], $D_3$ receptors [263] or $D_5$ receptors [264]. Additionally, dopamine receptors could oligomerize with adenosine $A_{2A}$ receptors [265]. The formation of such hetero-oligomeric receptor complexes changes the signalization cascades triggered by dopamine receptors and results in their different pharmacological properties [257]. These receptors are highly expressed in GABAergic medium spiny neurons (MSN) of the striatum, with $D_1$ and $D_2$ respectively enriched in direct striatonigral and indirect striatopallidal MSNs, but their expression was also detected in the olfactory bulb, cortex, and hippocampus [266-268]. A broad diapason of functions mediated by different subtypes of dopaminergic receptors includes, among others: voluntary movement, reward processing, learning, motivated behaviour, action selection, sleep, attention, and decision making, some of which are comorbid symptoms often associated with ASD [269].
**$D_1$ receptor** together with $D_5$, belongs to the $D_1$-like class of dopamine receptors. Interestingly, this class of dopamine receptors has 10-100-fold lower affinity for dopamine than $D_2$-like class, with $D_1$ receptors having the lowest dopamine affinity among all the dopaminergic receptors [270]. This suggests that $D_1$-like receptors are activated by high concentrations of dopamine phasic release, while $D_2$-like receptors might detect low tonic levels of dopamine [271]. These receptors couple to the $G\alpha_{s/olf}$ pathway that leads to phosphorylation of the dopamine- and cyclic-AMP-regulated phosphoprotein of 32 kDa protein (DARPP-32). Activated DARPP-32 inhibits the protein phosphatase 1 (PP1) responsible for histone dephosphorylation, thus resulting in the expression of genes in response to $D_1$ receptor stimulation [256, 270, 272]. Moreover, several studies have shown that $D_1$ receptors could activate the Ras-Raf-MEK-ERK signaling pathway, however only upon interaction with the glutamate NMDA receptors [273, 274]. The $D_5$ receptor, as well as $D_1$-$D_2$ receptor complexes, could also increase the levels of brain-derived neurotrophic factor (BDNF), which subsequently leads to activation of TrkB tyrosine receptor kinase and different signaling cascades in response to dopamine [256]. Alternatively, the $D_1$-like class of receptors could couple to the $G\alpha_{q/11}$, which results in increased intracellular calcium levels that are important for neurotransmitter release by exocytosis, and activation of the calcium/calmodulin-dependent protein kinase II (CaMKII) [256, 270, 272]. Interestingly, activation of the $D_1$ receptor also results in phosphorylation of the ribosomal protein rpS6 protein [275, 276], which results in increased Cap-dependent mRNA translation [277, 278]. The chemical antagonist flupentixol binds to $D_1$ receptor and also with higher affinity to $D_3$, $D_2$ and $D_5$ receptors. It is an approved antipsychotic drug for the treatment of schizophrenia [7] and at low doses it reduces the rate of deliberate self-harm injuries [279, 280]. Both $D_1$ and $D_5$ receptors are activated by an antagonist ecopipam which is currently in phase 2 clinical trials for the treatment of Tourette's syndrome, characterized by repetitive tics occurring involuntarily [281, 282]. $D_1$ are selectively activated by SKF-83959, a biased agonist that fail to recruit $\beta$-arrestin. In neurons, $D_1$-like receptors are highly concentrated in dendritic spines and postsynaptic densities of neurons where they regulate the functions of the postsynaptic compartments [266]. Besides, its well-known functions in memory, attention, locomotion, and impulse control [283], $D_1$ receptor may have a role, although controversial, in social behaviour [284, 285]. Administration of $D_1$ receptor antagonist SCH23390 ameliorated stereotyped behaviour in mice lacking the tyrosine hydroxylase that catalyse dopamine synthesis in dopaminergic neurons [286]. Pharmacological activation of $D_1$ receptors induces autism-like phenotype in WT mice [287], indicating that the $D_1$ receptor could be targeted for the treatment of ASD. Additionally, a study with a novel Wistar rat model bearing a mutation in the *Drd1* gene further provided a link between $D_1$ receptor and ASD [288] (**Table 3**). The Ile116Ser mutation in the $D_1$ receptor leads to impaired coupling of the receptor with G proteins and reduced $D_1$ receptor cell surface expression. The mutant rats exhibited ASD-like social symptoms, reduced social interaction (sociability and social novelty) in adult rats, lack of female urine preference in males and reduced ultrasonic vocalizations in mutant pups while calling their mothers. However, no stereotyped behaviour was observed for this rat model.

In humans, $D_1$ receptors are encoded by the *DRD1* gene. *DRD1* haplotypes C-A-T (**Table 2**), which consists of three common polymorphisms rs265981 (-684T>C), rs4532 (-48G>A), and rs686 (*62C>T) located in the untranslated region are more prevalent in autistic male children from families having only male children diagnosed with autism. Interestingly, this haplotype was associated with severe impairments in social interaction, nonverbal communication and increased motor stereotypies [289]. Caucasian mothers carying rs265981 and rs686 SNPs in the *DRD1* gene were orienting away from their infants more than the mothers with a wildtype version of the gene [290].

**$D_2$ and $D_3$ receptors** together with $D_4$, belongs to the $D_2$-like class of dopamine receptors that couple to the $G\alpha_{i/o}$, lead to increased levels of BDNF and activation of TrkB and ERK1/2 kinases [256, 270, 272]. The $D_2$-like receptors through $G\beta\gamma$ subunits lead to the activation of GIRKs, inhibition of the L- and N-type calcium channels [256, 270, 272], and activation of PLC and CaMKII, and their downstream signalling. $D_2$ receptors could also regulate the Akt/glycogen synthase kinase 3 (GSK3) pathway via β-arrestin 2. More specifically, this pathway is inhibited by the action of dopamine since it leads to the inactivation of Akt [256, 272]. $D_3$ receptors activate Akt and mTOR signalling, thus controlling protein synthesis [291]. Currently, the only available approved treatments for ASD patients are antipsychotics, with aripiprazole, a $D_2$ antagonist (also 5-$HT_{1A}$ partial agonist and 5-$HT_{2A}$ antagonist), or risperidone, a $D_2$ antagonist (also 5-$HT_{2A}$ antagonist and a partial agonist for histamine $H_1$, $\alpha_1$ and $\alpha_2$-adrenergic receptors) to treat irritability, aggression and repetitive behaviour of autistic children and adolescents [292-295]. Additionally, two other $D_2$ antagonists, pimozide and olanzapine are clinically used antipsychotics for schizophrenia treatment [7] and also antagonize other GPCRs, such as $D_3$, and serotonin receptors, including 5-$HT_7$ and histamine $H_1$. These ligands are also used to treat Tourette's syndrome, showing potential for amelioration of speech impairment [296-299]. The antipsychotics cariprazine, a partial agonist for $D_3$ receptors and $D_2$ receptors with lower affinity, are approved drugs for the treatment of schizophrenia and bipolar disorder. Interestingly, in male rat models of ASD, whose mothers were treated with valproate during pregnancy, cariprazine treatment improved social behaviours in a dose-dependent way, which makes this ligand a promising treatment for ASD [300]. $D_2$ and $D_3$ receptors located both in post-synaptic and pre-synaptic compartments, at distinct location. $D_2$ receptor gene encodes two splicing isoforms, short $D_{2S}$ and long $D_{2L}$ differing by 29 amino acids in their IL3. $D_{2S}$ serve as auto-receptors regulating dopamine release, opening of calcium channels and dopamine synthesis and $D_{2L}$ as postsynaptic receptors [301, 302]. $D_2$ receptors are widely expressed in the CNS and are involved in locomotion, attention, sleep, memory, and learning. $D_3$ receptors localize predominantly in the striatum, septum, hippocampus and hypothalamus, where they are involved in locomotion, and habituation to novelty [283].

*Drd2* KO mice showed great impairments in social behaviour (both sociability and social novelty), impaired social olfaction and stereotyped behaviour (increased time spent grooming) [287] (**Table 3**). Interestingly, they found exacerbated phosphorylated ERK1/2 and CaMKIIα kinases in the dorsal striatum of these mice, indicating a link between the absence of the $D_2$ receptor and its downstream signalling and observed behavioural impairments. This phenotype seems exclusively mediated by the dorsal striatum as specific knock-down of $D_2$ receptors in this structure is sufficient to recapitulate all the mentioned behavioural impairments reported in *Drd2* KO mice [287]. Moreover, *Drd2* heterozygous mice subjected to early maternal separation stress, also display social interaction deficits and stereotyped behaviour [303], associated with decreased levels of proteins BDNF, TrkB, phosphorylated ERK1/2, and CREB in the dorsal striatum [303]. conversely, $D_2$ receptor overexpression in the striatum and olfactory tubercle also revealed impairment in sociability only in female mice and alteration of the ultrasound vocalization in sex-matched congeners [304]. Deletion of $D_3$ receptors in mice (*Drd3* KO) display hyperactive behaviour and are particularly prone to addictive behaviours and vulnerable to alcohol and drug abuse [302, 305, 306]. To our knowledge, *Drd3* KO phenotype on social or stereotyped behaviours has not been reported.

In humans, the *DRD2* gene (encoding the two $D_2$ receptor splicing isoforms $D_{2S}$ and $D_{2L}$) display the highest number of variants associated to autism for this family (**Table 2**). It contributes to the dopaminergic dysfunction in neurological disorders such as schizophrenia, ASD and ADHD [307-309]. Several SNPs in the *DRD2* gene have been associated with a higher risk of ASD. Carriers of the common rs1800498 (286-2730C>T) SNP are increased in autistic male children from families with only male children affected by ASD. This allele showed also a correlation with more severe deficits in social interaction and communication, but also with stereotyped behaviour [310]. Interestingly, risk genes for language impairment and dyslexia also contributes to language problems in children with ASD condition, suggesting a shared and complex genetic association related to behavioural traits [311]. Several rare variants of the *DRD2* gene have also been associated with other neurological disorders such as epilepsy [312] and Tourette disorder [313]. rs167771 (113876275G>A) SNPs located in the intronic region of the *DRD3* gene, has been shown to be associated with ASD risk [314, 315]. SNPs have been related to stereotyped behaviour called "insistence on sameness", very frequently observed in autistic patients [316]. Moreover, Staal and colleagues showed a correlation between rs167771 polymorphism in the *DRD3* gene, the volume of the striatum, and stereotyped behaviour [317] indicating a role of $D_3$ in the onset of stereotyped behaviours. Finally, a highly frequent missense variant (encoding Gly9Ser mutation in its N-terminus) may alter $D_3$ drug response [12, 13], which could be an issue regarding its pharmacological potential for ASD.
In conclusion, alterations in both $D_1$- and $D_2$-like classes of dopaminergic receptors result in autistic-like behaviour in animal models or increased risk of ASD in humans. However, $D_2$-class of dopamine receptors, more specifically $D_2$ receptors, seems to contribute more to the aetiology of ASD as more variants were reported and mouse models recapitulates core symptoms. Furthermore, approved ligands targeting $D_2$ receptors could be used for amelioration of autistic symptoms, especially repetitive behaviours. This receptor is indeed a promising target for ASD treatment. Nonetheless, other dopamine receptor might be of interest, for example, the less known $D_5$, which belongs to the human GPCR genes that display the highest number of distinct missense and loss of function variants in the general population [12, 13].

- **Serotoninergic GPCRs**

All serotonin (5-hydroxytryptamine, 5-HT) receptors (5-$HT_{1-7}$) are class A GPCRs, except 5-$HT_3$, which is a channel receptor. They are expressed in several brain regions including raphe nuclei, medial prefrontal cortex and cortical areas, cerebellum, ventral tegmental area, thalamus, hypothalamus and hippocampus and modulate cognition, memory, sleep, appetite, respiration, thermo-regulation and mood [318-320]. The Dup (15)q11q13 mouse model has low 5-HT levels in all brain regions and display impaired social interaction, decreased exploratory and locomotion activities, increased fear response and latency to feed in a novel environment [321, 322]. Brain imaging (PET) studies of these mice show decreased neuronal activity mostly in the dorsal raphe nucleus, where the serotonin somas are located [323-325]. Dysfunction in the serotonergic system have been observed in different neuropsychiatric disorders, including ASD [326] and activation of this system has been shown to restore social deficit in several ASD mouse models [327].
In addition, studies have shown that more than 25% of the ASD patients show increased serotonin blood levels [328-330], which is defined as the most robust biomarker in ASD. Though from neuroimaging to pharmacology studies, the relationship with peripheral 5-HT measures and risk of ASD is not yet clearly understood, as in the brain of autistic children, the amount of 5-HT is 77% lower compared to control children and is lower than in the blood [331-334]. Decreased 5-HT levels in the CNS, especially during developmental stages affect neuronal development and formation of neuronal networks [335-337], leading to synaptic or spine abnormalities, a hallmark of ASD. Currently, antipsychotics administration to treat irritability, aggression and repetitive behaviour in ASD patients targets two serotonin receptors, partially activating 5-$HT_{1A}$ (aripiprazole) and/or inhibiting 5-$HT_{2A}$ (aripiprazole and risperidone), in addition to dopamine $D_2$ receptors [292-294, 338]. Further, 5-$HT_7$ receptor ligands have been shown to improve social deficits and stereotyped behaviours in mouse models of ASD [339, 340]. Therefore, considering the importance of the serotoninergic system in ASD, we reviewed the potential of $5\text{-HT}_{1A\text{-}B}$, $5\text{-HT}_{2A}$, and $5\text{-HT}_7$ receptor.

**$5\text{-HT}_1$ receptors** encoded by 7 genes (*HTR1A-F*), are all coupled to $G\alpha_{i/o}$ proteins. Among these, only the serotonin *HTR1B gene* is listed in the SFARI gene database (**Table 2**). $5\text{-HT}_{1B}$ receptors are mainly expressed in axon terminals of the basal ganglia, especially in the globus pallidus and substantia nigra [341]. Interestingly, selective activation of $5\text{-HT}_{1B}$ receptors using CP94253 full agonist reduced only aggression in resident male mice, whereas non-selective serotonin receptor full agonist CGS-12066 also reduce social motivation and exploration [342, 343], potentially through $5\text{-HT}_6$ activation. In addition, anpirtoline full agonist (that also targets $5\text{-HT}_3$ channels) restored isolation-induced impairments, increased pain threshold and exert antidepressant effects in mice [344]. $5\text{-HT}_{1B}$ exerts a coherent effect on anxious-like behaviours as administration of CP94253 agonist or the most selective antagonists of $5\text{-HT}_{1B}$ receptors (SB 216641 and GR 127935) lead to anxiogenic or anxiolytic effect respectively in rodents [345, 346]. Non-selective serotoninergic Triptans agonists should be soon approved drugs, as successful results have been obtained in clinical trials for migraine [347-350]. In agreement with pharmacological studies, deletion of the $5\text{-HT}_{1B}$ receptor in mice (*Htr1B* KO; **Table 3**) leads to exacerbated impulsive and aggressive behaviour [351-353], and social interest [354]. In addition, they show a decrease in anxious-like and maternal behaviour [352, 355], improved cognitive flexibility without deficits in locomotion [356] and exhibit vulnerability of abuse drugs [357-360]. Only two variants have been associated to ASD (**Table 2**), including one in Brazilian ASD population [361]. Several SNPs in the *HTR1B* (rs6296, rs11568817, rs130058 and rs13212041) has been associated with aggressive behaviour and in anger and hostile-like behaviour in young men [362, 363]. In addition to *HTR1B,* a post-mortem study revealed decreased binding for $5\text{-HT}_{1A}$ in the thalamus, the posterior cingulate cortex and the fusiform gyrus in ASD patients [364-367].

**$5\text{-HT}_{2A}$ receptor** family comprised also $5\text{-HT}_{2B}$ and $5\text{-HT}_{2C}$ receptors, which are all coupled to $G\alpha_{q/11}$ proteins. The postsynaptic $5\text{-HT}_{2A}$ receptor is one of the most highly expressed serotonin receptors in the brain and modulates the activity of many neurotransmitters such as glutamate, dopamine, acetylcholine and noradrenaline. Mice lacking the $5\text{-HT}_{2A}$ receptor (*Htr2a* KO) display reduced anxiety-like behaviour, which was normalized by restoration of $5\text{-HT}_{2A}$ signalling [368]. Furthermore, chronic exposure to corticosterone amplified the anxiety-like phenotype and increased immobility in *Htr2a* KO mice [369, 370] (**Table 3**). The two atypical antipsychotics aripiprazole and risperidone for the treatment of irritability and repetitive behaviours in ASD patients antagonize the $5\text{-HT}_{2A}$ receptors [82, 295, 371, 372]. Administration of M100907, a selective $5\text{-HT}_{2A}$ receptor antagonists also reduce repetitive grooming in the BTBR mouse model of ASD [373]. Several studies showed that $5\text{-HT}_2$ receptors are associated to ASD especially $5\text{-HT}_{2A}$ receptors whose RNA is found in lower quantities in the platelets of patients with ASD, who also have gastrointestinal problems [374-377]. The serotonin 2A receptor gene (*HTR2A*) located on chromosome 13q14.2 contains several SNPs associated with ASD risk [378]. The most frequent is rs6311 (1438A>G) variant located upstream and within the promoter region leading to higher promoter activity, which could play a modulatory role in ASD risk [376, 379-381]. Several studies showed association between lower expression of $5\text{-HT}_2$ receptors and functioning in individuals with ASD [375, 376]. Imaging studies reveals lower densities of $5\text{-HT}_{2A}$ receptor in fusiform gyrus and cingulate cortex compared to control groups [365, 375, 382], which was confirmed by post-mortem studies, together with reduced $5\text{-HT}_{2A}$ levels in the thalamus [364-367]. All three $5\text{-HT}_2$ receptor genes are among the highest GPCRs genes prone to missense variants, with the highest number of distinct loss-of-function variants in the *HTR2B* gene and two of the highest frequent missense variants in the *HTR2A* and *HTR2C* genes, His452Tyr and Ser23Cys mutations respectively, predicted to alter efficacy of drug response targeting these receptors [12]. The latter could explain the positive effects of atypical antipsychotic 5-$HT_{2A}$ antagonists whereas 5-$HT_{2A}$ density is already reduced in brain of patients.

**5-$HT_7$ receptors** couple mainly to $G\alpha_s$ proteins with the 5-$HT_4$ and 5-$HT_6$. They are targeted by numerous specific ligands with nanomolar affinities: SB269970 and JNJ-18038683 antagonists, SB258719 inverse agonist, and LP-12, LP-44, E55888 agonists, with JNJ-18038683 antagonist currently in clinical trials for sleep, bipolar and major depression disorders, and E5588 agonist that shows beneficial effect *in vivo* on pain relief [82, 383, 384]. Recently, a selective β-arrestin biased agonist and $G\alpha_s$/cAMP pathway antagonist serodolin has been shown to decrease hyperalgesia in several mouse models of inflammatory pain through ERK and Src kinases activation [385]. 5-$HT_7$ receptors also couple to $G\alpha_{12}$ protein promoting neurite length and synaptogenesis, participating in CNS development and maturation, which is suppressed by a selective 5-$HT_7$ receptor antagonist SB269970 [386-388]. They expressed at the post-synapse in the thalamus, cortex, limbic system and hypothalamus [389]. This serotonin receptor is involved in cognition, sleep, pain, circadian rhythm, hypothermia and mood. Deletion of the 5-$HT_7$ receptor in mice (*Htr7* KO) leads to decreased obsessive behaviours, associated with antidepressant phenotype, reduced pain thresholds and increased sensitivity to epilepsy-induced seizures and anxious-like behaviours (**Table 3**) [385, 390-392]. Moreover, activation of 5-$HT_7$ receptors normalizes repetitive behaviours and synaptic plasticity in *Fmr1* mice, a mouse model of Fragile X syndrome [393]. It rescued repetitive and anxious-like behaviours, motor skills and memory in a mouse model of Rett syndrome associated with ASD using a partial agonist LP-211. Finally, its activation also reversed sociability deficits, anxiety and hyperactivity in rats exposed to prenatal valproic acid using a 5-$HT_{1A}$/5-$HT_7$ agonist 8-OH-DPAT [339, 340]. Conversely, administration of selective 5-$HT_7$ receptor antagonists SB-269970 also inhibits stereotypies in *Pacap* KO mice [394]. In humans, the *HTR7* gene encode 3 splicing isoforms of the 5-$HT_7$ receptor that differ in their C-terminal domain and coupling [8]. The *HTR7* gene is one of the most human GPCR genes prone to gene duplication and potential gain-of-function [12, 13], level of expression might explain why activation or inhibition of the 5-$HT_7$ were inconsistently used. So far, no significant SNP or CNV has been associated to ASD [395]. Despite some paradoxical effect that need to be clarified, 5-$HT_7$ receptor ligands have become a promising target for cognitive and affective disorders, including ASD [396-400].

All these data show that *HTR2A* gene also fulfil SFARI criteria as strong candidate (category 2) in the SFARI gene list together with *HTR1B* gene, whereas *HTR7* gene might belong to category 3 or higher if a positive association with ASD risk would be found. The serotonin system is very complex with 14 receptors. Serotonin is involved in autism aetiology and serotonin receptors are of great interest as therapeutical targets for ASD, thus targeted medication against these receptors while avoiding parallel modulation of the dopaminergic system ($D_2$ receptors) is required. In addition to these three receptors, other serotoninergic receptors could be of interest, such as the 5-$HT_6$. Administration of the selective 5-$HT_6$ receptor antagonist SLV ameliorates impairments in cognition and social interaction in relevant animal models of deficits in cognition [401]. *HTR6* gene display a specific and discrete profile of expression and is associated with autism susceptibility [402]. Lacivita and collaborators developed arylpiperazine derivatives, drugs targeting multiple 5-HT receptors, such as 5-$HT_7$/5-$HT_{1A}$ agonists and 5-$HT_2$ antagonist [403], two highly promising drugs to be tested to ameliorate the core symptoms of ASD.

- **β2-adrenoceptor**

Noradrenaline and adrenaline activate with different potency this class A GPCR family composed of $\alpha_{1A,B,D}$, $\alpha_{2A\text{-}C}$, $\beta_1$, $\beta_2$ and $\beta_3$-adrenoceptors. $\beta_2$-adrenoceptors through recruitment of $G\alpha_s$ protein and β-arrestin induce neuronal stimulation, while via $G\alpha_i$ protein, they inhibit neurotransmitter secretion and neuronal response [404]. Noradrenaline is more potent than adrenaline to activate $\beta_2$-adrenoceptors. Many selective $\beta_2$-adrenoceptor agonists have been developed (full agonists formoterol, salmetrol, zinterol and partial agonist salbutamol) with different efficacy and potency for this family of GPCRs, but only ICI 118551 is a selective antagonist [82]. For example, the approved antagonist propanolol binds to $\beta_2$-adrenoceptors, but also to $\beta_3$-adrenoceptors with lower efficacy. They are expressed on neurons at pre- and post-synapse throughout the brain [405, 406], particularly in the cortex and hippocampus. Whereas their peripheral role in cardiac pathologies has been well described, in the CNS, noradrenaline system controls cognition, memory and emotion and stress-induced behaviour, but the central effect of $\beta_2$-adrenoceptors is still poorly understood. Despite its vital function, genetic deletion of the $\beta_2$-adrenoceptor (*Adrb2* KO) leads to fertile and viable mice (**Table 3**). They display increased anxious-like behaviours and inhibited depression-like behaviours [407].

The *ADRB2* gene in one exon encodes the $\beta_2$-adrenoceptor. Two common SNPs (**Table 2**), Gly16Arg (rs1042713) and Glu27Gln (rs1042714) showed even greater frequencies among individuals with ASD condition [408, 409] than in neurotypical population [12, 13]. Many studies have shown a link between an increase of adrenergic neuronal activity in the locus coeruleus or plasma concentration of noradrenaline and aberrant attentional function and a decrease of interest in external stimuli in ASD individuals [404, 410-414], without affecting the size or number of noradrenergic neurons of the locus coeruleus in ASD post-mortem tissues. Moreover, Gly16Arg and Glu27Gln SNPs display increased isoproterenol-induced desensitization and efficacy respectively [415], suggesting a link between $\beta_2$-adrenoceptor overstimulation and the development of autistic disorders. Administration of MK-801, a NMDA receptor antagonist in zebrafish induce social interaction deficits and anxious-like behaviours, associated with an increase in the $\beta_2$-adrenoceptors density [416]. All these data converge towards upregulation of noradrenalin and $\beta_2$-adrenoceptor system in ASD aetiology [404, 417]. In addition, many studies suggested an association between prenatal exposure to $\beta_2$-adrenoceptor agonists and autism spectrum disorders [408, 418, 419]. The $\beta_2$/$\beta_3$-adrenergic antagonist, propranolol, improve verbal responses and social interactions and decrease anxiety in ASD patients [420-423].

Therefore, $\beta_2$-adrenoceptor is an interesting target for ASD and its pharmacological inhibition with propranolol, an approved drug, should normalize the core symptoms observed in patients. Further investigations should also address the potential interest of the other members of this family, such as *ADRA1D* among the most downregulated genes in the prefrontal cortex of patients [11]. Administration of clonidine, a non-selective and approved drug targeting α-adrenoceptors improves hyperarousal, hyperactivity and social relationships in individuals with ASD [424-428]. Moreover, carriers of two highly frequent mutations, Ser49Gly and Trp64Arg in *ADRB1* and ADRB3 genes respectively [12, 13], may alter receptor functions.

### *5) Other class A receptors*

- **Muscarinic acetylcholine $M_3$ receptors**

In addition to ionotropic receptors, acetylcholine activates five metabotropic class A GPCRs or muscarinic $M_1$-$M_5$ receptors. Whereas many ligands including allosteric modulators have been reported, only few chemical compounds are selective for a muscarinic receptor subtype. The approved drugs Carbachol or pilocarpine described as agonists for $M_3$ also bind to all receptor subtypes or Biperiden antagonist target $M_1$ receptors and $M_3$ with mild lower affinity [82]. So far, only $M_3$ receptors are listed in the SFARI gene database and act through coupling with $G\alpha_{q/11}$, and ERK kinase phosphorylation [429, 430]. They localize in the periphery, where they promote respiration, glandular secretion in salivary glands and smooth muscle contraction and in the CNS and peripheral nervous system. These receptors are present at both pre and postsynaptic sites in the cerebral cortex, striatum, hippocampus, thalamus, hypothalamus, midbrain and pons-medulla. $M_3$ receptors like other muscarinic receptors, play a role in modulation of excitatory transmission, neuronal development including cellular proliferation and survival, neuronal differentiation, food intake, learning and memory [431-433].
Deletion of $M_3$ receptors in mice *(Chrm3* knockout mice) or knock-in mice expressing a mutant receptor, which cannot be phosphorylated in the IL3 has only been described with a significant alteration in hippocampus-dependent contextual fear memory formation and decreased paradoxal sleep [433, 434] (**Table 3**).
The human *CHRM3* gene encoding for $M_3$ receptor is a complex gene harboured on chromosome 1 terminal region that spans over 550Kb, 7 exons with only exon 7 coding $M_3$ receptor, with 10 described and 21 predicted transcript variants. Deletion of this region (1q43), is associated with different clinical phenotypes including ASD, mental retardation and intellectual disability, seizures, microcephaly, congenital malformation [435-439]. Studies have shown that patients diagnosed with above mentioned conditions carry either *de novo* interstitial deletion of 473 Kb or duplication 763Kb at 1q43 region, which includes only the *CHRM3* gene [440, 441] (**Table 2**). Another study involving a child with mental retardation reveal a *de novo* deletion of 911 Kb at 1q43 region, which in addition to *CHRM3,* display deletion of *RPS7P5* and *FMN2* genes [442]. These facts suggest a potential involvement of $M_3$ receptors in ASD aetiology. Reduced cholinergic enzymatic and receptor activity was observed in cortical areas of ASD patients and reduced numbers of $M_1$ receptors were identified in hippocampus of temporal lobe epilepsy patients [443, 444]. Large cohort association studies revealed 2 tolerated missense variants (Ile475Phe and Ile502Val) and Gln588Ter leading to deletion of the C-terminus domain that are also reported in the general population [2, 12, 445].
Few studies reported the association between muscarinic receptors and ASD, alterations in muscarinic receptor expression in ASD tissues, none investigated the social phenotype or stereotyped behaviours induced by muscarinic receptor deletion in animals or the effect of muscarinic ligands in animal models of autism. Therefore, further evidences are needed to conclude on the potential interest of muscarinic receptors as therapeutical targets for ASD.

- **Adenosine receptors**

Adenosine receptors are divided into four different subtypes, namely $A_1$, $A_{2A}$, $A_{2B}$ and $A_3$ receptors [446], among which $A_1$ and $A_{2A}$ show the highest affinity for their endogenous ligand adenosine [447]. Theophylline/caffeine act as a non-selective antagonist for all receptor subtypes, although selective agonists and antagonists have been developed for each subtype.

Despite their interest, many ligands have failed in clinical trials due to severe side effects, only few has been approved such as $A_{2A}$ receptor agonist, Regadenoson [82, 448]. However, CGS21680 is one of the most selective and used $A_{2A}$ agonists to study the role of these receptors both *in vitro* and *in vivo*. In addition, adenosine receptors have important roles in different processes in the brain, such as neuroplasticity, sleep-wake cycle, locomotion, and cognition [449]. Both adenosine $A_{2A}$ and $A_3$ receptors are listed in the SFARI gene database.

**$A_{2A}$ receptors** primarily couple to $G\alpha_s$ but also to $G\alpha_{q/11}$ [450]. When expressed in pre-synaptic glutamatergic neurons, $A_{2A}$ receptors stimulate glutamate secretion in the striatum and the hippocampus, which might be modulated by hetero-oligomerization with $A_1$ receptors. Post-synaptic $A_{2A}$ receptors are highly expressed in $D_2$ GABAergic striatopallidal medium spiny neurons of the striatum [451-457], where they form functionally active hetero-oligomers with $D_2$ and $mGlu_5$ to modulate locomotor activity [452, 458-466], but also with other GPCRs, including $\delta$ opioid receptors and orphan GPR88 [467, 468]. In addition to neurons, $A_{2A}$ receptors in astrocytes of the striatum form hetero-oligomers with OTR where they control evoked glutamate release [469]. Therefore, $A_{2A}$ receptors are involved in the control of locomotion and anxious-like behaviours [470, 471], in addition to myelination by oligodendrocytes [463, 464, 466, 472-474]. The $A_{2A}$ receptors, especially in the striatum, might be particularly of interest for ASD. During brain development, GABAergic synapses, which release adenosine and ATP in addition to GABA, are the first synapses to be formed and are crucial for the construction of the neural network. Activation of adenosine $A_{2A}$ receptors is necessary and sufficient to prune GABAergic synapses during development [475]. In contrast, any impairment in $A_{2A}$ signalling or expression may result in GABAergic synapse impairment and cognitive deficits in adults, as observed in animals administrated with $A_{2A}$ antagonists during development [475, 476].

Mice deleted for $A_{2A}$ receptors in mice (*Adora2a* KO) display motor impairment and anxious-like behaviours [449, 477] (**Table 3**). Further evidence points for the therapeutic potential of $A_{2A}$ receptors in preclinical mouse models of ASD. Spontaneous stereotypies observed in animal models, result from unbalanced cortical glutamatergic and GABAergic afferences (glutamate hyperactivity) on the striatum and decreased activation of the efferent subthalamic nucleus [478-480]. This is consistent with an excessive functional connectivity in striatal-cortical circuits, for review see [481]. Administration of CGS21680, a selective $A_{2A}$ receptor agonist, normalizes aberrant vertical repetitive behaviours, alone in the BTBR mice, a mouse model of idiopathic autism, and in combination with CPA $A_1$ agonist in C58 inbred mice [480, 482, 483]. Therefore, A2A (and $A_1$) receptors activate $D_2$ (and inactive $D_1$) medium spiny neurons, which in turn restores activation of the subthalamic nucleus.

In humans, *ADORA2A* gene encodes seven splicing transcript variants, which differ in their 5' untranslated region and one $A_{2A}$ receptor protein. One study has associated four SNPs in the *ADORA2A* gene, rs2236624 (21460T>C), rs35060421 (23351 dupTTTTTT) and rs3761422 (12108T>C) in introns and synonymous rs5751876 (1083T>C), to men with autism condition, with an history of severe anxiety [451] (**Table 2**). In addition, carrier of only one C allele of the rs5751876 SNP linked the ADORA2A gene with panic syndrome [484, 485]. In addition, *ADORA2A* and *ADORA2B* genes belongs to the GPCR genes that display the most duplications in the general population [12, 13]. Therefore, even if the same study already tested it in the small cohort of patients [451], copy number variants linked to ASD might also be identified in the future.

**$A_3$ receptors** couples to $G\alpha_{i/o}$ proteins [450]. In contrast to $A_{2A}$ that have a restricted pattern of expression, $A_3$ receptors are widely expressed in the cortex, hippocampus and cerebellum [446]. $A_3$ receptors has been shown to promote cell surface expression of the serotonin transporter SERT via the protein kinase G (PKG) [486]. Mice deleted for $A_3$ receptors (*Adora3* KO) seem to be prone to anxiety and behavioural despair [449, 477] (**Table 3**). *ADORA3* gene encodes 4 transcript variants and four different $A_3$ receptors and is interleaved with *TMIGD3* gene, sharing the same first exon. Two variants coding for Leu90Val and Val171Ile $A_3$ mutant receptors have been linked to autism [486]. As these two mutations probably impair adenosine binding, $A_3$ would not regulate serotonin transporter, leading to increased levels of serotonin observed in ASD patients (see 4-serotonin section).

In conclusion, whereas more evidences are needed for $A_3$ receptors, $A_{2A}$ receptors show the best potential for ASD via their mechanism of action, the existence of reported pathogenic variants and their potential as treatment for repetitive behaviours and anxiety. Whereas many $A_{2A}$ agonists have been tested in clinical trials, most of them failed due to severe peripheral side effects such as dyspnea, bradycardia, increased urinemia and creatinemia, and central excitotoxicity. In one hand, an alternative strategy would be to consider other members of this family, such as *ADORA1* gene which is one of the most downregulated GPCR genes in the prefrontal cortex of ASD patients [11] and its administration in combination to $A_{2A}$ agonists improve stereotyped behaviours [480]. One the other hand, a better specificity could be achieved by targeting $A_{2A}$ hetero-oligomers without the side effects, such as $D_2$-$A_{2A}$-$mGlu_5$.

- **Angiotensin $AT_2$ receptor**

Angiotensin receptors are divided in $AT_1$ and $AT_2$ subtypes. They are activated by different maturation of angiotensinogen peptides, angiotensin II and III. Selective agonists and antagonists have been developed for these two family members, with EMA401 $AT_2$ antagonists stopped due to potential long-term hepatotoxicity in phase 2 clinical trials for chronic pain [82, 487]. $AT_2$ receptors couple only to $G\alpha_{i/o}$, leading to activation of the neuronal nitric oxide synthase and production of second messenger cGMP, which induces neurite outgrowth and elongation. Additionally, $AT_2$ receptors are able to activate signalling cascades through G protein-independent coupling. More specifically, after stimulation of the $AT_2$ receptor, $AT_2$ receptor-interacting protein (ATIP) interacts with the tyrosine phosphatase SHP-1, and translocates to the nucleus in such a complex where it activates ubiquitin-conjugating enzyme methyl methanesulfonate sensitive 2 (MMS2), thus underpinning neuronal differentiation [488]. $AT_1$ receptor is expressed in several different tissues where it mediates well known functions of angiotensin such as vasoconstriction, cellular growth and proliferation. However, $AT_2$ receptor is located only in limited tissues and brain areas, such as medulla oblongata, septum, amygdala and cerebellum, while its functions remain elusive. Consequently, only $AT_2$ receptor and angiotensin system is implicated in different neurological disorders such as schizophrenia, autism and Parkinson and Alzheimer diseases [489, 490] and in the SFARI list.

Deletion of $AT_2$ receptor (*Agtr2* KO) in mice results in an increased number of cells in different regions, such as the piriform cortex, hippocampus, amygdala, and different thalamic nuclei [491]. Additionally, $AT_2$ receptor knock-out mice show impaired drinking response to water deprivation, reduction in spontaneous movement, and reduced exploratory behaviour [492, 493](**Table 3**). Up to date, there are no data indicating impairments in social or stereotyped behaviour of this mouse model. Interestingly, the administration of $AT_2$ receptor selective agonist C21/M024 improved spatial memory in a mouse model of Alzheimer's disease [494]. However, despite the fact that $AT_2$ receptor ligands might be used to improve cognition, for the moment, there are no specific ligands targeting this receptor that could be used in the treatment of ASD.

The *AGTR2* gene in X chromosome encode two transcript variants and one protein, the $AT_2$ receptor. Eight rare and loss-of-function SNPs in the *AGTR2* gene have been associated to ASD (**Table 2**) with seven in the coding sequence and one translocation. A rare stop gained variant, rs1569536545 (Gln253Ter), truncated the protein in TM6 has been found in Vietnamese ASD children [495]. The variants 572G>A (Gly191Glu), rs387906503 (Phe134Leu-Ter) and rs121917810 (Gly21Val) have been identified in patients suffering from X-linked mental retardation that also showed autistic behaviour [496, 497]. These mutants affect ligand binding, lead to a frameshift in TM3 and cause a substitution in a conserved residue in the N-terminus. Finally, one deleterious variant rs1049717729 leading to a Tyr82His substitution in TM2 has been associated with obsessive-compulsive disorder [498], potentially affecting folding of the receptor.

In conclusion, several studies and polymorphisms in *AGTR2* gene have been associated with increased risk of ASD in humans, indicating the potential involvement of this receptor in ASD aetiology. However, autistic-like behaviour was not investigated in $AT_2$ receptor-linked animal models, thus preventing us to conclude on their validity. Additionally, vague functions of this receptor as well as the absence of specific ligands, which could be used in autism treatment indicate that the $AT_2$ receptor might be an interesting target in autism treatment if further studies are pursued.

- **Cannabinoid $CB_1$ receptor**

Cannabinoid class A receptors are composed of $CB_1$ and $CB_2$ that share 48% amino acid sequence similarity. Whereas $CB_2$ is expressed in the periphery, mostly associated with the immune system and inflammation, $CB_1$ mediates all the central effect of cannabis. Cannabis derivates, cannabidiol (negative allosteric modulator of $CB_1$) or delta-9-tetrahydrocannabinol ($CB_1$ and $CB_2$ partial agonist) are approved therapeutic agents for many neurological applications [82], also act on GPR55 and GPR18. Agonists such as anandamide (N-arachidonoyl-ethanolamine, AEA) or the endogenous ligand 2-arachidonoylglycerol (2-AG) activates both the $CB_1$ and $CB_2$ receptors, revealing a coupling to $G\alpha_{i/o.}$ Activation of $CB_1$ triggers different signalling pathways in the CNS, which involves $K^+$ and $Ca^{2+}$ channels regulation and neurotransmitter release at both excitatory and inhibitory pre-synapses [499]. Activation of $CB_1$ receptors also result in regulation of ERK1/2 and phosphatidylinositol-3 kinases (PI3K) [500]. The endogenous cannabinoid system regulates the dopamine circuits, which are crucial for reward processes linked to addiction [501]. In the brain, $CB_1$ receptors, mostly the reference isoform [8], are expressed on presynaptic terminals and are enriched in brain structures such as frontal cortex, hippocampus, basal ganglia, amygdala, hypothalamus, midbrain, cerebellum. They regulate the synaptic transmission through neurotransmitter modulation and cannabinoid system have been shown to regulate different behavioural and physiological functions.

Deletion of $CB_1$ receptor (*Cnr1* KO) in mice display deficits in social behaviour and social communication, two core symptoms related to ASD [502] (**Table 3**). *Cnr1* KO mice also exhibit anxiogenic, context-dependent social aggressive behaviour, depressive behaviour and working memory deficits compared to control mice. Further *Cnr1* KO mice exhibited improved learning in active avoidance behaviour and social memory suggesting a role for $CB_1$ receptors in different behavioural components [503-508]. *Cnr2* KO mice display deficits in social memory in an age-dependent manner, contextual fear memory but enhanced spatial working memory and aggressive behaviour compared to control mice [509-512]. Indeed, endocannabinoid mediated signalling has been employed to improve social behaviour including social anxiety and reward in BTBR and *Fmr1* KO mice [513].

$CB_1$ receptors are encoded by the *CNR1* gene in humans. Three independent studies have reported a positive association between variants in this gene and ASD (**Table 2**). In fact, the two SNPs rs806380 (in intron 1) and rs806377 associated with developing cannabis dependence symptoms, have been also associated with differential gaze duration towards striatal response to the happy faces in children with ASD [514-517]. These results suggest that modulation of $CB_1$ receptors might cause a social reward deficit, an aspect linked to ASD. Further, reduced *CNR1* gene and protein expression was observed in ASD post-mortem brain samples [518, 519]. Cannabidiol or endocannabinoid mix targeting $CB_1$ receptors are currently in phase 1 clinical trials for ASD [23, 82] (https://clinicaltrials.gov/). Few states in the United States of America have authorized cannabis to treat ASD in certain cases associated with self-injurious or aggressive behaviour. The delta-9-THC (dronabinol) in a single case study, have shown to improve the hyperactivity, irritability and stereotyped behaviour in a boy with ASD [520]. However, clinical trial with a combination of cannabidiol and delta-9-tetrahydrocannabinol (from plant extract or purified products) results in mitigated results, requiring further extensive testing. However, these combinations were well tolerated and results in no side effects [23].

In conclusion, $CB_1$ receptors with their involvement in social impairment and potentially repetitive behaviours related to ASD, cannabinoids with ongoing research developments in clinical trials, could be useful therapeutics treating core and associated symptoms in ASD.

- **Chemokine $CX_3CR1$ receptor**

Chemokine receptors are a vast family of class A GPCRs involved in the immune system. The C-X3-C motif chemokine receptor 1 ($CX_3CR1$ or GPR13) receptor binds the secreted and membrane-bound chemokine CX3CL1 (also called fractalkine or neurotactin). Both chemokine isoforms similarly activate $CX_3CR1$, leading to $G\alpha_{i/o}$ protein coupling and β-arrestin recruitment [521]. Studies reported dysregulation of NFκB, CREB and ERK phosphorylation in absence of $CX_3CR1$ [522]. Apart for its role in lymphocytes and monocytes chemotaxis and HIV-1 co-receptor, the highest level of $CX_3CR1$ receptor expression is observed on microglia, where it is activated by CX3CL1 release from neurons upon inflammatory response and during synaptic maturation and pruning [523-525]. This highlights its potential major role in neuron-glia mutual interaction. Deletion of $CX_3CR1$ in mice (*Cx3cr1* KO) lead to social interaction deficits, subordination and increased grooming [525, 526] (**Table 3**), in addition to decreased functional brain connectivity from the prefrontal cortex, an ASD feature. Moreover, in animals exposed to social isolation, levels of *Cx3cr1* transcripts were increased in the prefrontal cortex, nucleus accumbens and hippocampus [527].

The *CX3CR1* gene is located in human chromosome 3 encoding for 4 transcript variants and two protein isoforms that differ in their N-terminus domain [8]. One publication has reported three rare missense deleterious mutations in the $CX_3CR1$ receptors associated to schizophrenia and ASD [528] (**Table 2**) that are also present in the neurotypical population [12, 13].

Based on its interesting crosstalk between neurons and microglial cells and the effect of its deletion or deleterious mutations, activation of this GPCR might be promising to treat ASD. However, further proof of its potential effect in preclinical models are needed and drug development might be necessary to prevent the unwanted effect of $CX_3CR1$ receptors activation in immune cells.

### *6) Orphan and olfactory receptors*

Orphan and olfactory GPCRs have highly interesting therapeutical targets considering their specific function, localisation in the CNS and their potential dysregulation in ASD aetiology. Due to the lack of any clear identified endogenous ligand or function, their study remains challenging. Except their classification by sequence homology to the class A of GPCRs, their molecular activation mechanism is also unknown with no available structure for the orphan GPR37 and GPR85, neither the 4 olfactory OR1C1, OR2M4, OR2T10, OR52M1 receptors. They have been identified as category 2 strong candidate from the SFARI and GPR37 and GPR85 are the top dysregulated GPCR genes in ASD tissues [11].

- **GPR37**

GPR37 or parkin-associated endothelin-like receptor (Pael-R) and its homologue glial GPR37 like 1 (GPR37L1) are closely related to endothelin-related GPCRs. Studies reported potential natural peptide ligands for GPR37, head activator from hydra, neuroprotective and glioprotective prosaposin and prosaptide, regenerating islet-derived family member 4 (REG4), and the neuroprotectin D1 [82, 529-532]. Whereas its real endogenous ligand remains to be determined or confirmed from those candidates, all studies converge towards a $G\alpha_{i/o}$ protein coupling, $Ca^{2+}$ intracellular mobilisation through probably $G\beta\gamma$ protein-induced GIRK channels, phosphorylation of ERK and Akt kinases and *Fosb* immediate-early gene regulation [529-534]. GPR37 is characterized by a poor export from ER to plasma membrane in heterologous cell lines, which is rescued by deletion of its long N-terminus domain, through oligomerization with the adenosine $A_{2A}$ or dopamine $D_2$ receptors or interaction with synthenin 1 via its PDZ-binding domain [535, 536]. GPR37 mostly express in the brain and exert different function depending on its (sub-)cellular localisation in oligodendrocytes (white matter), in neurons of the cerebellum, striatum and hippocampus [530, 537, 538]. It has also been reported in macrophages, and not in microglia, where it regulates their phagocytosis and inflammatory processes [530]. In oligodendrocytes, GPR37 is up-regulated during their differentiation where it inhibits late-stage differentiation and their myelination [534]. Gpr37 is located in dopaminergic axon terminals of the substantia nigra where it controls dopamine release through a direct interaction with the dopamine transporter DAT [539]. Interestingly, Parkinson disease and ASD seem to share some overlapping mechanisms. Pathogenic mutations in the *PRKN* gene coding for the parkin E3 ubiquitin ligase causing juvenile forms of Parkinson disease, lead to accumulation of misfolded GPR37 in the ER, causing stress and autophagy [537, 540]. Moreover, aberrant copy number variants of *PRKN* variants have been associated to ASD (also identified as a strong candidate in SFARI gene database) and *Prkn* KO mice display social and communication deficits and stereotyped behaviours [541].
Deletion of GPR37 (*Gpr37* KO) in mice lead to obsessive compulsive behaviour, decreased locomotion, reduced colon motility and abnormal sensory system (sound and smell) [539, 542-544] (**Table 3**). They may display increased anxious-like behaviours, which is not consistent on the tests, sex, housing conditions and labs. Conversely, transgenic mice overexpressing *Gpr37* display increased methamphetamine-induced stereotyped behaviours, motor coordination and locomotion [542]. Mouse models targeting *Gpr37* do not fully recapitulates primary symptoms as social behaviour remains to be investigated, but rather display large variety of comorbid symptoms of ASD. Both mouse models have altered striatal signalling, a feature of ASD, for review see Li and Pozzo-Miller [481] and drastic dopamine defects in the striatum. Interestingly, ASD-associated mutations in the dopamine transporter DAT [545], a direct interactor of GPR37, show similar alteration in dopamine transmission in the striatum than *Gpr37* mice.

*GPR37* gene spans over two exons, with one transcript variant and one receptor. Several evidences point towards *GPR37* gene as strong candidate for ASD. *GPR37* belongs to the genes identified in the first human autism genetic locus (AUTS1) on chromosome 7q31–33. Four variants in the *GPR37* gene have been associated with ASD [546] (**Table 2**) leading to receptor truncation in TM2 (Phe312Del) and three substitutions in GPR37 intracellular domain, Ile469Met in IL3 and Arg558Gln and Thr589Met in the C-terminus. Arg558Gln mutation, which potentially destabilize the 8$^{th}$ helix of the C-terminus, has been shown to prevent interaction of GPR37 with the scaffolding multiple PDZ domain protein 1 (MUPP1) and contactin-associated protein-like 2 (CASPR2) causing dendritic alteration [547]. *GPR37* has been found to be one of the most downregulated GPCR transcript in post-mortem tissues from prefrontal cortex of ASD patients [11].

Therefore, these multiple evidences confirm that GPR37 might be an interesting target for ASD. However, selective ligands should be developed and tested in preclinical models to further strengthen its potential for ASD.

- **GPR85**

GPR85/SREB2 belongs to the super-conserved receptor expressed in brain (SREB) family with orphan GPR27 and GPR173. Both rodent to human GPR85 display the exact same sequence of 370 amino acids. No natural ligand has been discovered so far and compounds reported as inverse agonists, which act on the 3 SREB receptors revealed a potential coupling to $G\alpha_s$ protein [548]. The PDZ-binding domain in GPR85 C-terminus allows direct or indirect interactions with Shank3 or PSD95 scaffolding protein, and indirectly to neuroligin through PSD95 [549, 550]. Despite its expression in all type of neurons, it is enriched in the adult hippocampus (especially dentate gyrus), subventricular zone of the amygdala and in Purkinje cells of the cerebellum [548, 551]. In adult hippocampus, it regulates negatively neurogenesis and dendritic morphology, thereof controlling brain size and cognitive abilities in spatial tasks [552].

The deletion of GPR85 (*Gpr85* KO) in mice display increased neurogenesis associated with enlarged brain size and increased cognitive abilities in spatial tasks [551, 552](**Table 3**). Conversely, mice overexpressing *Gpr85* in forebrain neurons show core symptoms of ASD, social interaction deficits and restrictive behaviours, in addition to cognitive inabilities and abnormal sensorimotor gating and reduced dendritic arborisation [551, 552].

In humans, *GPR85* gene is a relatively complex gene spanning over 4 exons and coding for 3 transcript variants and 7 predicted transcript variants with alternative splicing in exons 2-4 coding for the 3' untranslated region, suggesting specific regulation of these mRNAs. However, all isoforms encode for the extremely conserved receptor, encoded only in the first exon. Two substitution mutations in human *GPR85* gene have been identified in Japanese ASD patients [549] (**Table 2**). Met152Thr in TM3 and Val221Leu in IL3 potentially impair respectively receptor activation and/or proper receptor folding and recruitment of signalling proteins respectively. Both mutants and scaffolding partner PSD95 showed increased accumulation in the endoplasmic reticulum and reduced dendrite length, suggesting receptor misfolding. In addition, two SNPs in intron 2 and in exon 3 have been described as risk alleles for schizophrenia [551], which probably affect their levels of expression. Consistent with its expression profile [551, 552], these SNPs (rs56080411 and s56039557) associated with decreased hippocampal formation, dysregulated pattern of activation in the left amygdaloid complex, the right inferior frontal gyrus and the right occipital cortex and fusiform gyrus in emotion and memory paradigms [553]. Furthermore, *GPR85* has been shown to be one of the top downregulated GPCR mRNA in post-mortem tissues from ASD patients [11]. Those data are also coherent with increased Gpr85 mRNA levels in the striatum and medial prefrontal cortex of Shank3 overexpressing mice [550], and decreased *GPR85* splicing events in the cortex of splicing factor A2BP1-related ASD patients compared to control samples [554].
Despite studies on GPR37 remain sparse, data from mice and patients highlight the contribution and potential of GPR85 to a ASD and schizophrenia continuum of neurodevelopmental psychiatric disorders with social interaction deficits.

- **Olfactory receptors**

In humans, 387 genes encode olfactory receptors (OR) genes and 462 olfactory pseudogenes. Apes have roughly the same number of OR whereas rodents display more than 1100 ORs and 200-450 OR pseudogenes. ORs, entirely encoded by a single exon, are subdivided in aquatic ancestry class I receptors clustered on human chromosome 1 (OR1-15) and the largest terrestrial ancestry class II (OR51-56), located on other chromosomes, except 20 and Y [555]. ORs detect odorant volatile molecules, although most of them remain orphan. In the olfactory neurons, they couple to $G\alpha_{olf}$ protein, which activates adenylyl cyclase 3 and cAMP production resulting in calcium influx through cyclic nucleotide-gated channels. Since their discovery, growing evidences showed OR expression outside the olfactory epithelium, primarily in testis and, then in most tissues. OR and $G\alpha_{olf}$ are widely detected in the human and rodent CNS, in dopaminergic neurons of the substantia nigra, in pyramidal neurons of the cortex, in the hypothalamus, striatum, brainstem and hippocampus for example [556]. In contrast to olfactory neurons, which express a single OR, ORs and OR pseudogenes are co-expressed in brain cells, similarly to other GPCRs. Often qualified as 'ectopic' expression outside olfactory epithelium, their expression is rather conserved among species [557, 558] and their roles in development, chemotaxis, tissue injury and regeneration start to be deciphered.
In addition to the involvement of different ORs in Parkinson disease, Alzheimer disease and schizophrenia [556], three class II OR1C1, OR2M4 and OR2T10 and one class I OR52M1 ORs are listed in the SFARI gene database and are associated with ASD risk (**Table 2**). Except the highly conserved OR52M1, the other three are conserved only in apes and for some also in cows. No animal models have been developed for these receptors. Four different studies revealed 5 CNVs (4 copy number loss and 1 copy number gain) and two deleterious missense variants rs1278811157 (Leu40Phe), rs554796998 (Tyr123Ter) have been associated to ASD risk [559-562] in the *OR1C1* gene (previously named *HSTPCR27*, *OR1-42*, *HsOR1.5.10*, *ORL3029*, *ORL211*). One study reported two SNPs in the 3'UTR of the *OR2M4* gene (previously named *OR1-55, HSHTPCRX18, OST710, HsOR1.5.32, ORL203*) in Taiwanese ASD patients that may alter mRNA expression and three SNPs in the 3 Kb downstream region of the same OR and *OR2M3* genes [563]. This association is rather weak as they are located downstream of the genes and are already all identified in the general population [215]. On deleterious variant and one truncation in the C-terminus were reported for the *OR2T10* gene (*OR1-64*, *HsOR1.5.46*) [215, 564]. Two independent studies supported the potential of the *OR52M1* gene (previously known as *OR11-11*, *HsOR11.3.10*) with four different deleterious SNPs in the same location and one stop gained mutation in TM2 [215, 560]. However, none of these genes were detected in OR transcriptomic data of the human whole brain [558]. OR1C1 was highly enriched in testis tissues and detected in olfactory epithelia, OR2M4 was only enriched in olfactory epithelia, OR2T10 was enriched in kidney and OR52M1 was barely detected in any tissue. However, these ORs may have discrete and/or low expression in the brain, explaining their lack of detection. Resolution of the whole GPCR landscape within brain structures and cells would provide more evidences for these ORs.

In conclusion, genomic data showed some association of these 4 OR genes to ASD risk, which are the strongest for OR1C1. Nonetheless, other ORs might be more interesting for ASD as over hundred OR and OR pseudogenes were detected in the human transcriptomic data. In fact, deleterious variants in many other OR genes were also identified in ASD association studies [560, 562]. Further investigations are required to understand if specific OR are involved is ASD, or if dysregulated or pathogenic variants of OR genes are shared among neurological disorders and in particular schizophrenia. These receptors have a great potential for ASD in a far future. Their exact localisation, expression and function in the CNS remains to be identified as well as the development of selective drugs or the identification of their natural ligands.

## GENERAL CONCLUSIONS and FUTURE DIRECTIONS

Two main gaps remain in ASD research, the identification of robust and convergent targets and the discovery of new drugs to improve core symptoms of ASD. Recent efforts lead to the identification of convergent signalling pathways [2, 3, 5]. GPCRs, being per nature upstream receptors, control these downstream signalling and thus meet the criteria of convergent targets for ASD. As an example of their potential, we identified that 15% of the genes listed in the SFARI are involved in GPCR activity (ligand binding and release or direct and downstream effectors), including the 23 GPCRs we discussed in this review. Except olfactory receptors for which this information is not available, we reported that all the 19 GPCRs have been shown to control at least one of the key convergent signalling that are altered in ASD. Therefore, activation or inhibition of these targets would restore this signalling, highlighting their therapeutical potential.

We also listed clear evidences of their involvement in ASD aetiology. Based on genetic studies (number of independent reports and pathogenic variants that are not identified in the general population; **Table 2**) and evidence in post-mortem tissues, their function and specific localisation and their behavioural predictive validity in animals (**Table 3**), we further classify these 23 GPCRs regarding their involvement in ASD aetiology. For now, we identified the OTR, $V_{1A}$, $mGlu_5$, $D_2$, 5-$HT_{2A}$, $CB_1$ and GPR37 receptors that fulfil most, if not all, criteria, with OTR, $CB_1$ and $V_{1A}$ that display a clear potential for social behaviour and $D_2$, 5-$HT_{2A}$, and GPR37 for stereotyped and repetitive behaviours, and $mGlu_5$ eventually for both. Furthermore, based on recent findings in animal studies, OTR, $CB_1$ and 5-$HT_{2A}$ might also be interesting targets for the two core symptoms. In this review, we also listed variants in introns, untranslated region or coding regions that may explain the decreased levels of expression of these GPCRs observed in patients [11], particularly $mGlu_7$, 5-$HT_{2A}$, $CB_1$, GPR37 and GPR85 receptors. Most variants in the receptors are in the coding sequence affecting protein folding, ligand binding or activation of downstream signalling pathways. A good example is the orphan GPR85, which is strictly conserved between humans and rodents, thus any SNP or copy number variants may lead to profound effect, such as brain size control.

Overall, we wonder about the inclusion of relatively few (only 23) GPCRs in the list and only in the second category. Here, we highlight that OTR, $CB_1$ and $V_{1A}$ receptors should deserve a higher ranking than only strong candidate genes in the SFARI database and 5-$HT_{2A}$ should be listed in the SFARI list. GPCRs is the largest family of receptors, comprising over 800 active GPCR genes including for most of them, several splicing variants and hundreds of pseudogenes. In addition, genomic data have shown that hundreds of variants are present in the repertoires of the most studied GPCR genes [12, 13]. Further, increasing amounts of evidences showed the functional relevance of GPCR crosstalk within a given cell, such as $D_2$, $A_{2A}$ and $mGlu_5$ to control motor function for example [452, 458-466]. In fact, independently of their physical interaction, GPCRs are not individual entity but should rather be considered as a bunch of GPCRs and isoforms working together in a given cell to orchestrate the different signals and regulate downstream signalling and cellular processes. Studies, excluding olfactory receptors, showed that up to 100-300 of GPCRs could express in the same brain structures or cell types, with the highest diversity in the striatum, cortex and hypothalamus [8, 565]. Therefore, the role of GPCRs for autism research has only begun and rather than a single gene, GPCRs should be considered as one global functional entity of GPCRs expressed in a given cell to study their genetic contribution to ASD aetiology.

The second main gap in ASD research is the development of safe and efficient drugs that pass the placebo effect observed in clinical trials. Due to their intrinsic nature, GPCRs fulfil the criteria of valuable therapeutical targets for ASD: they contribute to the polygenic aetiology, pathogenic variants are recessive, they are therapeutically rescuable and they are upstream receptors that control convergent mechanisms affected in patients. Therefore, as an example for further investigations on the 800 GPCRs, we also review the therapeutical potential of these 23 GPCRs and categorize them as 'high', 'moderate' and 'low' candidates, based on drugs that are already approved or tested for a relative disorder (e.g., schizophrenia), their beneficial effects in animal models of ASD or in clinical trials, their pharmacogenomic and their safety. We identified that $mGlu_5$, $GABA_B$, $D_2$, 5-$HT_{2A}$, 5-$HT_7$, $CB_1$ are 'high' therapeutical candidates and OTR, $V_{1A}$, $mGlu_7$, $D_1$, 5-$HT_{1B}$ and $M_3$ as 'moderate' candidates. We excluded $\beta_2$-adrenoceptors, $A_{2A}$, and $AT_2$ as toxicity or severe side effects have been reported in clinical trials. Despite their strong potential, $mGlu_5$ and 5-$HT_{2A}$ display the highest number of variants, which might impair their responsiveness to drugs [12] and explain why $mGlu_5$-targeted clinical trials have failed. Finally, despite their strong potential in the future, orphan and olfactory receptors belong to the 'low' category as no natural ligands or drugs have been clearly identified. We also ranked $CX3CR_1$, $V_{1B}$, $D_3$ and $A_3$ as 'low' candidates, as no selective ligand has been developed and further investigations remains to be conducted to understand if they should be activated or inhibited and their real contribution for sociability and stereotyped or repetitive behaviours. Therefore, considering, their involvement in ASD aetiology, their therapeutical potential and results of clinical trials, out of 23 genes, we conclude that $D_2$, 5-$HT_{2A}$, $CB_1$, OTR, $V_{1A}$ and GPR37 are promising target for clinical development. $D_2$, 5-$HT_{2A}$ and $CB_1$ are already ongoing for irritability, repetitive behaviours and aggressive and self-injury behaviours, but could be tested on other core symptoms, especially $CB_1$ on social scales. In a near future, when specific ligand would be developed, OTR, $V_{1A}$ and GPR37 should be tested as well. Recently, new class of compounds, the biologicals (e.g., antibody fragments) targeting GPCRs appeared [14] or new high throughput screening methods, such as DNA-based bar-coded chemical libraries [566] will boost the identification of new drugs to target GPCRs, including orphan and olfactory receptors. Interestingly, biologicals can be easily encoded and vectorized in viral particles for long-term effects without re-administration of the compounds.

These compounds display all types of pharmacological profiles, but can also be used as chaperones, or target oligomers of GPCRs. Considering that GPCRs do not function as single receptor but rather as global GPCRs expressed in a cell, a similar approach might also apply for future drug development. The use of several drugs or a drug targeting multiple GPCRs within these global GPCR entities might be relevant for ASD. Such drugs already exist. For example, the approved drugs for irritability, aggression and repetitive behaviours in ASD, aripiprazole, is a $D_2$ antagonist, a partial 5-$HT_{1A}$ agonist and 5-$HT_{2A}$ antagonist, and risperidone is a $D_2$ antagonist, a 5-$HT_{2A}$ antagonist and a partial agonist for histamine $H_1$ and $\alpha_1$ and $\alpha_2$-adrenergic receptors. In addition, arylpiperazine derivatives has been developed to target multiple 5-HT receptors [403]. Based on their fine tune pharmacology (biased ligands, oligomerisation, global GPCRs entities) and their diversity, GPCRs represent the greatest therapeutical options for ASD and lead to successful clinical trials.

**Data and Software Availability**

No data are associated with this article. All 1045 genes were downloaded from Simons Foundation Autism Research Initiative (SFARI; SFARI-Gene_genes_05-05-2022 release) genes database (https://gene.sfari.org/). Gene scoring are defined as syndromic and category 1, 2 and 3 respectively accounting for high confidence, strong candidate and suggestive evidence (https://gene.sfari.org/about-gene-scoring/). We used *Cytoscape v3.9.1 StringApp v1.7.1* [1,2] to compare the SFARI genes list to the Gene Ontology [3,4] term '*G protein-coupled receptor signaling pathway*' (GO:0007186, v2022-03-14), the REACTOME [5] pathway '*Signaling by GPCR*' (10.3180/REACT_14797.1, v81 2022-06-16) and the KEGG [6] (v103 2022-07) pathway '*cAMP signaling pathway*' (hsa04024 ) to retrieve SFARI GPCR genes or SFARI genes related to GPCR activity/signaling (93 genes retrieved). We added all 113 SFARI GPCR-linked genes found in the KEGG pathways '*Oxytocin signaling pathway*' (hsa04921), '*Dopaminergic 1CNRCNRCN*' (hsa04728), '*Retrograde endocannabinoid signaling*' (hsa04723) and '*Renin-angiotensin system*' (hsa04614) as all these signaling pathways are solely activated by GPCR, with the exception of *GABRA3*, *GABRA4*, *GABRB2*, *GABRB3* and *GABRG3* from the '*Retrograde endocannabinoid signaling*' pathway that we considered off-target for our analysis (20 genes were retrieved). We then added SFARI genes from part of the KEGG pathways '*Glutamatergic synapse*' (hsa04724), '*Cholinergic synapse*' (hsa04725), '*Serotoninergic synapse*' (hsa04726), '*Adrenergic signaling in cardiomyocytes*' (hsa04261), '*Chemokine signaling pathway*' (hsa04062), '*Synaptic vesicle cycle*' (hsa04721) and '*Olfactory transduction*' (hsa04740) if those genes were related to GPCR activation (+11 genes retrieved, 124 total). Finally, using our knowledge, we manually added 22 genes not present in these pathways/term (*ADA*, *AKAP9*, *CASK*, *CTNNB1*, *CYFIP1*, *EGR3*, *EIF3G*, *EIF4E*, *EIF4G1*, *FMR1*, *FOXP1*, *FOXP2*, *MTOR*, *PRKD1*, *PRKD2*, *TSC1*, *TSC2*; for literature references see **Table 1**). Overall, at least 152 out of the 1045 SFARI genes are involved in GPCR related cellular processes. Identical nomenclature and classification have been used for all GPCRs throughout the study [7].

**Author Contributions**

All authors provide substantial contributions: AA and LP for Conceptualization, AA, CG, AD, XL, VB, PC and LP for Data Curation, AA, PC and LP for Formal Analysis, LP for Funding Acquisition and Project Administration, LP and AA for Supervision, AA, CG, AD, XL, VB and LP for Writing – Original Draft Preparation, AA, PC and LP for Writing – Review & Editing

**Competing Interests**

No competing interests were disclosed.

**Grant Information**
This project has received funding from the European Research Council (ERC) under the European Union's Horizon 2020 research and innovation programme (grant agreement No. 851231). LPP, AD and CG acknowledge the LabEx MabImprove (grant ANR-10-LABX-53-01) for their financial support of their PhD co-fund.

**Abbreviations**

5-HT 5-hydroxytryptamine
ASD autism spectrum disorders
cAMP Cyclic adenosine monophosphate
CNS central nervous system
CNV copy number variants
CREB cAMP response element binding protein
DAG diacylglycerol
ER Endoplasmic Reticulum
ERK Extracellular signal-regulated kinases
FXS Fragile X syndrome
GIRK G protein-coupled inwardly-rectifying potassium channels
GPCRs G protein-coupled receptors
$IP_3$ inositol 1,3,4- triphosphate
KO knockout
KI knock-in
KD knock-down
OR Olfactory receptor
PKA protein kinase A
PLC Phospholipase C
RTT Rett syndrome
SNP Single nucleotide polymorphism
Tg Transgenic
UTR untranslated region
WT wildtype

**Table 1. List of 152 genes involved in GPCR activity and signalling processes extracted from the last SFARI release**

| Ensembl-id | Gene-symbol | Gene-name | Figure based term | KEGG, GO | Gene-score | Syndromic | References |
|---|---|---|---|---|---|---|---|
| ENSG00000159640 | ACE | angiotensin I converting enzyme | metabolic enzyme | 8 | 2 | 0 | |
| ENSG00000087085 | ACHE | Acetylcholinesterase (Yt blood group) | metabolic enzyme | 11 | 2 | 0 | |
| ENSG0000196839 | ADA | adenosine deaminase | metabolic enzyme | | 2 | 0 | PMID: 29497379 |
| ENSG00000132437 | DDC | dopa decarboxylase | metabolic enzyme | 5, 12 | 2 | 0 | |
| ENSG00000189221 | MAOA | monoamine oxidase A | metabolic enzyme | 5, 12 | 2 | 0 | |
| ENSG00000069535 | MAOB | monoamine oxidase B | metabolic enzyme | 5, 12 | 2 | 0 | |
| ENSG00000073756 | PTGS2 | prostaglandin-endoperoxide synthase 2 | metabolic enzyme | 4, 6, 12 | 2 | 0 | |
| ENSG00000106688 | SLC1A1 | solute carrier family 1 (neuronal/epithelial high affinity glutamate transporter, system Xag), member 1 | neurotransmitter transporter | 7, 1 | 2 | 0 | |
| ENSG00000110436 | SLC1A2 | Solute carrier family 1 (glial high affinity glutamate transporter), member 2 | neurotransmitter transporter | 7, 1 | | 1 | |
| ENSG00000157103 | SLC6A1 | Solute carrier family 6 (neurotransmitter transporter), member 1 | neurotransmitter transporter | 7 | 1 | 0 | |
| ENSG00000142319 | SLC6A3 | Solute carrier family 6 (neurotransmitter transporter), member 3 | neurotransmitter transporter | 5, 7 | 2 | 0 | |
| ENSG00000108576 | SLC6A4 | solute carrier family 6 (neurotransmitter transporter, serotonin), member 4 | neurotransmitter transporter | 7, 12 | 2 | 0 | |
| ENSG00000042753 | AP2S1 | adaptor related protein complex 2 subunit sigma 1 | synaptic vesicle cycle | 7 | 1 | 0 | |
| ENSG00000185344 | ATP6V0A2 | ATPase H+ transporting V0 subunit a2 | synaptic vesicle cycle | 7 | 2 | 0 | |
| ENSG00000004468 | CD38 | CD38 molecule | synaptic vesicle cycle | 4 | 2 | 0 | |
| ENSG00000070371 | CLTCL1 | clathrin, heavy chain-like 1 | synaptic vesicle cycle | 7 | 2 | 0 | |
| ENSG00000081189 | MEF2C | myocyte enhancer factor 2C | synaptic vesicle cycle | 4 | 1 | 0 | |
| ENSG00000132639 | SNAP25 | Synaptosomal-associated protein, 25kDa | synaptic vesicle cycle | 7 | 2 | 0 | |
| ENSG00000106089 | STX1A | Syntaxin 1A (brain) | synaptic vesicle cycle | 7 | 2 | 0 | |
| ENSG00000136854 | STXBP1 | Syntaxin binding protein 1 | synaptic vesicle cycle | 7 | 1 | 0 | |
| ENSG00000067715 | SYT1 | synaptotagmin 1 | synaptic vesicle cycle | 7 | | 1 | |
| ENSG00000130477 | UNC13A | unc-13 homolog A | synaptic vesicle cycle | 7 | 2 | 1 | |
| ENSG00000220205 | VAMP2 | vesicle associated membrane protein 2 | synaptic vesicle cycle | 7 | | 1 | |
| ENSG00000101405 | OXT | oxytocin/neurophysin I prepropeptide | ligand | 2, 3, 4 | 2 | 0 | |
| ENSG00000125084 | WNT1 | Wingless-type MMTV integration site family, member 1 | ligand | 2 | 2 | 0 | |
| ENSG00000128271 | ADORA2A | adenosine A2a receptor | GPCR | 1, 2, 3 | 2 | 0 | |
| ENSG00000282608 | ADORA3 | Adenosine A3 receptor | GPCR | 1, 2 | 2 | 0 | |
| ENSG00000169252 | ADRB2 | adrenergic, beta-2-, receptor, surface | GPCR | 1, 2, 3, 13 | 2 | 0 | |
| ENSG00000180772 | AGTR2 | angiotensin II receptor, type 2 | GPCR | 1, 2, 8, 13 | 2 | 0 | |
| ENSG00000166148 | AVPR1A | arginine vasopressin receptor 1A | GPCR | 1, 2 | 2 | 0 | |
| ENSG00000198049 | AVPR1B | arginine vasopressin receptor 1B | GPCR | 1, 2 | 2 | 0 | |
| ENSG00000133019 | CHRM3 | cholinergic receptor muscarinic 3 | GPCR | 1, 2, 11 | 2 | 0 | |
| ENSG00000118432 | CNR1 | cannabinoid receptor 1 (brain) | GPCR | 1, 2, 6 | 2 | 0 | |
| ENSG00000168329 | CX3CR1 | Chemokine (C-X3-C motif) receptor 1 | GPCR | 1, 2, 9 | 2 | 0 | |
| ENSG00000184845 | DRD1 | Dopamine receptor D1 | GPCR | 1, 2, 3, 5 | 2 | 0 | |
| ENSG00000149295 | DRD2 | Dopamine receptor D2 | GPCR | 1, 2, 3, 5 | 2 | 0 | |
| ENSG00000151577 | DRD3 | dopamine receptor D3 | GPCR | 1, 2, 5 | 2 | 0 | |
| ENSG00000136928 | GABBR2 | gamma-aminobutyric acid type B receptor subunit 2 | GPCR | 1, 2, 3 | 2 | 1 | |
| ENSG00000170775 | GPR37 | G protein-coupled receptor 37 | GPCR | 1, 2 | 2 | 0 | |
| ENSG00000164604 | GPR85 | G protein-coupled receptor 85 | GPCR | 1 | 2 | 0 | |
| ENSG00000168959 | GRM5 | glutamate metabotropic receptor 5 | GPCR | 1, 2, 6, 10 | 2 | 0 | |
| ENSG00000196277 | GRM7 | Glutamate receptor, metabotropic 7 | GPCR | 1, 2, 10 | 2 | 0 | |
| ENSG00000135312 | HTR1B | 5-hydroxytryptamine (serotonin) receptor 1B | GPCR | 1, 2, 3, 12 | 2 | 0 | |
| ENSG00000221888 | OR1C1 | olfactory receptor, family 1, subfamily C, member 1 | GPCR | 1, 14 | 2 | 0 | |
| ENSG00000171180 | OR2M4 | Olfactory receptor, family 2, subfamily M, member 4 | GPCR | 1, 14 | 2 | 0 | |
| ENSG00000184022 | OR2T10 | olfactory receptor family 2 subfamily T member 10 | GPCR | 1, 14 | 2 | 0 | |
| ENSG00000197790 | OR52M1 | Olfactory receptor, family 52, subfamily M, member 1 | GPCR | 1, 14 | 2 | 0 | |
| ENSG00000180914 | OXTR | oxytocin receptor | GPCR | 1, 2, 3, 4 | 2 | 0 | |
| ENSG00000127955 | GNAI1 | G protein subunit alpha i1 | G protein/RGS | 1, 2, 3, 4, 5, 6, 9, 10, 11, 12, 13 | 1 | 0 | |
| ENSG00000087460 | GNAS | GNAS complex locus | G protein/RGS | 1, 2, 3, 4, 5, 10, 12, 13 | 2 | 0 | |
| ENSG00000172354 | GNB2 | G protein subunit beta 2 | G protein/RGS | 1, 2, 5, 6, 9, 10, 11, 12 | | 1 | |
| ENSG00000182901 | RGS7 | regulator of G-protein signaling 7 | G protein/RGS | 1, 2 | 2 | 0 | |
| ENSG00000138031 | ADCY3 | adenylate cyclase 3 | enzyme effector | 1, 2, 3, 4, 6, 9, 10, 11, 13, 14 | 2 | 0 | |
| ENSG00000173175 | ADCY5 | Adenylate cyclase 5 | enzyme effector | 1, 2, 3, 4, 5, 6, 9, 10, 11, 12, 13 | 2 | 0 | |
| ENSG00000134780 | DAGLA | diacylglycerol lipase alpha | enzyme effector | 1, 2, 6 | 2 | 0 | |
| ENSG00000154678 | PDE1C | phosphodiesterase 1C | enzyme effector | 2, 14 | 2 | 0 | |
| ENSG00000182621 | PLCB1 | phospholipase C, beta 1 (phosphoinositide-specific) | enzyme effector | 1, 2, 4, 5, 6, 9, 10, 11, 12, 13 | 2 | 0 | |
| ENSG00000141837 | CACNA1A | Calcium channel, voltage-dependent, P/Q type, alpha 1A subunit | channel effector | 5, 6, 7, 10, 11, 12 | 1 | 1 | |
| ENSG00000148408 | CACNA1B | calcium voltage-gated channel subunit alpha1 B | channel effector | 5, 6, 7, 11, 12 | 2 | 0 | |
| ENSG00000151067 | CACNA1C | calcium channel, voltage-dependent, L type, alpha 1C subunit | channel effector | 3, 4, 5, 6, 10, 11, 12, 13 | 1 | 0 | |
| ENSG00000157388 | CACNA1D | calcium channel, voltage-dependent, L type, alpha 1D | channel effector | 1, 3, 4, 5, 6, 10, 11, 12, 13 | 2 | 0 | |
| ENSG00000102001 | CACNA1F | calcium channel, voltage-dependent, alpha 1F | channel effector | 3, 4, 6, 11, 12, 13 | 2 | 0 | |
| ENSG00000153956 | CACNA2D1 | calcium voltage-gated channel auxiliary subunit alpha2delta 1 | channel effector | 4, 13 | 2 | 0 | |
| ENSG00000157445 | CACNA2D3 | Calcium channel, voltage-dependent, alpha 2/delta subunit 3 | channel effector | 4, 13 | 1 | 0 | |
| ENSG00000067191 | CACNB1 | calcium voltage-gated channel auxiliary subunit beta 1 | channel effector | 4, 13 | 3 | 0 | |
| ENSG00000165995 | CACNB2 | Calcium channel, voltage-dependent, beta 2 subunit | channel effector | 4, 13 | 2 | 0 | |
| ENSG00000170289 | CNGB3 | cyclic nucleotide gated channel beta 3 | channel effector | 3 | 2 | 0 | |
| ENSG00000150995 | ITPR1 | inositol 1,4,5-trisphosphate receptor type 1 | channel effector | 2, 4, 5, 6, 10, 11, 12 | 2 | 0 | |
| ENSG00000144285 | SCN1A | sodium channel, voltage-gated, type I, alpha subunit | channel effector | 5 | 1 | 0 | |
| ENSG00000137672 | TRPC6 | Transient receptor potential cation channel, subfamily C, member 6 | channel effector | 2 | 2 | 0 | |
| ENSG00000070808 | CAMK2A | calcium/calmodulin dependent protein kinase II alpha | kinase/2$^{nd}$ messenger target | 1, 2, 3, 4, 5, 11, 13, 14 | 2 | 1 | |
| ENSG00000058404 | CAMK2B | calcium/calmodulin dependent protein kinase II beta | kinase/2$^{nd}$ messenger target | 2, 3, 4, 5, 11, 13, 14 | | 1 | |
| ENSG00000152495 | CAMK4 | calcium/calmodulin dependent protein kinase IV | kinase/2$^{nd}$ messenger target | 2, 3, 4, 11 | 2 | 0 | |
| ENSG00000121879 | PIK3CA | phosphatidylinositol-4,5-bisphosphate 3-kinase catalytic subunit alpha | kinase/2nd messenger target | 2, 3, 9, 11 | 3 | 0 | |
| ENSG00000105851 | PIK3CG | phosphoinositide-3-kinase, catalytic, gamma polypeptide | kinase/2nd messenger target | 1, 2, 4, 9, 11, 13 | 2 | 0 | |
| ENSG00000105647 | PIK3R2 | phosphoinositide-3-kinase regulatory subunit 2 | kinase/2nd messenger target | 2, 3, 9, 11 | | 1 | |
| ENSG00000188191 | PRKAR1B | protein kinase cAMP-dependent type I regulatory subunit beta | kinase/2nd messenger target | 2 | 2 | 0 | |
| ENSG00000154229 | PRKCA | protein kinase C alpha | kinase/2nd messenger target | 2, 4, 5, 6, 10, 11, 12, 13 | 2 | 0 | |
| ENSG00000166501 | PRKCB | protein kinase C beta | kinase/2nd messenger target | 2, 4, 5, 6, 9, 10, 11, 12 | 2 | 0 | |
| ENSG00000091428 | RAPGEF4 | Rap guanine nucleotide exchange factor (GEF) 4 | kinase/2nd messenger target | 1, 3, 13 | 2 | 0 | |
| ENSG00000147044 | CASK | calcium/calmodulin dependent serine protein kinase | Scaffolding partner | | 1 | 0 | bécamel, embo J, 2002, 10.1093/emboj/21.10.2332 |
| ENSG00000132535 | DLG4 | discs large MAGUK scaffold protein 4 | Scaffolding partner | 10 | 1 | 0 | |
| ENSG00000170579 | DLGAP1 | DLG associated protein 1 | Scaffolding partner | 10 | 2 | 0 | |
| ENSG00000152413 | HOMER1 | Homer homolog 1 (Drosophila) | Scaffolding partner | 1, 10 | 2 | 0 | |
| ENSG00000161681 | SHANK1 | SH3 and multiple ankyrin repeat domains 1 | Scaffolding partner | 10 | 2 | 0 | |
| ENSG00000162105 | SHANK2 | SH3 and multiple ankyrin repeat domains 2 | Scaffolding partner | 10 | 1 | 0 | |
| ENSG00000251322 | SHANK3 | SH3 and multiple ankyrin repeat domains 3 | Scaffolding partner | 1, 10 | 1 | 0 | |
| ENSG00000075624 | ACTB | actin beta | cytoskeleton | 4 | 1 | 1 | |
| ENSG00000127914 | AKAP9 | A kinase (PRKA) anchor protein 9 | downstream cellular process | | 2 | 0 | Sehrawat, blood, 2011, doi.org/10.1182/blood-2010-02-268870 |
| ENSG00000135046 | ANXA1 | Annexin A1 | downstream cellular process | 1, 2 | 2 | 0 | |
| ENSG00000104728 | ARHGEF10 | Rho guanine nucleotide exchange factor 10 | downstream cellular process | 2 | 2 | 0 | |
| ENSG00000131089 | ARHGEF9 | Cdc42 guanine nucleotide exchange factor (GEF) 9 | downstream cellular process | 2 | 1 | 0 | |
| ENSG00000163399 | ATP1A1 | ATPase Na+/K+ transporting subunit alpha 1 | downstream cellular process | 3, 13 | 2 | 1 | |
| ENSG00000105409 | ATP1A3 | ATPase Na+/K+ transporting subunit alpha 3 | downstream cellular process | 3, 13 | 2 | 1 | |
| ENSG00000157087 | ATP2B2 | ATPase, Ca++ transporting, plasma membrane 2 | downstream cellular process | 3, 13 | 2 | 0 | |
| ENSG00000157764 | BRAF | v-raf murine sarcoma viral oncogene homolog B | downstream cellular process | 3, 9, 12 | 1 | 0 | |
| ENSG00000168036 | CTNNB1 | catenin beta 1 | downstream cellular process | | 1 | 0 | He, biomed and pharma, 2020, doi.org/10.1016/j.biopha.2020.110851 |
| ENSG00000145864 | GABRB2 | gamma-aminobutyric acid type A receptor subunit beta2 | downstream cellular process | 12 | 1 | 0 | |
| ENSG00000166206 | GABRB3 | gamma-aminobutyric acid (GABA) A receptor, beta 3 | downstream cellular process | 12 | 1 | 0 | |
| ENSG00000101958 | GLRA2 | glycine receptor, alpha 2 | downstream cellular process | 1 | 2 | 0 | |
| ENSG00000155511 | GRIA1 | glutamate ionotropic receptor AMPA type subunit 1 | downstream cellular process | 3, 5, 6, 10 | 2 | 0 | |
| ENSG00000120251 | GRIA2 | glutamate ionotropic receptor AMPA type subunit 2 | downstream cellular process | 3, 5, 6, 10 | 1 | 0 | |
| ENSG00000125675 | GRIA3 | glutamate ionotropic receptor AMPA type subunit 3 | downstream cellular process | 3, 5, 6, 10 | | 1 | |
| ENSG00000163873 | GRIK3 | glutamate ionotropic receptor kainate type subunit 3 | downstream cellular process | 1, 10 | 2 | 0 | |
| ENSG00000176884 | GRIN1 | Glutamate receptor, ionotropic, N-methyl D-aspartate 1 | downstream cellular process | 3, 1 | 1 | 0 | |
| ENSG00000183454 | GRIN2A | glutamate receptor, ionotropic, N-methyl D-aspartate 2A | downstream cellular process | 3, 5, 10 | 1 | 0 | |
| ENSG00000273079 | GRIN2B | glutamate receptor, inotropic, N-methyl D-apartate 2B | downstream cellular process | 3, 5, 10 | 1 | 0 | |
| ENSG00000152402 | GUCY1A2 | guanylate cyclase 1 soluble subunit alpha 2 | downstream cellular process | 4 | 2 | 0 | |
| ENSG00000174775 | HRAS | v-Ha-ras Harvey rat sarcoma viral oncogene homolog | downstream cellular process | 2, 4, 9, 11, 12 | 1 | 0 | |
| ENSG00000205726 | ITSN1 | intersectin 1 | downstream cellular process | 2 | 2 | 0 | |
| ENSG00000184408 | KCND2 | potassium voltage-gated channel subfamily D member 2 | downstream cellular process | 12 | 2 | 0 | |
| ENSG00000168280 | KIF5C | Kinesin family member 5C | downstream cellular process | 5 | 2 | 1 | |
| ENSG00000113889 | KNG1 | kininogen 1 | downstream cellular process | 2 | 3 | 0 | |
| ENSG00000102882 | MAPK3 | mitogen-activated protein kinase 3 | downstream cellular process | 2, 3, 4, 6, 9, 10, 11, 12, 13 | 2 | 0 | |
| ENSG00000198793 | MTOR | mechanistic target of rapamycin kinase | downstream cellular process | | 1 | 0 | Jewell JL, elife, 2019, DOI: 10.7554/eLife.43038 ; Meffre 2006; Sodhi 2006; |
| ENSG00000065534 | MYLK | myosin light chain kinase | downstream cellular process | 4 | 3 | 0 | |
| ENSG00000172915 | NBEA | neurobeachin | downstream cellular process | 2 | 1 | 0 | |
| ENSG00000128609 | NDUFA5 | NADH dehydrogenase (ubiquinone) 1 alpha subcomplex, 5, 13kDa | downstream cellular process | 6 | 2 | 0 | |
| ENSG00000149269 | PAK1 | p21 (RAC1) activated kinase 1 | downstream cellular process | 2, 3, 9 | | 1 | |
| ENSG00000198523 | PLN | phospholamban | downstream cellular process | 1, 3, 13 | 2 | 0 | |
| ENSG00000164050 | PLXNB1 | plexin B1 | downstream cellular process | 2 | 2 | 0 | |
| ENSG00000131771 | PPP1R1B | Protein phosphatase 1, regulatory (inhibitor) subunit 1B | downstream cellular process | 2, 3, 5 | 2 | 0 | |
| ENSG00000113575 | PPP2CA | protein phosphatase 2 catalytic subunit alpha | downstream cellular process | 2, 5, 13 | | 1 | |
| ENSG00000137713 | PPP2R1B | protein phosphatase 2 regulatory subunit A, beta | downstream cellular process | 2, 5, 13 | 2 | 0 | |
| ENSG00000112640 | PPP2R5D | Protein phosphatase 2, regulatory subunit B', delta | downstream cellular process | 2, 5, 13 | 1 | 0 | |
| ENSG00000138814 | PPP3CA | protein phosphatase 3 catalytic subunit alpha | downstream cellular process | 2, 4, 5, 10 | 3 | 1 | |
| ENSG00000124126 | PREX1 | Phosphatidylinositol-3,4,5-trisphosphate-dependent Rac exchange factor 1 | downstream cellular process | 1, 2, 9 | 2 | 0 | |
| ENSG00000184304 | PRKD1 | Protein kinase D1 | downstream cellular process | | | 1 | Steinberg, mol pharmaco, 2012, 10.1124/mol.111.075986 |
| ENSG00000105287 | PRKD2 | protein kinase D2 | downstream cellular process | | 2 | 0 | Steinberg, mol pharmaco, 2012, 10.1124/mol.111.075986 |
| ENSG00000136238 | RAC1 | Rac family small GTPase 1 | downstream cellular process | 1, 3, 9 | | 1 | |
| ENSG00000079841 | RIMS1 | Regulating synaptic membrane exocytosis 1 | downstream cellular process | 6, 7 | 1 | 0 | |
| ENSG00000176406 | RIMS2 | regulating synaptic membrane exocytosis 2 | downstream cellular process | 1 | 3 | 1 | |
| ENSG00000152214 | RIT2 | Ras-like without CAAX 2 | downstream cellular process | 1 | 2 | 0 | |
| ENSG00000198963 | RORB | RAR related orphan receptor B | downstream cellular process | 1 | 1 | 0 | |
| ENSG00000156395 | SORCS3 | sortilin related VPS10 domain containing receptor 3 | downstream cellular process | 1 | 2 | 0 | |
| ENSG00000100485 | SOS2 | SOS Ras/Rho guanine nucleotide exchange factor 2 | downstream cellular process | 2, 9 | | 0 | |
| ENSG00000038382 | TRIO | Trio Rho guanine nucleotide exchange factor | downstream cellular process | 2 | 1 | 0 | |
| ENSG00000134160 | TRPM1 | transient receptor potential cation channel subfamily M member 1 | downstream cellular process | 1 | 2 | 0 | |
| ENSG00000165699 | TSC1 | tuberous sclerosis 1 | downstream cellular process | | 1 | 0 | PMID: 21559457 |
| ENSG00000103197 | TSC2 | tuberous sclerosis 2 | downstream cellular process | | 1 | 0 | PMID: 21559457 |
| ENSG00000273749 | CYFIP1 | cytoplasmic FMR1 interacting protein 1 | translation | | 2 | 0 | PMID: 21842222 |
| ENSG00000130811 | EIF3G | eukaryotic translation initiation factor 3 subunit G | translation | | 1 | 0 | PMID: 34803767 |
| ENSG00000151247 | EIF4E | eukaryotic translation initiation factor 4E | translation | | 2 | 0 | PMID: 18276584 |
| ENSG00000114867 | EIF4G1 | eukaryotic translation initiation factor 4 gamma 1 | translation | | 3 | 0 | PMID: 24711644; Tréfier, front. Endocrinol. 2018, doi.org/10.3389/fendo. 2018.00017 |
| ENSG00000102081 | FMR1 | fragile X mental retardation 1 | translation | | 1 | 0 | Pacey, mol cell neuroscience, 2011, doi.org/10.1016/j.mcn. 2010.12.005 ; Tréfier, front. Endocrinol. 2018, doi.org/10.3389/fendo. 2018.00017 |
| ENSG00000071242 | RPS6KA2 | ribosomal protein S6 kinase, 90kDa, polypeptide 2 | translation | 2 | 2 | 0 | Tréfier, front. Endocrinol. 2018, doi.org/10.3389/fendo. 2018.00017 |
| ENSG00000177189 | RPS6KA3 | Ribosomal protein S6 kinase, 90kDa, polypeptide 3 | translation | 2 | 2 | 1 | Tréfier, front. Endocrinol. 2018, doi.org/10.3389/fendo. 2018.00017 |
| ENSG00000005339 | CREBBP | CREB binding protein | Transcription factor | 3 | 1 | 0 | |
| ENSG00000179388 | EGR3 | early growth response 3 | transcription factor | | 2 | 0 | Tan, geno prot bioinfo, 2003, 10.1016/s1672-0229(03)01022-2 |
| ENSG00000114861 | FOXP1 | forkhead box P1 | transcription factor | | 1 | 0 | PMID: 20709693 |
| ENSG00000128573 | FOXP2 | forkhead box P2 | transcription factor | | 1 | 0 | PMID: 20709693 |
| ENSG00000100393 | EP300 | E1A binding protein p300 | transcription regulation | 3 | 1 | 0 | |

All genes involved in GPCR activity and signalling processes were extracted from the last SFARI gene release (05-05-2022) using our knowledge (references) and KEGG Gene Ontology terms: (1) '*G protein-coupled receptor signaling pathway*', (2) REACTOME '*Signaling by GPCR*', (3) '*cAMP signaling pathway*', (4) '*Oxytocin signaling pathway*', (5) '*Dopaminergic synapse*', (6) '*Retrograde endocannabinoid signaling*', (7) '*Synaptic vesicle cycle'*, (8) '*Renin-angiotensin system*', (9) '*Chemokine signaling pathway*', (10) '*Glutamatergic synapse*', (11) '*Cholinergic synapse*', (12) '*Serotoninergic synapse*', (13) '*Adrenergic signaling in cardiomyocytes*' and (14) '*Olfactory transduction*'.

**Table 2 List of the 23 GPCR and oxytocin gene and their variants associated to ASD**

| Gene | Variant | Inheritance pattern | Allele change | Reference SNP | Variant localisation | Variant type | Residue change | 1000 genome project GPCRdb | Functional annotation | References |
|---|---|---|---|---|---|---|---|---|---|---|
| *ADORA2A* | common | Unknown | 12108T>C (b) | rs3761422 | intron1 | - | - | | expression | Freitag CM , et al. (2009) |
| *ADORA2A* | common | Unknown | 21460T>C (b) | rs2236624 | intron2 | - | - | | expression | Freitag CM , et al. (2009) |
| *ADORA2A* | common | Unknown | 1083T>C (b) | rs5751876 | C-terminus | synonymous | Tyr361= | x | expression | Freitag CM , et al. (2009) |
| *ADORA2A* | common | Unknown | *453-458dupTTTTTT (b) | rs35060421 | 3'UTR | - | - | x | expression | Freitag CM , et al. (2009) |
| | | | | | | | | | | |
| *ADORA3* | rare | Familial | 268C>G | rs77883500 | TM3 | missense | Leu90Val | x | tolerated variant; ligand binding | Campbell NG , et al. (2013) |
| *ADORA3* | rare | De novo; Familial | 511G>A | rs139935750 | ECL2 | missense | Val171Ile | x | tolerated variant; ligand binding | Campbell NG , et al. (2013) |
| | | | | | | | | | | |
| *ADRB2* | common | Unknown | 5285A>G | rs1042713 | N-term | missense | Gly16Arg | x | deleterious variant; trafficking | Connors SL , et al. (2006); Cheslack-Postava K , et al. (2007) |
| *ADRB2* | common | Unknown | 5318C>G | rs1042714 | TM1 | missense | Glu27Gln | x | tolerated variant | Connors SL , et al. (2006); Cheslack-Postava K , et al. (2007) |
| | | | | | | | | | | |
| *AGTR2* | rare | De novo | | | | translocation | - | | expression | Vervoort VS , et al. (2002) |
| *AGTR2* | rare | Unknown | 62G>T | rs121917810 | N-term | missense | Gly21Val | x | tolerated variant; trafficking | Vervoort VS , et al. (2002) |
| *AGTR2* | rare | De novo | 244T>C | rs1049717729 | TM2 | missense | Tyr82His | | deleterious variant | Cappi C , et al. (2019) |
| *AGTR2* | rare | Unknown | 402del | rs387906503 | TM3 | frameshift | Phe134Leu-Ter | x | loss of function | Vervoort VS , et al. (2002) |
| *AGTR2* | rare | Familial | 572G>A | | ECL2 | missense | Gly191Glu | | deleterious variant; ligand binding | Takeshita E , et al. (2012) |
| *AGTR2* | rare | Familial | 757C>T | rs1569536545 | TM6 | stop gained | Gln253Ter | | loss of function | Tran KT et al. (2020) |
| *AGTR2* | rare | Unknown | 971G>A | rs35474657 | H8 | missense | Arg324Gln | x | tolerated variant | Vervoort VS , et al. (2002) |
| *AGTR2* | rare | Unknown | 1009A>G | rs121917811 | C-term | missense | Ile337Val | x | tolerated variant | Vervoort VS , et al. (2002) |
| | | | | | | | | | | |
| *AVPR1A* | common | Unknown | g.63154312T>C | rs7294536 | 5KB upstream | - | - | | expression | Kim SJ , et al. (2002); Yang SY , et al. (2010); Francis SM , et |
| | | | | | | | | | | al. (2016); Procyshyn TL , et al. (2016) |
| *AVPR1A* | common | Unknown | -2554-60A>G | rs10877969 | 2KB upstream | - | - | | expression | Yang SY , et al. (2010); Tansey KE , et al. (2011); Wassink TH , et al. (2004); Kantojrvi K , et al. (2015); Bachner-Melman R , et al. (2005); Yang SY , et al. (2017); Francis SM , et al. (2016) |
| *AVPR1A* | rare | Unknown | GATA>AATA (127498R (a)) | rs11174815 | 2KB upstream | - | - | | expression | Kim SJ , et al. (2002) |
| *AVPR1A* | common | Unknown | -3476A>G | rs3759292 | 2KB upstream | - | - | | expression | Yang SY , et al. (2010) |
| *AVPR1A* | common | Unknown | g.63154312T>C (−1502 bp in paper) | rs7294536 | 2KB upstream | - | - | | expression | Yang SY , et al. (2017) |
| *AVPR1A* | common | Unknown | -2623A>G | rs3759292 | 2KB upstream | - | - | | expression | Yang SY , et al. (2010) |
| *AVPR1A* | common | Unknown | g.63153459T>C (−649 bp in paper) | rs10877969 | 2KB upstream | - | - | | expression | Yang SY , et al. (2017) |
| *AVPR1A* | common | Unknown | g.63153316G>A | rs11174815 | 2KB upstream | - | - | | expression | Tansey KE , et al. (2011) |
| *AVPR1A* | common | Unknown | -242C>T | rs3741865 | 5'UTR | - | - | | expression | Tansey KE , et al. (2011) |
| *AVPR1A* | rare | Unknown | 128085Y(a) C>? | | 5'UTR | - | - | | expression | Kim SJ , et al. (2002) |
| *AVPR1A* | rare | Unknown | 128626K(a) T>? | | 5'UTR | - | - | | expression | Kim SJ , et al. (2002) |
| *AVPR1A* | rare | Unknown | 128932Y(a) C>? | | 5'UTR | - | - | | expression | Kim SJ , et al. (2002) |
| *AVPR1A* | rare | Unknown | 129523M(a) A>? | | 5'UTR | - | - | | expression | Kim SJ , et al. (2002) |
| *AVPR1A* | rare | Unknown | 129908S(a) C>? | | 5'UTR | - | - | | expression | Kim SJ , et al. (2002) |
| *AVPR1A* | rare | Unknown | 16G>A (130011R (a)) | rs2228154 | N-term | missense | Gly6Ser | x | post-translational modification (glycosylation) | Kim SJ , et al. (2002) |
| *AVPR1A* | rare | De novo | 236C>T | rs1387200841 | TM1 | missense | Thr79Met | | deleterious variant | Takata A , et al. (2018) |
| *AVPR1A* | rare | Unknown | 130286Y(a) C>? | rs1565879126 | TM2 | synonymous | Ala99= | | expression | Kim SJ , et al. (2002) |
| *AVPR1A* | rare, common | Unknown | 408T>C (130403Y(a)) | rs1042615 | TM3 | synonymous | Phe136= | (+Phe136Leu, Phe136Cys) | expression | Kim SJ , et al. (2002); Wassink TH , et al. (2004) |
| *AVPR1A* | common | Unknown | 417T>C | rs758045345 | TM3 | synonymous | Ala139= | (+Ala139Pro) | expression | Wassink TH , et al. (2004) |
| *AVPR1A* | rare | Familial | 971-2A>C | | intron1 | splice site | - | | gain/loss of function | Ohashi K et al. (2021); Wu H , et al. (2019) |
| *AVPR1A* | common | Unknown | ? | | intron1 | | - | | expression | Yirmiya N , et al. (2006) |
| *AVPR1A* | rare | Unknown | A>? (130478R(a)) | rs2228155 | ICL2 | synonymous | Gln161= | | expression | Kim SJ , et al. (2002) |
| *AVPR1A* | rare | De novo | 966G>A | | ECL3 | stop gained | Trp322Ter | | loss of function | C Yuen RK et al. (2017) |
| *AVPR1A* | rare | Familial | 1130A>T | rs138652515 | C-term | missense | Asp377Val | | deleterious variant | Ohashi K et al. (2021) |
| *AVPR1A* | common | Unknown | ? | | gene variant | | - | | | Pappa I , et al. (2015) |
| *AVPR1A* | common | Unknown | g.63164134G>A | rs7307997 | intergenic | | - | | expression | Kantojrvi K , et al. (2015) |
| | | | | | | | | | | |
| *AVPR1B* | common | Unknown | *2833= | rs28405931 | intergenic | | - | | expression | Chakrabarti B , et al. (2009) |
| *AVPR1B* | common | Unknown | 195G>C | rs35369693 | ICL1 | missense | Lys65Asn | x | post-translational modification (SUMOylation/Ubiquitination) | Francis SM , et al. (2016); Dempster EL , et al. (2007) |
| *AVPR1B* | common | Unknown | 1091G>A | rs28632197 | C-term | missense | Arg364His | x | post-translationa | Francis SM , et al. (2016) |
| | | | | | | | | | l modification (methylation) | |
| *AVPR1B* | common | Unknown | *117A>G | rs33985287 | 3'UTR | | - | | expression | Dempster EL , et al. (2007) |
| | | | | | | | | | | |
| *CHRM3* | rare | Unknown | duplication of 763.3 Kb | | - | copy number gain | - | | potential gain of function | Cheng X , et al. (2019) |
| *CHRM3* | rare | Unknown, De novo | deletion of 473 and 911 Kb | | - | copy number loss | - | | loss of function | Petersen AK , et al. (2012); Perrone MD , et al. (2011) |
| *CHRM3* | rare | De novo | 1423A>T | | ICL3 | missense | Ile475Phe | | tolerated variant; signaling effector recruitment | Li J , et al. (2017) |
| *CHRM3* | rare | De novo | 1504A>G | rs1680193624 | TM6 | missense | Ile502Val | | tolerated variant | De Rubeis S , et al. (2014) |
| *CHRM3* | rare | Familial | 1762C>T | rs764058084 | C-term | stop gained | Gln588Ter | x | loss of function | Li J , et al. (2017) |
| *CHRM3* | common | Unknown | 239046758C>A | rs72769124 | intergenic | | - | | expression | Pardias AF , et al. (2018) |
| | | | | | | | | | | |
| *CNR1* | rare | Familial | - | | - | copy number gain | - | | potential gain of function | Girirajan S , et al. (2013) |
| *CNR1* | common | Unknown | -1452A>G | | 2KB upstream | - | - | | expression | Chakrabarti B , et al. (2006) |
| *CNR1* | common | Unknown | -318A>C | | intron1 | - | - | | expression | Chakrabarti B , et al. (2006) |
| *CNR1* | common | Unknown | -207+2592A>C | rs6454674 | intron1 | - | - | | expression | Chakrabarti B , et al. (2006) |
| *CNR1* | common | Unknown | -207+1046A>C | rs6454674 | intron1 | - | - | | expression | Chakrabarti B , et al. (2006) |
| *CNR1* | common | Unknown | -206-1198A>G | | intron1 | - | - | | expression | Chakrabarti B , et al. (2006) |
| *CNR1* | common | Unknown | -206-7128T>C | rs806380 | intron1 | - | - | | expression | Chakrabarti B , et al. (2006) |
| *CNR1* | common | Unknown | -80+2592A>C | | intron1 | - | - | | expression | Chakrabarti B , et al. (2006) |
| *CNR1* | common | Unknown | -79-3667A>G | | intron1 | - | - | | expression | Chakrabarti B , et al. (2006) |
| *CNR1* | common | Unknown | -79-9597T>C | | intron1 | - | - | | expression | Chakrabarti B , et al. (2006) |
| *CNR1* | common | Unknown | -64+8023T>C | rs806380 | intron1 | - | - | | expression | Chakrabarti B , et al. (2006) |
| *CNR1* | common | Unknown | -64+2536A>C | rs6454674 | intron1 | - | - | | expression | Chakrabarti B , et al. (2006) |
| *CNR1* | common | Unknown | -64+2592A>C | rs6454674 | intron1 | - | - | | expression | Chakrabarti B , et al. (2006) |
| *CNR1* | common | Unknown | -64+3365A>C | rs6454674 | intron1 | - | - | | expression | Chakrabarti B , et al. (2006) |
| *CNR1* | common | Unknown | -63-3667A>G | | intron1 | - | - | | expression | Chakrabarti B , et al. (2006) |
| *CNR1* | common | Unknown | -63-9597T>C | rs806380 | intron1 | - | - | | expression | Chakrabarti B , et al. (2006) |
| *CNR1* | rare | Unknown | 20G>A | rs777561684 | N-term | missense | Gly7Asp | | tolerated variant; post-translational modification; trafficking | Smith DR , et al. (2017) |
| *CNR1* | rare | Unknown | 80A>G | rs768086951 | N-term | missense | Asp27Gly | x | tolerated variant; post-translational modification; trafficking | Smith DR , et al. (2017) |
| *CNR1* | rare | Unknown | 169C>G | rs1245725620 | N-term | missense | Pro57Ala | | tolerated variant; post-translational modification; trafficking | Smith DR , et al. (2017) |
| *CNR1* | rare | Unknown | 277G>A | rs202238406 | N-term | missense | Glu93Lys | x | tolerated variant; post-translational modificatio | Smith DR , et al. (2017) |
| | | | | | | | | | n; trafficking | |
| *CNR1* | rare | Unknown | 443G>A | rs1274654924 | ICL1 | missense | Arg148His | | post-translational modification (methylation) | Smith DR , et al. (2017) |
| *CNR1* | rare | Unknown | 581G>C | rs1777050772 | TM3 | missense | Gly194Ala | | deleterious variant | Smith DR , et al. (2017) |
| *CNR1* | rare | Unknown | 622T>C | rs1360859098 | TM3 | missense | Phe208Leu | | deleterious variant | Smith DR , et al. (2017) |
| *CNR1* | rare | Unknown | 725C>T | rs1777038251 | TM4 | missense | Thr242Ile | | deleterious variant | Smith DR , et al. (2017) |
| *CNR1* | rare | Unknown | 745G>A | rs146828007 | TM4 | missense | Val249Met | x | deleterious variant | Smith DR , et al. (2017) |
| *CNR1* | rare | Unknown | 916G>A | rs568621412 | TM5 | missense | Val306Ile | x | deleterious variant | Smith DR , et al. (2017) |
| *CNR1* | rare | Unknown | 1015A>G | | TM6 | missense | Ile339Val | | deleterious variant | Smith DR , et al. (2017) |
| *CNR1* | common | Unknown | 1176G>A | | TM7 | synonymous | Val392= | | expression | Chakrabarti B , et al. (2006) |
| *CNR1* | rare | Unknown | 1225C>T | rs149335541 | H8 | missense | Arg409Trp | x | deleterious variant | Smith DR , et al. (2017) |
| *CNR1* | rare | Unknown | 1256C>A | rs78783387 | C-term | missense | Ala419Glu | x | signaling effector recruitment; post-translationa | Smith DR , et al. (2017) |
| | | | | | | | | | l modification (phospho-mimetic mutant) | |
| *CNR1* | common | Unknown | 1260G>A | rs1049353 | C-term | synonymous | Gln420= | | expression | Chakrabarti B , et al. (2006) |
| *CNR1* | rare | Unknown | 1278G>T | | C-term | missense | Met426Ile | | signaling effector recruitment | Smith DR , et al. (2017) |
| *CNR1* | rare | Unknown | 1306G>A | rs370487985 | C-term | missense | Ala436Thr | x | signaling effector recruitment; post-translational modification (extra phosphorylation) | Smith DR , et al. (2017) |
| *CNR1* | rare | Unknown | 1348A>G | rs951061482 | C-term | missense | Ile450Val | (+Ile450Thr) | signaling effector recruitment | Smith DR , et al. (2017) |
| *CNR1* | rare | Unknown | 1357A>G | rs1776980237 | C-term | missense | Thr453Ala | | signaling effector recruitment; post-translational modificatio | Smith DR , et al. (2017) |
| | | | | | | | | | n (deletion phosphorylation) | |
| *CNR1* | common | Unknown | 1359G>A | rs1049353 | C-term | synonymous | Val454= | | expression | Chakrabarti B , et al. (2006) |
| | | | | | | | | | | |
| *CX3CR1* | rare, common | Unknown, Familial | 163G>A | rs750585901 | TM1 | missense | Ala55Thr | x | deleterious variant | Ishizuka K , et al. (2017) |
| *CX3CR1* | rare | Unknown, Familial | 335G>C | rs201442030 | TM3 | missense | Gly112Ala | x | deleterious variant | Ishizuka K , et al. (2017) |
| *CX3CR1* | rare | Unknown | 414G>T | rs758302878 | ICL2 | missense | Met138Ile | x | signaling effector recruitment | Ishizuka K , et al. (2017) |
| | | | | | | | | | | |
| *DRD1* | common | Unknown | -48G>A | rs4532 | 5'UTR | - | - | x | expression | Bobb AJ , et al. (2005); Hettinger JA, et al. (2008) |
| *DRD1* | common | Unknown | -684T>C | rs265981 | 5'UTR | - | - | x | expression | Bobb AJ , et al. (2005); Hettinger JA, et al. (2008) |
| *DRD1* | common | Unknown | *62C>T | rs686 | 3'UTR | - | - | x | expression | Bobb AJ , et al. (2005); Hettinger JA, et al. (2008) |
| | | | | | | | | | | |
| *DRD2* | common | Unknown | 71572C>T | rs2242592 | 5KB downstream variant | - | - | x | expression | Eicher JD and Gruen JR (2014) |
| *DRD2* | common | Unknown | -32+2047G>A | rs4581480 | intron1 | - | - | | expression | Eicher JD and Gruen JR (2014) |
| *DRD2* | common | Unknown | -32+21324G>A | rs4581480 | intron1 | - | - | | expression | Eicher JD and Gruen JR (2014) |
| *DRD2* | rare | De novo | 152G>A | | TM1 | missense | Gly51Asp | | deleterious variant | Deciphering Developmental Disorders Study (2014) |
| *DRD2* | common | Unknown | 286-2730C>T | rs1800498 | intron2 | - | - | x | expression | Hettinger JA , et al. (2012) |
| *DRD2* | common | Unknown | 290-2737C>T | | intron2 | - | - | | expression | Hettinger JA , et al. (2012) |
| *DRD2* | rare | Unknown | 1079G>A | rs138873192 | ICL3 | missense | Arg360His | x | signaling effector recruitment ; post-translational modification (deletion methylation ) | Klassen T , et al. (2011) |
| *DRD2* | rare | De novo | 1151G>A | | TM6 | stop gained | Trp384Ter | | loss of function | Willsey AJ , et al. (2017) |
| *DRD2* | rare | Unknown | 1228G>A | rs758683320 | TM7 | missense | Ala410Thr | x | deleterious variant | Klassen T , et al. (2011) |
| *DRD2* | rare | De novo | 1238G>A | | TM7 | stop gained | Trp413Ter | | loss of function | Willsey AJ , et al. (2017) |
| *DRD2* | common | Unknown | 113522272C>T | rs2514218 | intergenic | | - | | expression | Pardias AF , et al. (2018) |
| | | | | | | | | | | |
| *DRD3* | common | Unknown | 383+2327C>T | rs167771 | intron2 | - | - | | expression | Staal WG , et al. (2015); de Krom M , et al. (2008) |
| *DRD3* | rare | Unknown | 448G>A | rs1442197591 | TM4 | missense | Val150Met | | deleterious variant | Klassen T , et al. (2011) |
| *DRD3* | rare | Unknown | 987A>G | rs61735073 | TM6 | synonymous | Gln329= | | expression | Klassen T , et al. (2011) |
| | | | | | | | | | | |
| *GPR37* | common | Unknown | 699T>C | rs62638681 | N-term | synonymous | Leu16= | | expression | Fujita-Jimbo E , et al. (2012) |
| *GPR37* | rare | Unknown | 891T>C | rs149569764 | N-term | synonymous | Phe80= | | expression | Fujita-Jimbo E , et al. (2012) |
| *GPR37* | common | Unknown | 1311C>T | rs61744942 | N-term | synonymous | Ala220= | | expression | Fujita-Jimbo E , et al. (2012) |
| *GPR37* | rare | Familial | 1585_1587del | | TM2 | inframe deletion | Phe312del | | deleterious variant | Fujita-Jimbo E , et al. (2012) |
| *GPR37* | common | Unknown | 1698C>T | rs724356 | TM3 | synonymous | Thr349= | | expression | Fujita-Jimbo E , et al. (2012) |
| *GPR37* | common | Unknown | 1980G>C | rs3735270 | TM5 | synonymous | Leu443= | | expression | Fujita-Jimbo E , et al. (2012) |
| *GPR37* | rare | Familial | 2058C>G | rs200556919 | ICL3 | missense | Ile469Met | | signaling effector recruitment | Fujita-Jimbo E , et al. (2012) |
| *GPR37* | rare | Familial | 2324G>A | rs201551401 | H8 | missense | Arg558Gln | | tolerated variant | Fujita-Jimbo E , et al. (2012) |
| *GPR37* | common | Unknown | 2417C>T | rs149031046 | C-term | missense | Thr589Met | | signaling effector recruitment ; post- | Fujita-Jimbo E , et al. (2012) |
| | | | | | | | | | translational modification (deletion phosphorylation) | |
| | | | | | | | | | | |
| *GPR85* | common | Unknown | -171+414G>C | | intron2 | | - | | expression | Matsumoto M , et al. (2008) |
| *GPR85* | common | Unknown | -170-687G>C | | intron2 | | - | | expression | Matsumoto M , et al. (2008) |
| *GPR85* | rare | Familial | 1033T>C | VAR_074204 | TM4 | missense | Met152Thr | | deleterious variant | Fujita-Jimbo E , et al. (2015) |
| *GPR85* | rare | Familial | 1239G>T | VAR_074205 | ICL3 | missense | Val221Leu | | signaling effector recruitment | Fujita-Jimbo E , et al. (2015) |
| *GPR85* | common | Unknown | 2949A>G | | 3'UTR | | - | | expression | Matsumoto M , et al. (2008) |
| | | | | | | | | | | - |
| *GABBR2* | rare | De novo | 399C>T | rs754203594 | N-term | synonymous | Gly133= | | expression | Cappi C , et al. (2019) |
| *GABBR2* | rare | De novo | 1699G>A | rs922847767 | TM3 | missense | Ala567Thr | | deleterious variant | Chen S et al. (2021); Lopes F , et al. (2016); Yoo Y , et al. (2017); Takata A , et al. (2018); Alonso-Gonzalez A et al. (2021) |
| *GABBR2* | rare | De novo | 2077G>T | rs1554689320 | TM6 | missense | Gly693Trp | | deleterious variant | Hamdan FF , et al. (2017) |
| *GABBR2* | rare | De novo | 2084G>T | rs1554689319 | TM6 | missense | Ser695Ile | | deleterious variant | EuroEPINOMICS-RES Consortium , et al. (2014) |
| *GABBR2* | rare | De novo | 2114T>A | rs1554689315 | TM6 | missense | Ile705Asn | | deleterious variant | EuroEPINOMICS-RES Consortium , et al. (2014) |
| | | | | | | | | | | - |
| *GRM5* | common | Unknown | - | | - | copy number loss | - | | loss of function | Elia J , et al. (2011) |
| *GRM5* | rare | Unknown | 911+3G>A | | intron1 | | - | | expression | Kelleher RJ 3rd , et al. (2012) |
| *GRM5* | rare | Unknown | 5T>C | rs78375107 | N-term | missense | Val2Ala | x | tolerated variant; trafficking | Klassen T , et al. (2011) |
| *GRM5* | rare | Unknown | 87T>C | | N-term | synonymous | Ala29= | | expression | Kelleher RJ 3rd , et al. (2012) |
| *GRM5* | rare | Unknown | 381A>G | | N-term | synonymous | Val127= | | expression | Klassen T , et al. (2011) |
| *GRM5* | rare | Unknown | 412C>T | rs538043568 | N-term | missense | Arg138Cys | x | deleterious variant; trafficking; disulfide bridge; ligand binding | Klassen T , et al. (2011) |
| *GRM5* | rare | De novo | 523A>G | rs1941663554 | N-term | missense | Thr175Ala | | deleterious variant; trafficking; post-translational modification (deletion glycosylation); ligand binding | Guo H , et al. (2018) |
| *GRM5* | rare | Unknown | 657A>C | rs76850190 | N-term | synonymous | Thr219= | | expression | Klassen T , et al. (2011) |
| *GRM5* | rare | Unknown | 727G>T | rs199573699 | N-term | missense | Ala243Ser | | deleterious variant; trafficking; post-translational modification (extra glycosylation) | Kelleher RJ 3rd , et al. (2012) |
| *GRM5* | rare | Unknown | 846G>A | rs200298073 | N-term | synonymous | Thr282= | (+Thr282Met) | expression | Kelleher RJ 3rd , et al. (2012) |
| *GRM5* | rare | Unknown | 2630+10G>A | | intron2 | | - | | expression | Kelleher RJ 3rd , et al. (2012) |
| *GRM5* | rare | Unknown | 1167A>G | rs2306153 | N-term | synonymous | Thr389= | | expression | Kelleher RJ 3rd , et al. (2012) |
| *GRM5* | rare | Unknown | 1206C>T | rs61745770 | N-term | synonymous | Asn402= | | expression | Klassen T , et al. (2011) |
| *GRM5* | rare | Unknown | 1358C>T | rs61741175 | N-term | missense | Thr453Met | x (+Thr453Arg) | deleterious variant; trafficking; post-translational modification (glycosylation); ligand binding | Klassen T , et al. (2011) |
| *GRM5* | rare | Unknown | 1417G>C | rs75266285 | N-term | missense | Glu473Gln | x | deleterious variant; trafficking; ligand binding | Kelleher RJ 3rd , et al. (2012) |
| *GRM5* | rare | Unknown | 1949G>A | rs763815567 | TM3 | missense | Gly650Asp | (+Gly650Val) | deleterious variant | Klassen T , et al. (2011) |
| *GRM5* | rare | De novo | 2036_2038del | | ICL2 | inframe deletion | Lys679del | (+Val709Ile) | tolerated variant; post-translational modification (SUMOylati | Iossifov I , et al. (2012) |
| | | | | | | | | | on/ubiquitin ation) | |
| *GRM5* | rare | Unknown | 2127T>A | rs139898033 | TM4 | synonymous | Val709= | | expression | Kelleher RJ 3rd , et al. (2012) |
| *GRM5* | rare | Unknown | 2379T>C | rs764573716 | TM6 | synonymous | Phe793= | | expression | Kelleher RJ 3rd , et al. (2012) |
| *GRM5* | rare | Unknown | 2652G>A | | C-term | synonymous | Ala884= | | expression | Kelleher RJ 3rd , et al. (2012) |
| *GRM5* | rare | Unknown | 3027G>A | | C-term | synonymous | Ser1009= | | expression | Kelleher RJ 3rd , et al. (2012) |
| *GRM5* | rare | Unknown | 3503T>C | rs142032384 | C-term | missense | Leu1168Pr o | | signaling effector recruitment | Kelleher RJ 3rd , et al. (2012) |
| | | | | | | | | | | - |
| *GRM7* | rare | De novo | - | | - | copy number loss | - | | loss of function | Liu Y , et al. (2015) |
| *GRM7* | common | Unknown | - | | - | copy number loss | - | | loss of function | Elia J , et al. (2011) |
| *GRM7* | common | Unknown | 222C>T | rs3749380 | N-term | synonymous | Asn74= | | expression | Ohtsuki T , et al. (2008) |
| *GRM7* | rare | Familial | 461T>C | rs111416729 8 | N-term | missense | Ile154Thr | | tolerated variant; trafficking; post-translationa l modificatio n (glycosylatio | Charng WL , et al. (2016) |
| | | | | | | | | | n); ligand binding | |
| *GRM7* | common | Unknown | 519+18164C>T | rs17031835 | intron1 | | - | | expression | Ganda C , et al. (2009) |
| *GRM7* | common | Unknown | 736+53390G>A | rs1508724 | intron1 | | - | | expression | Kandaswamy R , et al. (2014) |
| *GRM7* | common | Unknown | 736+57886C>T | rs11710946 | intron1 | | - | | expression | Kandaswamy R , et al. (2014) |
| *GRM7* | common | Unknown | 736+63078A>G | rs6769814 | intron1 | | - | | expression | Kandaswamy R , et al. (2014) |
| *GRM7* | common | Unknown | 1033+45994C>G | rs12491620 | intron1 | | - | | expression | Shibata H , et al. (2009) |
| *GRM7* | common | Unknown | 1175-9823T>C | rs779867 | intron1 | | - | | expression | Yang Y and Pan C (2012); Noroozi R , et al. (2016) |
| *GRM7* | common | Unknown | 1376-3623C>T | rs6782011 | intron1 | | - | | expression | Yang Y and Pan C (2012) |
| *GRM7* | common | Unknown | 1515+34967T>G | rs1450099 | intron1 | | - | | expression | Shibata H , et al. (2009) |
| *GRM7* | rare | Unknown | 1558G>A | rs758720116 | N-term | missense | Ala520Thr | | tolerated variant; trafficking; post-translational modification (glycosylatio | Woodbury-Smith M et al. (2022) |
| | | | | | | | | | n); ligand binding | |
| *GRM7* | rare | Familial | 1757G>A | rs1695081736 | N-term | stop gained | Trp586Ter | | loss of function | Reuter MS , et al. (2017) |
| *GRM7* | rare | De novo | 1865G>A | rs775656438 | ICL1 | missense | Arg622Gln | | tolerated variant; post-translational modification (methylation) | Sanders SJ , et al. (2012) |
| *GRM7* | rare | Familial | 1972C>T | rs1114167300 | TM3 | missense | Arg658Trp | (+Arg658Gln) | deleterious variant | Charng WL , et al. (2016) |
| *GRM7* | rare | Familial | 2024C>A | rs1114167301 | TM3 | missense | Thr675Lys | | deleterious variant | Charng WL , et al. (2016) |
| *GRM7* | common | Unknown | 2451+45740C>T | rs3792452 | intron2 | | - | | expression | Park S , et al. (2013) |
| *GRM7* | rare | Familial | 2671G>A | rs746833089 | C-term | missense | Glu891Lys | x | signaling effector recruitment | Mitani T et al. (2021) |
| *GRM7* | common | Unknown | *401A>T | rs56173829 | 3'UTR | | - | | expression | Kandaswamy R , et al. (2014) |
| *GRM7* | common | Unknown | *472A>T | rs56173829 | 3'UTR | | - | | expression | Kandaswamy R , et al. (2014) |
| | | | | | | | | | | |
| *HTR1B* | common | Unknown | -182insCC | rs130057 | 2KB upstream variant | | - | x | expression | Orabona GM , et al. (2008) |
| *HTR1B* | rare | Unknown | 139C>T | | TM1 | missense | Pro47Ser | x | deleterious variant | Klassen T , et al. (2011) |
| | | | | | | | | | | |
| *OR1C1* | rare | Unknown | 44644, 577691, 1464677, 44643bp deletion | | copy number loss | - | | | loss of function | Bucan M , et al. (2009) |
| *OR1C1* | rare | Unknown | 1,399,900bp gain | | copy number gain | - | | | gain of function | PMID: 34664255 |
| *OR1C1* | rare | Unknown | 120G>C | rs1278811157 | TM1 | missense | Leu40Phe | x | deleterious variant | Woodbury-Smith M et al. (2022) |
| *OR1C1* | rare | Familial | 369T>A | rs554796998 | ICL2 | stop gained | Tyr123Ter | | loss of function | Ruzzo EK , et al. (2019) |
| | | | | | | | | | | |
| *OR2M4* | common | Unknown | *1488 | rs10888329 | 3'UTR | | - | x | expression | Kuo PH , et al. (2015) |
| *OR2M4* | common | Unknown | *4034 | rs6672981 | 3'UTR | | - | x | expression | Kuo PH , et al. (2015) |
| *OR2M4* | common | Unknown | 248247755G>C | rs4642918 | 3 Kb downstream | | - | x | expression | Kuo PH , et al. (2015) |
| *OR2M4* | common | Unknown | 248247787C>? | rs4397683 | 3 Kb downstream | | - | x | expression | Kuo PH , et al. (2015) |
| | | | | | | | | | | |
| *OR2T10* | rare | De novo | 795C>A | rs1660034711 | ECL3 | stop gained | Tyr265Ter | | loss of function | Iossifov I et al. (2014) |
| *OR2T10* | rare | Familial | 910del 7 aa | | C-term | frameshift variant | Met304Cys -Ter | | signaling effector recruitment | Codina-Sol M , et al. (2015) |
| | | | | | | | | | | |
| *OR52M1* | rare | De novo | 244A>T | rs140198502 0 | TM2 | stop gained | Lys82Ter | | loss of function | Iossifov I et al. (2014) |
| *OR52M1* | rare | Unknown | 456C>G | rs369159292 | TM4 | missense | Leu152= | | expression | Woodbury-Smith M et al. (2022) |
| *OR52M1* | rare | Unknown | 457C>T | rs145046460 | TM4 | missense | Arg153Trp | | deleterious variant | Woodbury-Smith M et al. (2022) |
| *OR52M1* | rare | Unknown | 458G>A | rs138834789 | TM4 | missense | Arg153Gln | | deleterious variant | Woodbury-Smith M et al. (2022) |
| *OR52M1* | rare | Unknown | 461G>A | | TM4 | missense | Gly154Asp | | deleterious variant | Woodbury-Smith M et al. (2022) |
| *OR52M1* | rare | Unknown | 463G>C | rs61747524 | TM4 | missense | Val155Leu | | deleterious variant | Woodbury-Smith M et al. (2022) |
| | | | | | | | | | | |
| *OXT* | common | Unknown | g.3067849G>A | rs6084258 | 2 KB upstream | | - | | expression | Francis SM , et al. (2016) |
| *OXT* | common | Unknown | g.3067891G>A | rs11697250 | 2 KB upstream | | - | | expression | Francis SM , et al. (2016) |
| *OXT* | common | Unknown | -90+120G>C | rs4813625 | 5'UTR | | - | | expression | Francis SM , et al. (2016) |
| *OXT* | rare | Familial | g.3072280T>C | rs115770980 2 | intron1 | splice site | - | | gain/loss of function | Ruzzo EK , et al. (2019) |
| *OXT* | common | Unknown | *450G>A | rs2770378 | 3'UTR | | - | | expression | Hovey D , et al. (2014); Chakrabarti B , et al. (2009) |
| *OXT* | common | Unknown | g.3081821G>T | rs2740204 | intergenic | | - | | expression | Francis SM , et al. (2016); Yrigollen CM , et al. (2008) |
| | | | | | | | | | | |
| *OXTR* | rare | Familial | | | | copy number loss | - | | loss of function | Gregory SG , et al. (2009) |
| *OXTR* | common | Unknown | -7103C>G | rs2270465 | upstream gene | | | | expression | Wermter AK , et al. (2009) |
| *OXTR* | common | Unknown | g.8770725T>C | rs2268498 | 2 KB upstream | | - | | expression | Montag C , et al. (2017) |
| *OXTR* | common | Unknown | -1710A>G | | 2 KB upstream | | - | | expression | Montag C , et al. (2017) |
| *OXTR* | common | Unknown | -238-131C>T | rs4564970 | intron2 | | - | | expression | Tansey KE , et al. (2010) |
| *OXTR* | common | Unknown | 652G>A | rs4686302 | TM5 | missense | Ala218Thr | x | tolerated variant | Francis SM , et al. (2016) |
| *OXTR* | common | Unknown | 690C>T | rs237902 | TM5 | synonymous | Asn230= | | expression | Ocakolu FT , et al. (2018) |
| *OXTR* | common | Unknown | 922+1417C>T | rs2268495 | intron3 | | - | | expression | Liu X , et al. (2010) |
| *OXTR* | common | Unknown | 922+4581T>C | rs53576 | intron3 | | - | | expression | Haram M , et al. (2015); Tops M , et al. (2011); Wu S , et al. (2005); |
| | | | | | | | | | | Baribeau DA , et al. (2017); Parker KJ , et al. (2014) |
| *OXTR* | common | Unknown | 922+6469A>G | rs237889 | intron3 | | - | | expression | Kranz TM , et al. (2016) |
| *OXTR* | common | Unknown | 922+667T>C | rs237897 | intron3 | | - | | expression | Kranz TM , et al. (2016) |
| *OXTR* | common | Unknown | 922+6724C>T | rs2254298 | intron3 | | - | | expression | Yang S , et al. (2017); Wu S , et al. (2005); Jacob S , et al. (2007); Liu X , et al. (2010); Francis SM , et al. (2016); Baribeau DA , et al. (2017); Parker KJ , et al. (2014); LoParo D and Waldman ID (2014) |
| *OXTR* | common | Unknown | 922+6906A>T | rs2268494 | intron3 | | - | | expression | Lerer E , et al. (2007) |
| *OXTR* | common | Unknown | 923-2132C>T | rs237887 | intron3 | | - | | expression | Liu X , et al. (2010); Baribeau DA , et al. (2017); LoParo D and |
| | | | | | | | | | | Waldman ID (2014) |
| *OXTR* | common | Unknown | 923-3676G>A | rs4686301 | intron3 | | - | | expression | Lerer E , et al. (2007) |
| *OXTR* | common | Unknown | 923-5488G>A | rs2268491 | intron3 | | - | | expression | Liu X , et al. (2010); LoParo D and Waldman ID (2014) |
| *OXTR* | common | Unknown | 923-5930A>G | rs2268493 | intron3 | | - | | expression | Francis SM , et al. (2016); Di Napoli A , et al. (2014); Campbell DB , et al. (2011); Yrigollen CM , et al. (2008) |
| *OXTR* | rare | Familial | 1126C>G | rs35062132 | C-term | missense | Arg376Gly | x | post-translational modification (methylation); signaling effector recruitment | Ma WJ , et al. (2013); Ohashi K et al. (2021) |
| *OXTR* | common | Unknown | *118C>A | rs1042778 | 3'UTR | | - | | expression | Lerer E , et al. (2007); Francis SM , et al. (2016); |
| | | | | | | | | | | Campbell DB , et al. (2011) |
| *OXTR* | common | Unknown | *939G>T | rs6770632 | 3'UTR | | - | | expression | Lerer E , et al. (2007) |
| *OXTR* | common | Unknown | *1078C>T | rs237884 | 3'UTR | | - | | expression | Harrison AJ , et al. (2015) |
| *OXTR* | common | Unknown | g.8749760G>A | rs7632287 | downstream gene | | - | | expression | Tansey KE , et al. (2010); Harrison AJ , et al. (2015); Tansey KE , et al. (2010); Campbell DB , et al. (2011); LoParo D and Waldman ID (2014) |

All variants of the GPCR and oxytocin genes were primarily extracted from the SFARI database (https://gene.sfari.org), reviewed in literature, found when available the corresponding Reference SNP (rs) number, compared to variants listed in the 1000 genomes project and/or GPCRdb database (GPCRdb.org) [12, 13] and referenced using RefSeqGene in the table. Variants were analysed for their localization, variant type and variant potential effect on protein function or expression (GPCRdb.org) [567]. Globally, ASD-associated variants in the extracellular domain (N-term, ECL1-3) impact post-translational modification, ligand binding and/or trafficking to the plasma membrane, in the transmembrane domain may lead to receptor unfolding or ligand binding impairment, and in the intracellular domain (ICL1-3, C-term) impact post-translational modification and/or signalling effector recruitment. (a) accession number AC025525.3, (b) from RefSeqGene.

## Table 3. ASD-linked behavioural phenotype in rodent models of the GPCRs included in the study

| Face and Predictive (induction) validity | | | | | | Construct validity | Predictive (remission) validity | | | |
|---|---|---|---|---|---|---|---|---|---|---|
| Gene | Animal model | Social behaviour | Communication | Stereotyped behaviour | Other behaviour or comorbid symptoms | | Ligands | GPCR structures | references | PMID |
| *ADORA 2A* | Knock-out mouse | NA | NA | NA | reduced spontaneous activity more pronounced during the dark phase; general motor impairment (accelerating rotarod), abnormal circadian rhythm; improved spatial memory (elevated Y-maze); increased aggression (resident-intruder test); decreased behavioral despair (tail suspension test, forced swim test); prepulse inhibition deficit (PPI test) | striatal D1R and D2R expression increased for mice generated in the CD1 background | agonist CGS21680 attenuates abnormally increased self-grooming in the BTBR T+ Itpr3tf/J (BTBR) mouse model of autism | A1, A2A | Lewis et al., 2019; Amodeo et al., 2018 | 31054946; 29193861 |
| | Transgenic rats | NA | NA | NA | impaired working memory (water maze, 6 arm radial tunnel maze, novel object recognition test) | overexpressing AD2AR in the brain; lower expression levels of D2Rs and mGlu5Rs only in striatum | | | Giménez-Llort et al., 2007 | 16824773 |
| | | | | | | | | | | |
| *ADORA3* | Knock-out mouse | NA | NA | NA | reduced spontaneous activity only during the dark phase and more pronounced in females; increased anxiety (elevated plus maze, ligh-dark box), increased behavioral despair (tail suspension test, forced swim test) | NA | NA | A1, A2A | Deckert., 1998 | 11281962 |
| | | | | | | | | | | |
| *ADRB2* | Knock-out mouse | NA | NA | NA | no basal change in locomotion, but increased cocaine-induced locomotion; increase level of anxieous-like behaviour: increase preference for the dark compartment and closed arm in light/Dark box and elevated plus maze tests ; antidepressant-like behavior : decrease immobility in tail suspension test; tends to have reduced motor strength in the hanging wire test; increased reward | reduction of b2-AR expression in the hippocampus and mPFC | propranolol improve verbal responses and social interactions and decrease anxiety in ASD patients | a1B, a2A, a2B, a2C, β1-, β2-, β3-adrenergic | Zhu et al., 2017; Wisely et al., 2014 | 28348522; 24626633 |
| | | | | | process in cocaine-indued CPP | | | | | |
| | | | | | | | | | | |
| *AGTR2* | Knock-out mouse | NA | NA | NA | impaired drinking response to water deprivation for 40 hours; decrease in the activity of mice 6 hours after exposure to new environment (photobeam cage system); increased blood pressure; decreased number of times mice crossed from one square of the open field to another (open field test) | increased number of cells in piriform cortex, hippocampus, amygdala, and different thalamic regions | NA | AT1, AT2 | von Bohlen und Halbach et al, 2001; Hein et al, 1995; Ichiki et al, 1995 | 11384784; 7477266; 7477267 |
| | | | | | | | | | | |
| *AVPR1A* | Knock-out mouse | decrease in the social behaviour (social interaction test); impaired social novelty (social discrimination test), deficit in social recognition (social recognition task); impaired social novelty (social discrimination test), deficit in social recognition (social recognition task); social aggression, anxiety-like behavior and social recognition are unaffected; deficit in social recognition (social recognition test) | NA | decrease in marble burrying; increase in stereotyped behaviour for WT mice after AVP injection (scratching, grooming) but not for V1AR KO mice | recuded anxious-like behaviour (elevated plus maze); failure of the V1AR KO mice to investigate novel odors (habituation/dishabituation to odors), shorter latency to attack; reduced anxious-like behaviour (elevated plus maze, dark/light box, open-field arena) | NA | V1aR antagonist injection leads to social behavioral defictis that are rescued by V1aR activation in the lateral septum alone; significantly lower creatinine kinase levels in the blood of KO mice; huge increase in scratching and autogrooming after AVP administration compared to a vehicle in WT mice only. No effect in KO mice | OTR, V2 | Egashira et al., 2007; Bielsky et al., 2005; Wersinger et al., 2006; Bielsky et al., 2004 | 17227684; 16102534; 17083331; 14647484 |
| | CRISPR/Cas9 V1A receptors in Syrian hamster | NA | NA | NA | fail to induce flank marking after AVP or Phe2OVT injection but exacerbed flank marking when | NA | NA | NA | Taylor et al., 2022 | 35512092 |
| | | | | | exposed to the odors of a same-sex conspecific (Odor-stimulated flank marking test) and increased aggression | | | | | |
| | | | | | | | | | | |
| *AVPR1B* | Knock-out mouse | social novelty deficit (social recognition test); impaired respond to social cues with less time to investigate bedding and no preference for one stimulus (Bedding preference); reduced maternal aggression behaviour (maternal aggression test); deficits in sociability and social novelty (social recognition test) | decrease in vocalizations (Resident-Intruder test), impaired numbers of calls during the second separation (pup separation vocalizations) | NA | fewer entries into the open arms (elevated plus maze), on day 1 decrease in aggression behaviour and more mounting behaviour (social dominance); reduced aggressive behaviour (Resident-Intruder Paradigm, Neutral Arena Paradigm); reduced aggressive behaviour (aggression test), no anxiety; decrease in defensive behaviour (defensive test) and aggression, no anxiety; expression of Avpr1b in the CA2 alone is sufficient to enable the Avpr1b KO mouse to exhibit social aggression; impaired episodic memory (Object recognition: “what– | Increase in cFos immunoreactivity after social exposure in both WT and KO mice | Avpr1b replacement in CA2 restores aggression, d[Leu4,Lys8] Avp or D3PVP induced a significant potentiation of EPSCs (for d[Leu4,Lys8] AVP, only the AMPA receptor-dependent component) | OTR, V2 | Caldwell et al., 2010; Wersinger et al., 2002; Wersinger et al., 2004; Wersinger et al., 2007; Pagani et al., 2015; Scattoni et al., 2008; DeVito et al., 2009 | 20298692; 12399951; 15555506; 17284170; 24863146; 18005969; 19261862 |
| | | | | | where–when" memory for objects), impaired at recalling object–odor paired associates (Object–trace–odor task) | | | | | |
| | | | | | | | | | | |
| *CHRM3* | Knock-out mouse | NA | NA | NA | Contextual fear memory deficits; decreased paradoxal sleep | increased dopamine release in striatal slices | NA | M1, M2, M3, M4, M5 | Leaderbr and et al., 2016 | 27918283; 12151512; 16110248 |
| | KI mouse carrying 15 Ser to Ala substitutions in the IL3 phosphorylation serine clusters | NA | NA | NA | Contextual fear memory deficits | lack of receptor internalisation and beta-arrestin recruitment | | | Poulin et al., 2010 | 20439723 |
| | | | | | | | | | | |
| *CNR1* | Knock-out mouse | Sociability deficits (3-chambered test and direct social interaction), Reduced social interactions (Social interaction in neutral cage), Increased social investigation and social discrimination | Altered ultrasonic vocalizations (Maternal seperation and direct social interaction) | NA | Normal anxiety or Anxious behavior; startle reactivity; deficits in spatial reversal learning | NA | endocannabinoid improve social behaviour including social anxiety and reward in BTBR and Fmr1 KO mice | CB1, CB2 | Fyke et al., 2021; Varvel and Lichtman ., 2002; Haller et al., 2002; Haller et al., 2004; Litvin et al., 2013 | 34173729; 12023519; 12405999; 15078564; 23647582 |
| | | | | | | | | | | |
| *CX3CR1* | Knock-out mouse | Reduced exploration of mother contained in a tube and reduced time investigating a social stimuli; subordination in social interaction | NA | increased time spent grooming | no defect in olfaction or memory; no effect of sensorigating induced by social isolation | only in microglia; decreased brain connectivity from prefrontal cortext; NFκB and pCREB/CREB ratio in the HPC, and an increase CX3CL1 and p-p38/p38 ERK ratio in the hippocampus | chemokine CX3CL1 | CCR1, CCR2, CCR5, CCR6, CCR7, CCR9, CXCR2, CXCR4 | | 33013335, 34145941, 24487234, 31446159 |
| | | | | | | | | | | |
| *DRD1* | Mutant rat carrying Ile116Ser | Impairments in sociability (reciprocal social interaction chambers test). Significant impairments in sociability and social novelty (modified T-maze test with pups). Impaired social odor recognition (olfactory habituation/dishabituation test) | Reduced vocalizations by pups and increased vocalizations by adult mutants | NA | regular earing in home cage reduced, rearing decreased (open field test) | Impaired transmembrane insertion of the receptor | NA | D1, D2, D3, D4 | Homberg et al., 2016 | 27483345 |
| | | | | | | | | | | |
| *DRD2* | Knock-out mouse | Deficits in sociability (3-chambered test). Decreased social interaction (Reciprocal social interaction test). Impaired sociability and social novelty (U-field assay). Impaired social odor recognition (olfactory habituation/dishabituation test) | NA | Increased grooming, decreased digging and rearing (Motor stereotypies test). Decreased marble-burying (Marble-burying test) | NA | IncreasedpERK1/2 and pCAMKIIa levels in the dorsal striatum and increased $mGlu_5$ receptor levels. | risperidone and aripiprazole are approved drug to treat compulsive behaviours and irritability in ASD ; cariprazine for potential improvement of sociability | D1, D2, D3, D4 | Lee et al., 2018 | 29027111 |
| | heterozygous mouse with early life stress | Deficits in sociability (3-chambered test). Decreased social interaction (Home- | NA | Increased grooming (Open field) | NA | Downregulation of the BDNF-TrkB pathway caused by | | | Lee and Han., 2019 | 31308794 |
| | | cage social interaction test) | | | | decreased levels of BDNF, TrkB, phospho-ERK1/2, and phospho-CREB in the dorsal striatum. | | | | |
| | Transgenic D2 over expression mouse | Impaired sociability in female mice (Interaction in female-female pairs) | Impaired female vocalization (interaction in female-female pairs) | NA | NA | D2 receptor overexpression in the striatum, including NAc, CPu and the olfactory tubercle | | | Kabitzke et al., 2015 | 26176662 |
| | | | | | | | | | | |
| *DRD3* | Knock-out mouse | NA | NA | NA | increased cocaine self-administration, and increased cocaine-seeking behavior; increased locomotion in a novel environment, mutant mice are more active in the first 5 min of testing in locomotor activity chambers; increased locomotion and rearing (open field test) | up-regulation of dopamine transporters in the striatum | aripiprazole is approved to treat irritability in ASD; cariprazine for potential improvement of sociability | D1, D2, D3, D4 | Song et al., 2012; Xu et al., 1997; Accili et al., 1996 | 23045656; 9354330; 8700864 |
| | | | | | | | | | | |
| *GPR37* | Transgenic mouse | NA | NA | NA | NA | 5-6 times overexpression using PDGFbeta promoter | potential peptides : head activator from hydra, neuroprotective and glioprotective prosaposin and prosaptide, regenerating islet-derived family member 4 (REG4) and the neuroprotectin D1 | NA | Imai et al., | 17889953 |
| | Transgenic mouse | NA | NA | Increased methamphetamine-induced sensitivity index (piloerection, circling, jumping) | increased locomotion, increased motor coordination in rotarod | 12 times overexpression using PrP promoter; altered striatum signalling and reduced number of neurons, impaired dopamine release and D1 and D2 GPCR activity | | | Imai et al., | 17889953 |
| | Knock-out mouse | NA | NA | no effet in methamphetamine-induced sensitivity index (piloerection, circling, jumping); obsessive compulsive behavior in marble burrying test | decreased basal and psychostimulant-induced locomotion, no impairment in motor coordination in rotarod, reduced D1/D2 antagonists-induced catalepsy; reduced colon motility ; either decreased or increased anxious-like behavior in elevated plus maze, forced swim test, fear conditioning, light-dark box; either slight decrease or no effect in long-term memory in novel object recognition test (and LTP); abnormal sensorimotor gating as reduced PPI acoustic startle and improved olfactory discrimination; lack of psychostimulant-conditioned place preference | altered striatum signalling and reduced number of neurons, impaired dopamine release and D1 and D2 GPCR activity | | | Imai et al., 2007; Marazziti et al., 2007; Lopes et al., 2015; Zhang et al., 2020; Mandillo et al., 2013; Veenit et al., 2021; Marazziti et al., 2011 | 17889953; 17519329; 25824528; 32292338; 23574697; 35008836; 21372109 |
| | | | | | | | | | | |
| *GPR85* | Transgenic mouse | decreased social interaction (less trimmed whiskers and reciprocal social interaction test) | NA | Delayed spontaneaous alternation in delayed Y maze test | decrease freezing in fear conditioning, no deficit in motor coordination or locomotion; impaired memory; abnormal sensorimotor gating as decrease PPI acoustic startle inhibition compared to WT; lower discrimination in spatial pattern separation test | two times overexpression, only in forebrain neurons via CAMKII promoter; rduced dendritic arborisation | NA | NA | Chen et al., 2012; Matsumoto et al., 2008 | 22697179; 18413613 |
| | Knock-out mouse | NA | NA | Improved spontaneaous alternation in delayed Y maze test | tendency for increased freezing in fear conditioning; no deficit in motor coordination or locomotion; increase PPI acoustic startle inhibition only at 73dB compared to WT; higher discrimination in spatial pattern separation test | | | | Chen et al., 2012; Matsumoto et al., 2008 | 22697179; 18413613 |
| | | | | | | | | | | |
| *GABBR 2* | Knock-out mouse | NA | NA | NA | Anxious-like and anti-depressent behaviour | NA | Agonists (R) Baclofen improves sociability in fragile X syndrome and 16p11.2 micro delition mice. STX209 reverses the social behavioral deficits and other ASD related symptoms in valproic acid model | $GABA_B$ subunits | Momber eau et al., 2005; Partyka et al., 2007; Stoppel et al., 2018; Jiang et al., 2022 | 15706241; 18195467; 28984295; 35244195 |
| | | | | | | | | | | |
| *GRM5* | Knock-out mouse | decreased social interaction | NA | Digging and marble burrying deficits | Decreased anxiety and Hyperactivity in open field and hypoactivity in elevated zero maze test and unaltered motor coordination on accelerating rotarod; Sensorimotor gating deficits in pre pulse inhibition; Deficits in extension of aversive learning and spatial learning | NA | NA | mGlu1, mGlu2, mGlu3, mGlu4, mGlu5, mGlu7 | Xu et al., 2021; Brody et al., 2004; Xu et al., 2009 | 34029630; 14699440; 19321764 |
| | Viral KD in mouse dorsal hippocampus | NA | NA | NA | Impaired spatial learning | NA | NA | | Tan et al., 2015 | 25957750 |
| | | | | | | | | | | |
| *GRM7* | Knock-out mouse | Social memory deficits | NA | NA | Anxiolytic behavior, decreased aversive learning, motor deficits, increased seizures and anti-depressive behavior | NA | NA | mGlu1, mGlu2, mGlu3, mGlu4, mGlu5, mGlu7 | Cryan et al., 2003; Fisher et al., 2020 | 12814372; 32248644 |
| | Mutant mouse carrying Ile154Thr | NA | NA | NA | context fear learning deficits, motor coordination deficits and seizures | Post-transcriptional regulation of mGlu7 expression loss | NA | | Fisher et al., 2021 | 33476302 |
| | | | | | | | | | | |
| *HTR1B* | transgenic mouse (floxed tetO1B transgenic mouse model) | NA | NA | NA | exacerbated impulsive behaviour (DRL operant paradigm, Go/No-Go task) and aggressivity (isolation-induced aggression assay ) : rescue of receptor expression in adulthood with doxycycline reverses the impulsive, but not the aggressive, phenotype | 5-HT1BR mRNA loss | NA | NA | Zhuang et al., 1999 | 10432489 |
| | homologous recombination (of what?) | NA | NA | NA | exacerbated aggressive behaviour (Resident-intruder aggression test) | | NA | NA | Ramboz et al., 1996 | 8788525 |
| | homologous recombination (of what?) | NA | NA | NA | higher consumption of ethanol (alcohol preference drinking test), reduced ethanol sensitivity (grid test) | NA | NA | NA | Crabbe et al., 1996 | 8782828 |
| | Knock-out mouse | increased social interest (approach, sniffing : duration and frequency) during Resident-Intruder Paradigm; increased in time spent outside the nest (mother's activity test) | decrease in vocalizations (isolation test) | increase in jumping (isolation test), tendency to increase grooming; increase in intense stereotypies (head weaving and circling) after acute cocaine administration but decrease in intense stereotypies after chronic cocaine administration (openfield test) | increased aggressive behaviour and hyperactivity (resident intruder test), higher heart rate and locomotor activity (baseline 24h cycle); increase in overall motor activity (rising, rolling) without locomotion problems (isolation test), decrease in anxious-like behaviour (elevated plus maze); decrease of anxious-like behaviour and increase in time spent exploring the object (object interaction test), better performance in spatial memory and reach the | autoradiography with (1251-labeled cyanopindolol) which specifically bind to 5-HTR1B receptors; increased AP-1 transcription complex in KO mice with most of them containing FosB. Several Fos-related proteins including the 35K–37K chronic FRAs were more abundant and more immunoreact | NA | 5-HT1A, 5-HT1B, 5-HT1D, 5-HT1E, 5-HT1F | Bouwknecht et al., 2001; Brunner et al., 1999; Malleret et al., 1999; Rocha et al., 1998; Saudou et al., 1994 | 11297712; 10443785; 10407051; 9603521; 8091214 |
| | | | | | platform faster (water maze memory learning); higher "break points" (self-administration of drugs test), increase in lomotor activity after acute and chronic cocaine administration (openfield test); exacerbated impulsive and aggressive behaviour (Resident-intruder aggression test, lomotion test) | ive in KO than in WT particularly in the nucleus accumbens | | | | |
| | | | | | | | | | | |
| *HTR2A* | Knock-out mouse | NA | NA | NA | decrease in anxious-like behaviour with no change in depression-related and fear-conditioned behaviors (openfield, elevated plus maze, dark/light choice test, novelty suppressed feeding) : restored by re-expression of receptors; comorbid symptoms : lower paneth cell density and enterocyte height, thinner | Cortical restoration of 5HT2AR function normalizes anxiety in KO mice | selective antagonists M100907 reduce repetitive grooming in the BTBR mouse model of ASD | 5-HT2A, 5-HT2B, 5-HT2C | Weisstaub et al., 2006; Fiorica-Howells et al., 2002 | 16873667; 11960784 |
| | | | | | muscle layers and intestinal wall | | | | | |
| | KO CORT mouse (administered with corticosterone) | NA | NA | decrease in grooming duration (splash test) | increase in time spent immobile (tail suspension test), increase in integrated emotionality z-scores (calculated from performance in tail suspension, elevated plus maze and splash tests) | NA | NA | NA | Petit et al., 2014 | 24801750 |
| | | | | | | | | | | |
| *HTR7* | Knock-out mouse | NA | NA | reduced compulsive behaviour in marble burying test | reduced immobility in tail suspension and forced swin test; less time and less frequent episode of rapid eye movement sleep; resistant to serotonin-induced hypothermia; increased anxious-like behaviours in contextual fear conditioning, shock-probe burying, novelty-suppressed feeding; reduction in seizure threshold; increased locomotor coordination | circadian rhythm shifts in hypothalamic slice recordings | antagonist SB-269970 and serotonin uptake inhibitor citalopram affects the immobility | NA, other serotonin subtypes | Hedlund et al., 2003; 2005; 2007 Guscott et al., 2005; Roberts et al., 2004; Witkin et al., 2007; Liu et al., 2009; Balcer et al., 2019 | 12529502; 16018977; 17267119; 15755477; 15078565; 17485199; 19458153; 30543903 |
| | | | | | | | | | | |
| *OXT* | Knock-out mouse | social memory deficits (memory task, Morris water maze, two-trial Y-maze task), social recognition deficit (Social recognition tests); no deficits in general social approach behaviors | decrease in pup vocalizations after separation | decrease in grooming (open field test) | increase in anxious-like behaviour in males (Open field activity test); decrease in aggressive behaviour (resident intruder test, openfield test); increase in aggressive behaviour after day 3 of food deprivation and water restriction (semi-natural environment test, Feeding aggression, Intruder aggression), pups immediately killed and cannibalized (maternal behaviour); increase in aggression frequency (solation-induced aggression, resident–intruder aggression), males are less reactive during an acoustic startle test | Lack of cFos immunoreactivity in amygdala | Administration of OXT and OXT antagonist ornithine vasotocin analog (OTA) intraventricularly and/or in local injection in amygdala, rescues the social deficit | | Ferguson et al., 2000; 2001; Crawley et al., 2007; De Vries et al., 1997; Ragnauth et al., 2005; Winslow et al., 2000 | 10888874; 11588199; 17420046; 9181490; 15924555; 10753584 |
| | | | | | | | | | | |
| *OXTR* | Knock-out mouse | Severe social behavior deficits (Reciprocal social interaction or 3-chamber test); Impaired sociability (social choice paradigm) and social novelty (social recognition task) ; longer latency to retrieve the pups and spent less time crouching over the pups with pups found scattered around the cage (pups retrieval test) but normal latency to approach the pups and sniffing, impairment of social discrimination | Deficits in vocalizations in pups; males emitted significantly fewer calls (vocalization test) | Increased self-grooming | Anxiety, increased aggression ; increase in agression behaviour (tested males are placed in a new cage with a low aggressive interactor), impaired in learning (T-maze task); increase in locomotor activity (vocalization test), increase in aggressive behaviour (resident-intruder test); increase in aggressive behaviour (resident intruder test) | reduced GABAergic presynapses in hippocampal neurons | Oxytocin and vasopressin intracerebral administration rescues the social behavioral deficits, which is prevented by V1A receptor antagonist | OTR, V2 | Lee et al., 2008; Pobbe et al., 2012; Takayanagi et al., 2005; Sala et al., 2010; Takayanagi et al., 2005; Dhakar et al., 2012 | 18356275; 22100185; 16249339; 21306704; 16249339; 22609339 |
| | Heterozygous mouse | Sociability and social novelty deficits (3-chamber test) | NA | | No other beahvioural effects | NA | NA | | Sala et al., 2013 | 22967062 |
| | Knock-out in prairie voles using CRISPR/Cas 9 | deficit in social novelty but no social deficit (Three Chamber Test) or maternal behaviour issue (Alloparental Behavior Test, Retrieved pups test) | NA | increase in repetitive behaviour (Marble Burying Test ) | NA | autoradiography with the specific ligands [125I]-OVTA | NA | | Horie et al., 2018 | 30713102 |

The phenotypes observed with the different rodent models for these GPCRs are broadly classified according to the validity criteria such as face and predictive (induction), construct and predictive (remission) validities which indicates the effect of behavioural phenotypes, signalling changes, ligand binding and structures. Behavioural domains are further classified as core ASD symptoms including social behaviour and communication, stereotyped or repetitive behaviour and other comorbid changes.